\documentclass{aa}  

\usepackage{graphicx}
\usepackage{txfonts}
\usepackage{txfonts}
\usepackage{color}

\begin{document} 

   \title{Measuring stellar granulation during planet transits}
%\subtitle{Using three-dimensional \textsc{Stagger}-grid simulations}
\titlerunning{Characterizing and quantifying stellar granulation during planet transits}

  \author{A. Chiavassa \inst{1}, A. Caldas\inst{2}, F. Selsis\inst{2}, J. Leconte\inst{2}, P. Von Paris\inst{2}, P. Bord\'e\inst{2}, Z. Magic\inst{3,4}, R. Collet\inst{5,6}, M. Asplund\inst{5}}
%\authorrunning{A. Chiavassa et al.}
\authorrunning{A. Chiavassa et al.}
\institute{Laboratoire Lagrange, Universit\'e C\^ote d'Azur, Observatoire de la C\^ote d'Azur, CNRS,
Blvd de l'Observatoire, CS 34229, 06304 Nice cedex 4, France \\
\email{andrea.chiavassa@oca.eu} 
\and 
Laboratoire d'astrophysique de Bordeaux, Univ. Bordeaux, CNRS, B18N, all\'ee Geoffroy Saint-Hilaire, 33615 Pessac, France
\and
Niels Bohr Institute, University of Copenhagen, Juliane Maries Vej 30, DK--2100 Copenhagen, Denmark  
\and 
Centre for Star and Planet Formation, Natural History Museum of Denmark, University of Copenhagen, {\O}ster Voldgade 5-7, DK--1350 Copenhagen, Denmark
\and
Research School of Astronomy $\&$ Astrophysics, Australian National University, Cotter Road, Weston ACT 2611, Australia
\and
Stellar Astrophysics Centre, Department of Physics and Astronomy, Ny Munkegade 120,  Aarhus University, DK-8000 Aarhus C, Denmark}
 \date{...; ...}

% \abstract{}{}{}{}{} 
% 5 {} token are mandatory
 
  \abstract
    {Stellar activity and convection-related surface structures
might cause bias in planet detection and characterization that
use these transits. Surface convection simulations help to quantify the granulation signal.}
    {We used realistic three-dimensional (3D) radiative hydrodynamical (RHD) simulations from the \textsc{Stagger} grid and synthetic images computed with the radiative transfer code {{\sc Optim3D}} to model the transits of three prototype planets: a hot Jupiter, a hot Neptune, and a terrestrial planet.}
{We computed intensity maps from RHD simulations of the Sun and
a K-dwarf star at different wavelength bands from optical to far-infrared that cover the range of several ground- and space-based telescopes which observe exoplanet transits. We modeled the transit using synthetic stellar-disk images obtained with a spherical-tile imaging method and emulated the temporal variation of the granulation intensity generating random images covering a granulation time-series of 13.3 hours. We measured the contribution of the stellar granulation  on the light curves during the planet transit.}
    {We identified two types of granulation noise that act simultaneously during the planet transit: (i) the intrinsic change in the granulation pattern with timescale (e.g., 10 minutes for solar-type stars assumed in this work) is smaller than the usual planet transit ($\sim$hours as in our prototype cases), and (ii) the fact that the transiting planet occults isolated regions of the photosphere that differ in local surface brightness as a result of convection-related surface structures. First, we showed that our modeling approach returns granulation timescale fluctuations that are comparable with what has been observed for the Sun. Then, our statistical approach shows that the granulation pattern of solar and K-dwarf-type stars have a non-negligible effect of the light curve depth during the transit, and, consequentially on the determination of the planet transit parameters such as the planet radius (up to 0.90$\%$ and $\sim0.47\%$ for terrestrial and gaseous planets, respectively). We also showed that larger (or smaller) orbital inclination angles with respect to values corresponding to transit at the stellar center display a shallower transit depth and longer ingress and egress times, but also granulation fluctuations that are correlated to the center-to-limb variation: they increase (or decrease) the value of the inclination, which amplifies the fluctuations. The granulation noise appears to be correlated among the different wavelength ranges either in the visible or in the infrared regions.}
  {The prospects for planet detection and characterization with transiting methods are excellent with access to large amounts of data for stars. The granulation has to be considered as an intrinsic uncertainty (as a result of stellar variability) on the precise measurements of exoplanet transits of planets. The full characterization of the granulation is essential for determining the degree of uncertainty on the planet parameters. In this context, the use of 3D RHD simulations is important to measure the convection-related fluctuations. This can be achieved by performing precise and continuous observations of stellar photometry and radial velocity, as we explained with RHD simulations, before, after, and during the transit periods.}
  

   \maketitle
%
%________________________________________________________________

\section{Introduction}

\begin{table*}
\centering
\begin{minipage}[t]{\textwidth}
\caption{3D RHD simulations from \textsc{Stagger} grid.}             % title of Table
\label{simus}      % is used to refer this table in the text
\centering                          % used for centreing table
\renewcommand{\footnoterule}{} 
\begin{tabular}{c c c c c c c c c}        % centreed columns (4 columns)
\hline\hline                 % inserts double horizontal lines
$<T_{\rm{eff}}>$\footnote{Horizontal and temporal average of the emerging effective temperatures from \cite{2013A&A...557A..26M}} & [Fe/H]  & $\log g$ & $x,y,z$-dimensions & $x,y,z$-resolution   & $\rm{M}_{\star}$ & $\rm{R}_{\star}$ & granule size\footnote{approximate granulation size from \cite{2014arXiv1405.7628M} divided by the stellar radius. See also Fig.~\ref{mu1}}  & Number of tiles \footnote{$N_{\rm{tile}} = \frac{\pi \cdot \rm{R}_\star}{x,y\rm{-dimension}}$} \\
$[\rm{K}]$ & & [cgs]  & [Mm]  & [grid points]   & [$\rm{M}_\odot$] & [$\rm{R}_\odot$] & [$10^{-3}$] & over the diameter\\
\hline
5768 (Sun) & 0.0 & 4.4 &  7.76$\times$7.76$\times$5.20 & 240$\times$240$\times$240 & 1.0 & 1.00 &  4.5  & 286\\
4516 (K~dwarf) & 0.0 & 4.5 & 4.00$\times$4.00$\times$3.17 & 240$\times$240$\times$240 & 0.7\footnote{Fig.~1 of \cite{2013A&A...557A..26M}} & 0.78 & 5.1 & 427\\
\hline\hline                          % inserts single horizontal line
\end{tabular}
\end{minipage}
\end{table*}

Among the different methods used to detect exoplanets, the transit method is a very successful technique: 1147 planets and 3787 transit candidates have been confirmed with it \citep[as of November 2015 from http://exoplanets.org, ][]{2011PASP..123..412W}. A transit event occurs when the planet crosses the line of sight between the star and the observer, thus occulting part of the star. This creates a periodic dip in the brightness of the star. The typical stellar light blocked is $\sim1\%$, 0.1$\%$, and 0.01$\%$ for Jupiter-, Neptune- and Earth-like planets transiting in front of a Sun-like star, respectively \citep{1984Icar...58..121B}, making the detection very challenging, in particular for Earth-like planets. During the transit, the flux decrease is proportional to the squared ratio of planet and stellar radii. For sufficiently bright stars, the mass can also be measured from the host star's radial velocity semi-amplitude \citep{2012A&A...538A...4M}. When
the mass and radius of an exoplanet are known, its mean density can also be deduced and provide useful information for the physical formation processes. Today and in the near future, the prospects for planet detection and characterization with the transiting methods are excellent with access to a large amount of data coming, for instance, from the NASA missions Kepler \citep{2010Sci...327..977B} and TESS \citep[Transit Exoplanet Survey Satellite,][]{2010AAS...21545006R}, or from the ESA missions PLATO 2.0 \citep[PLAnetary Transits and Oscillation of stars,][]{2014ExA...tmp...41R} and CHEOPS \citep[CHaracterizing ExOPlanet Satellite, ][]{2013EPJWC..4703005B}. \\

Space- and ground-based telescopes used for transit photometry require high photometric precision to provide accurate planetary radii, masses, and ages. Moreover, transit photometry also needs continuous time series data over an extended period of time. Earth-sized planets are the most challenging targets: if the radius of the Earth is approximately 1/100 that of the Sun, then a transit of the Sun by Earth blocks $\sim10^{-4}$ of the solar flux, in addition to the challenge of the limited number of photons arriving from a faint star. For all these reasons, it is necessary to go to space to monitor the target fields continuously with minimal interruptions.\\

However, with improved photometric precision, additional sources of noise that are due to the presence of stellar surface inhomogeneities such as granulation, will become relevant, and the overall photometric noise will be less and less dominated by pure photon shot noise. The Sun's total irradiance varies on all timescales relevant for transit surveys, from minutes to months \citep{2004A&A...414.1139A}. In particular, granulation analysis of SOHO quiet-Sun data shows that the photometric variability ranges from 10 to 50 part-per-million (ppm) \citep{2002ApJ...575..493J,1997SoPh..170....1F}. The granulation was observed for the first time on the Sun by \cite{1801RSPT...91..265H}, but \cite{1864MNRAS..24..161D} coined the term granules. The granulation pattern is associated with heat transport by convection, on horizontal scales on the order of a thousand kilometers \citep{2009LRSP....6....2N}. The bright areas on the stellar surfaces, the granules, are the locations of upflowing hot plasma, while the dark intergranular lanes are the locations of downflowing cooler plasma. Additionally, the horizontal scale on which radiative cooling drives the convective motions is linked with the granulation diameter \citep{1990A&A...228..155N}. Stellar granulation manifests either on 
spatially resolved (e.g., images of the solar disk) or unresolved observables such as spectral line profiles in terms of widths, shapes, and strengths. The best observational evidence comes from unresolved spectral lines because they combine important properties such as velocity amplitudes and velocity-intensity correlations, which produce line broadening. This is interpreted as the Doppler shifts arising from the convective flows in the solar photosphere and solar oscillations \citep{2000A&A...359..729A,2009LRSP....6....2N}. Similarly, correlations of velocity and temperature cause characteristic asymmetries of spectral lines as well as net blueshifts for main-sequence stellar types \citep{1987A&A...172..211D,2005oasp.book.....G}.

The purpose of this work is to study the impact of stellar granulation on the transit shape and retrieved planetary parameters (e.g., radius). We considered three prototypes of planets with different sizes and transit time lengths corresponding to a hot Jupiter, a  hot Neptune, and a terrestrial planet. We used theoretical modeling of stellar atmospheres where the multidimensional radiative hydrodynamic equations are solved and convection emerges naturally. These simulations take surface inhomogeneities (i.e., the granulation pattern) and velocity fields into account in a self-consistent manner. They cover a substantial portion of the Hertzsprung-Russell diagram \citep{2013A&A...557A..26M,2009MmSAI..80..711L,2013ApJ...769...18T}, including the evolutionary phases from the main sequence over the turnoff up to the red giant branch for low-mass stars.

\begin{figure}
   \centering
   \begin{tabular}{c}  
                        \includegraphics[width=0.9\hsize]{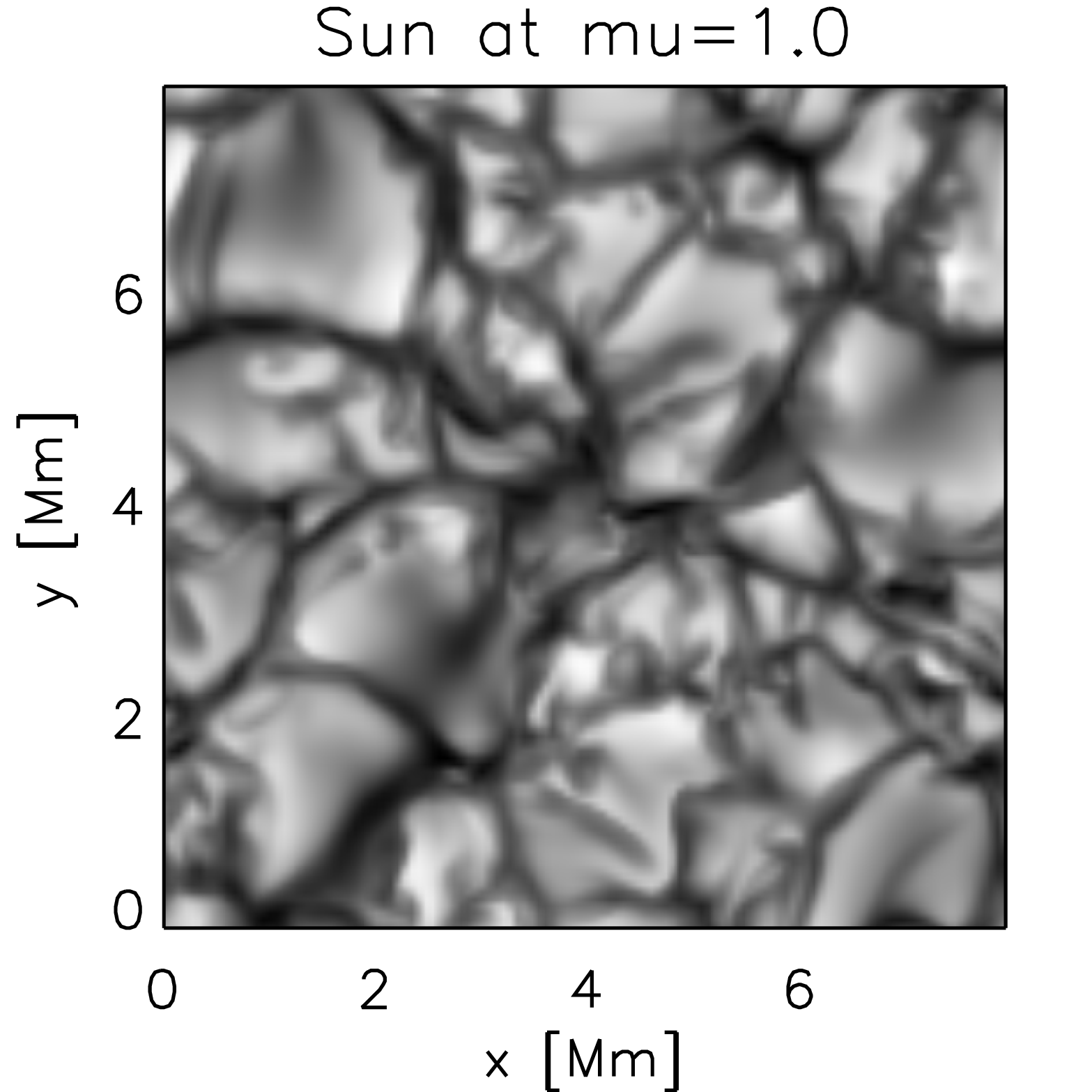}   \\
                     \hspace*{-2.96cm}   \includegraphics[width=0.45\hsize]{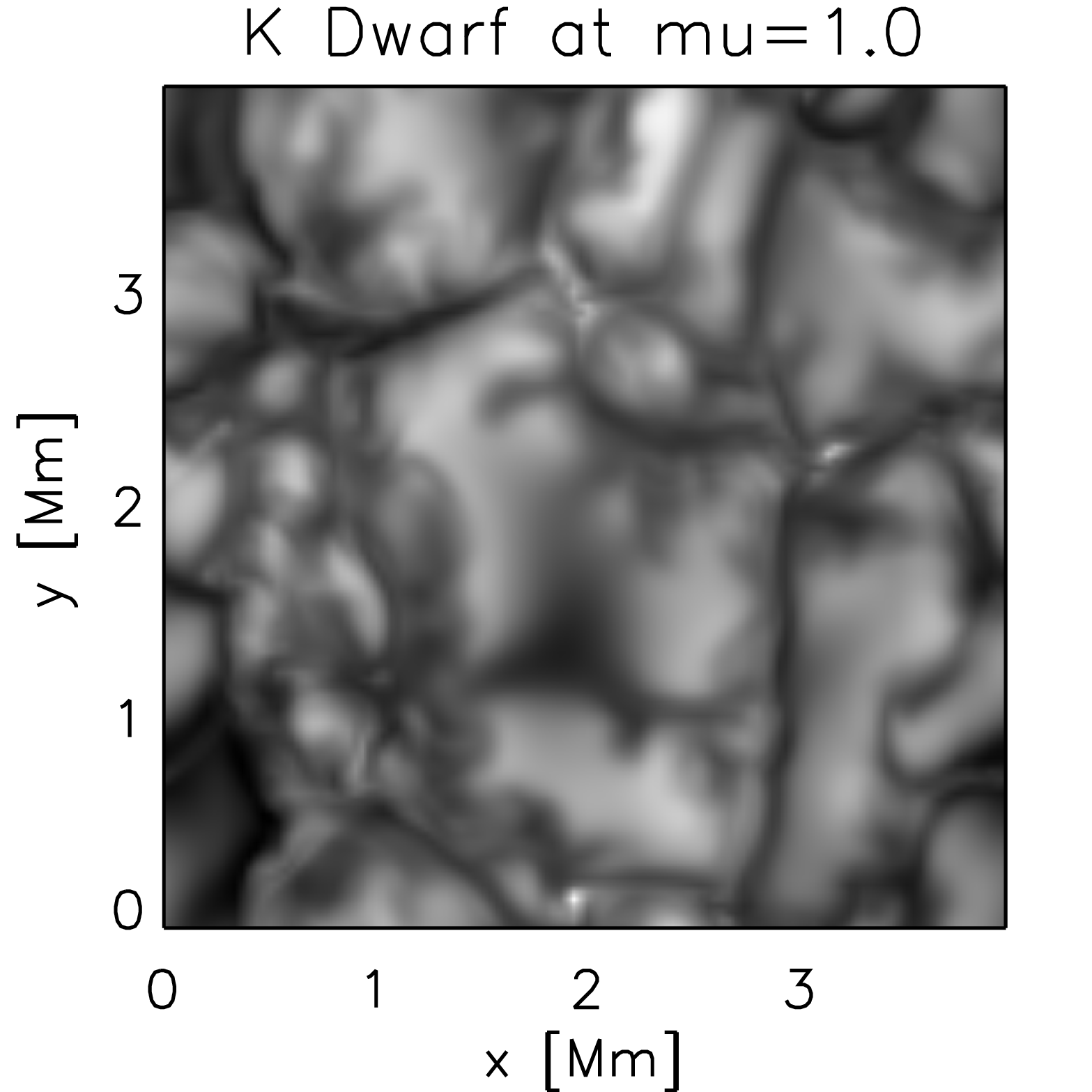} 
               \end{tabular}
      \caption{Intensity maps computed at [7620-7640] $\AA$ (Table~\ref{wavelengths}) of the 3D RHD simulations of Table~\ref{simus} and for the vertical direction ($\mu=1.0$). The intensity ranges from [$1.56$--$2.76]\times10^6$\,erg\,cm$^{-2}$\,s$^{-1}$\,{\AA}$^{-1}$ for the Sun (top) and from [$0.68$--$1.10]\times10^6$\,erg\,cm$^{-2}$\,s$^{-1}$\,{\AA}$^{-1}$ for the K~dwarf (bottom). The size ratio between the two images corresponds approximatively to the numerical box sizes.}
        \label{mu1}
   \end{figure}

\section{Stellar atmospheres and radiative transfer}

\begin{figure*}
   \centering
   \begin{tabular}{cc}  
                        \includegraphics[width=0.4\hsize]{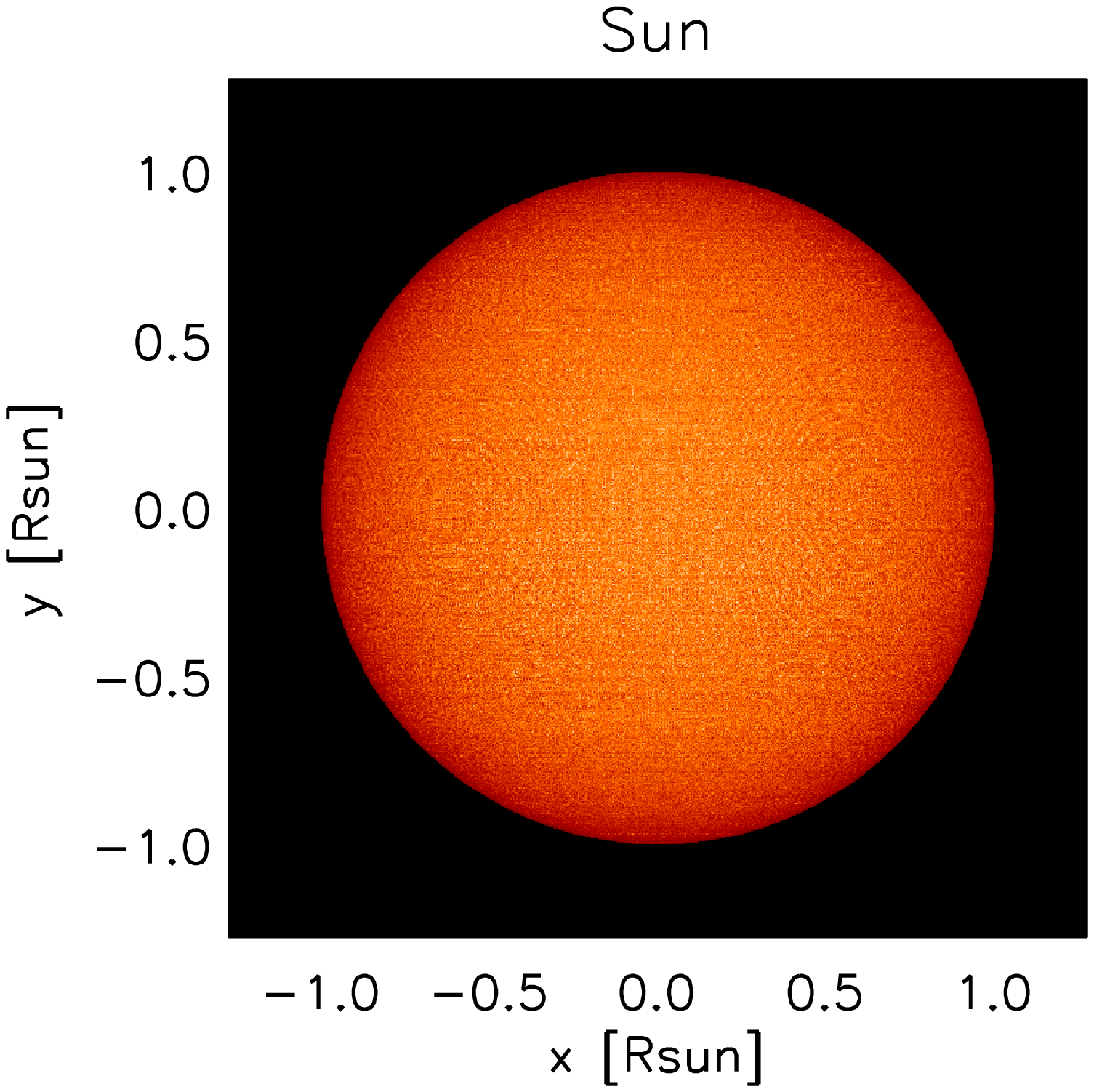}   
                        \includegraphics[width=0.4\hsize]{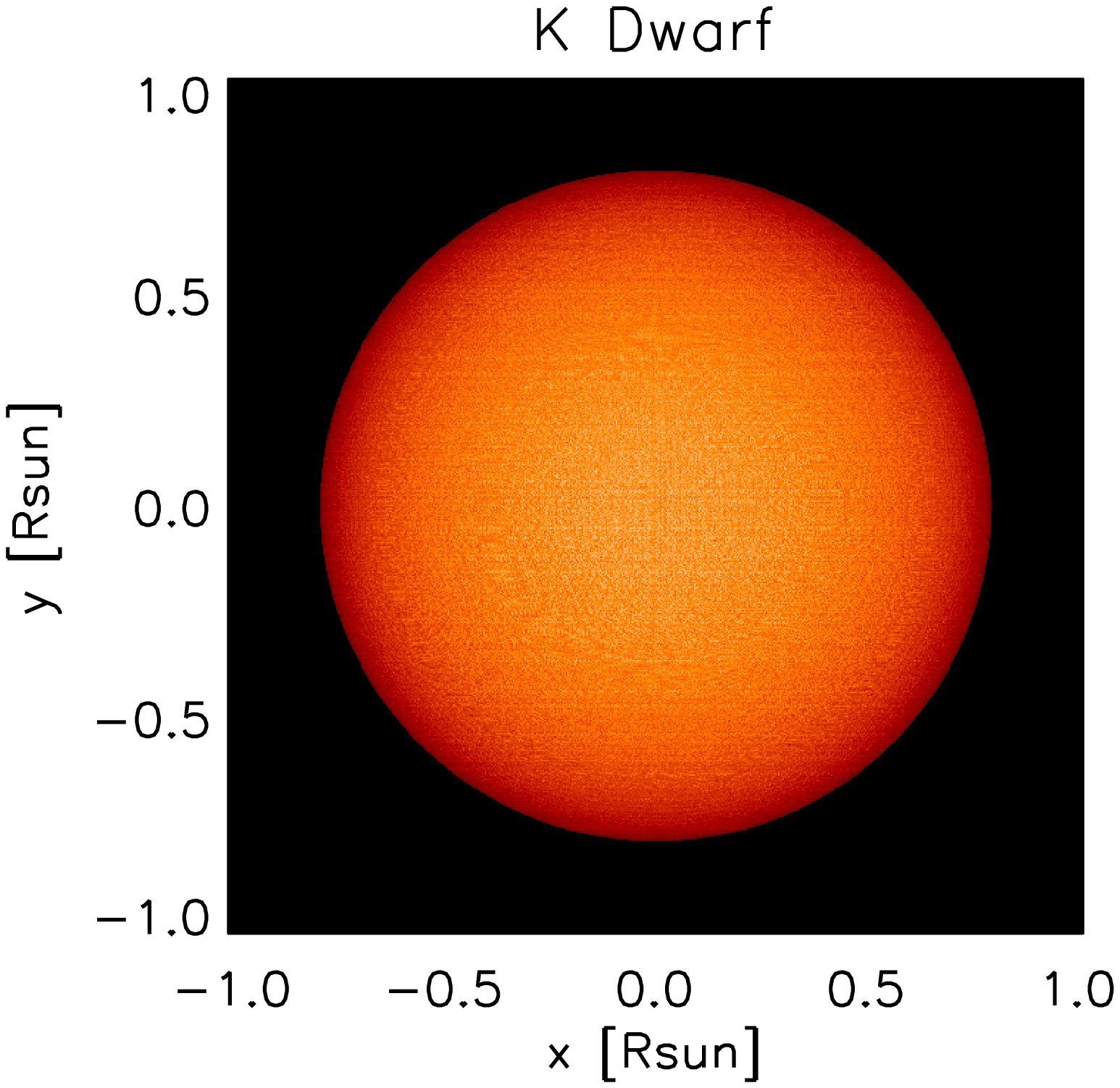} \\
        \end{tabular}
      \caption{Representative synthetic solar-disk image computed at [7620-7640] $\AA$ (Table~\ref{wavelengths}) of the 3D RHD simulations of Table~\ref{simus}. The intensity range is  [$0.0$--$2.86]\times10^6$\,erg\,cm$^{-2}$\,s$^{-1}$\,{\AA}$^{-1}$ for the Sun and [$0.0$--$1.82]\times10^6$\,erg\,cm$^{-2}$\,s$^{-1}$\,{\AA}$^{-1}$ for the K-dwarf star. We generated 80 different synthetic solar-disk images to account for a granulation time-series of 800 minutes (13.3 hours).}
        \label{starnoplanet}
   \end{figure*}

We used the simulations (Table~\ref{simus}) from the \textsc{Stagger}
grid of realistic 3D RHD simulations of stellar convection for cool stars \citep{2013A&A...557A..26M}. This grid is computed using the \textsc{Stagger} code (originally developed by Nordlund $\&$ Galsgaard 1995\footnote{http://www.astro.ku.dk/$\sim$kg/Papers/MHD\_code.ps.gz}, and continuously improved over the years by its user community). In a Cartesian box located around the optical surface (i.e., $\tau\sim1$), the code solves the time-dependent equations for conservation of mass, momentum, and energy coupled to a realistic treatment of the radiative transfer.  The simulation domains are chosen large enough to cover at least ten pressure scale heights vertically and to allow for about ten granules to develop at the surface; moreover, there are periodic boundary conditions horizontally and open boundaries vertically. At the bottom of the simulation, the inflows have a constant entropy, and the whole bottom boundary is set to be a pressure node for p-mode oscillations.  The simulations employ realistic input physics:  the equation of state is an updated version of the one described
by \cite{1988ApJ...331..815M}, and the radiative transfer is calculated for a large number over wavelength points merged into 12 opacity bins \citep{1982A&A...107....1N,2000ApJ...536..465S,2013A&A...557A..26M}. They include continuous absorption opacities and scattering coefficients from \cite{2010A&A...517A..49H} as well as line opacities described in \cite{2008A&A...486..951G}, which in turn are based on the VALD-2 database \citep{2001ASPC..223..878S} of atomic lines. The abundances employed in the computation are the solar chemical composition by \cite{asplund09}.

Theses simulations have been used to compute synthetic images with the pure-LTE radiative transfer code \textsc{Optim3D} \citep{2009A&A...506.1351C}. The code takes into account the Doppler shifts that are due to convective motions. The radiative transfer equation is solved monochromatically using pre-tabulated extinction coefficients as a function of temperature, density, and wavelength. \textsc{Optim3D} uses lookup tables with the same chemical compositions as the 3D RHD simulations as well as the same extensive atomic and molecular continuum and line opacity data as the latest generation of MARCS models \citep{2008A&A...486..951G}. The microturbulence is assumed to be zero \citep[i.e., the non-thermal Doppler broadening of spectral lines is the consequence of the self-consistent velocities in the simulations, ][]{2000A&A...359..729A} and the temperature and density ranges spanned by the tables are optimized for the values encountered in the RHD simulations. The detailed methods used in the code are explained in \cite{2009A&A...506.1351C,2010A&A...524A..93C}. 

\section{Stellar image disks to model the transits}
   
   \begin{figure}
   \centering
   \begin{tabular}{ccc}  
                        \includegraphics[width=1.00\hsize]{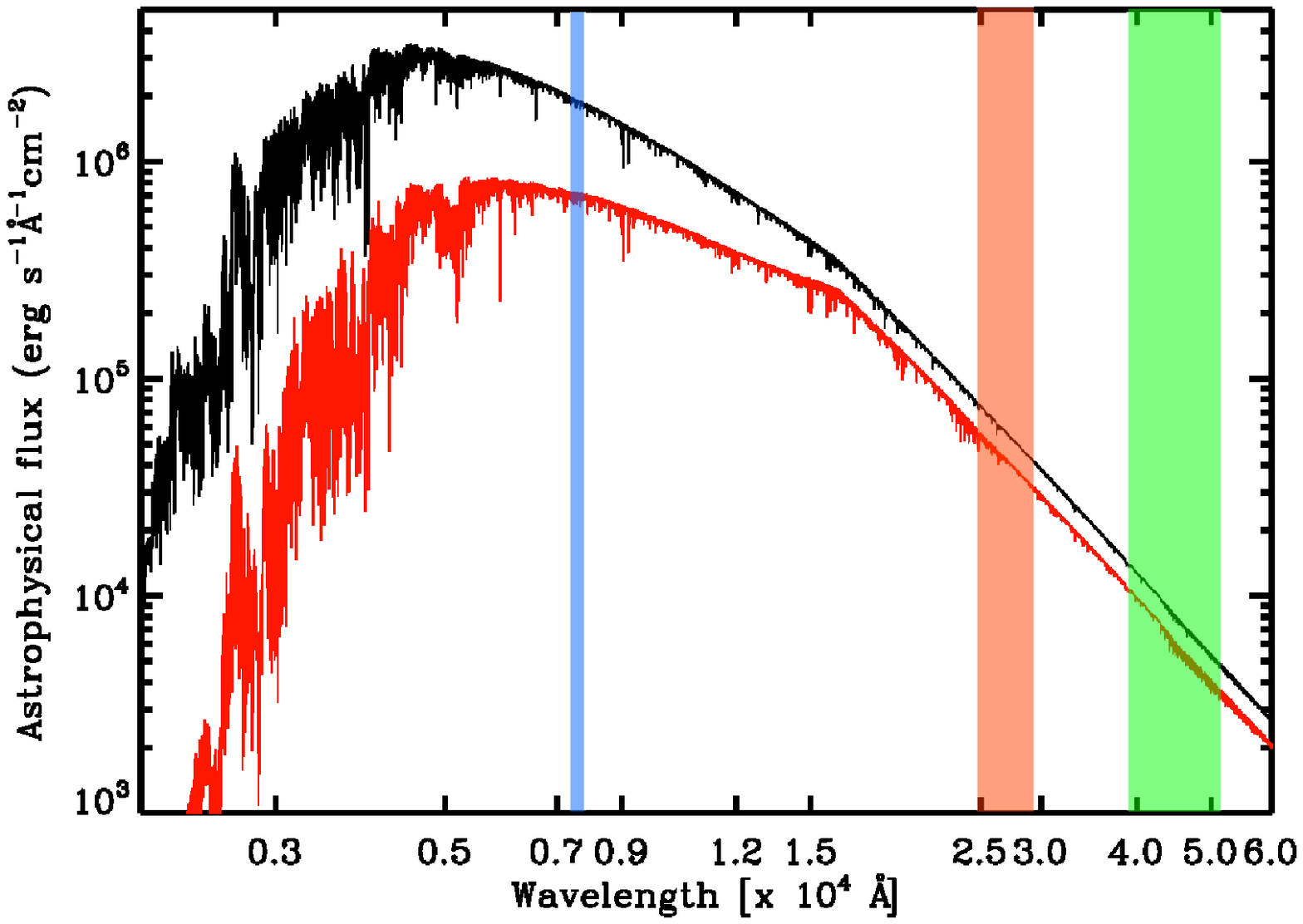} 
   \end{tabular}
         \caption{Synthetic spectra computed for the Sun (black curve) and the K~dwarf (red curve) from the optical to the infrared with a constant resolution of $\lambda/\Delta\lambda$=20000. The blue shading indicates the [7500-7700] \AA\ region used in Table~\ref{wavelengths}, the red shading the [25000-29000] \AA, and the green shading the [39000-51000] \AA. Note the use of the astronomical flux (i.e., the flux divided by a factor $\pi$) such that the values of the flux and intensity are the same. Note also the logarithmic x- and y-axis scale.} 
        \label{spectra}
   \end{figure}

We employed the tiling method explained in \cite{2010A&A...524A..93C} and used in \cite{2012A&A...540A...5C,2014A&A...567A.115C,2015A&A...576A..13C}. We used \textsc{Optim3D} to compute intensity maps \citep[see an illustrative image for $\mu$=1.0 in Fig.~\ref{mu1} and more examples in ][]{2012A&A...540A...5C} at different integrated wavelength bands (Table~\ref{wavelengths}) covering spectral regions in the optical, which are characterized by a higher density of transition lines toward the infrared part of the spectrum (Fig.~\ref{spectra}).

\begin{table}
\centering
%\begin{minipage}[t]{\textwidth}
\caption{Integrated wavelength bands computed.}             % title of Table
\label{wavelengths}      % is used to refer this table in the text
\centering                          % used for centreing table
\renewcommand{\footnoterule}{} 
\begin{tabular}{c c c }        % centreed columns (4 columns)
\hline\hline                 % inserts double horizontal lines
Wavelength  & Number of   & $\Delta\lambda$   \\
   range, [\AA ]             &  bands          & per band [\AA ] \\
\hline
7500-7700 & 10 & 20  \\
25000-29000 & 16 & 250 \\
39000-51000 & 1 & 12000 \\
\hline\hline                          % inserts single horizontal line
\end{tabular}
        
%\end{minipage}
\end{table}

\begin{figure}
   \centering
   \begin{tabular}{ccc}  
                        \includegraphics[width=0.5\hsize]{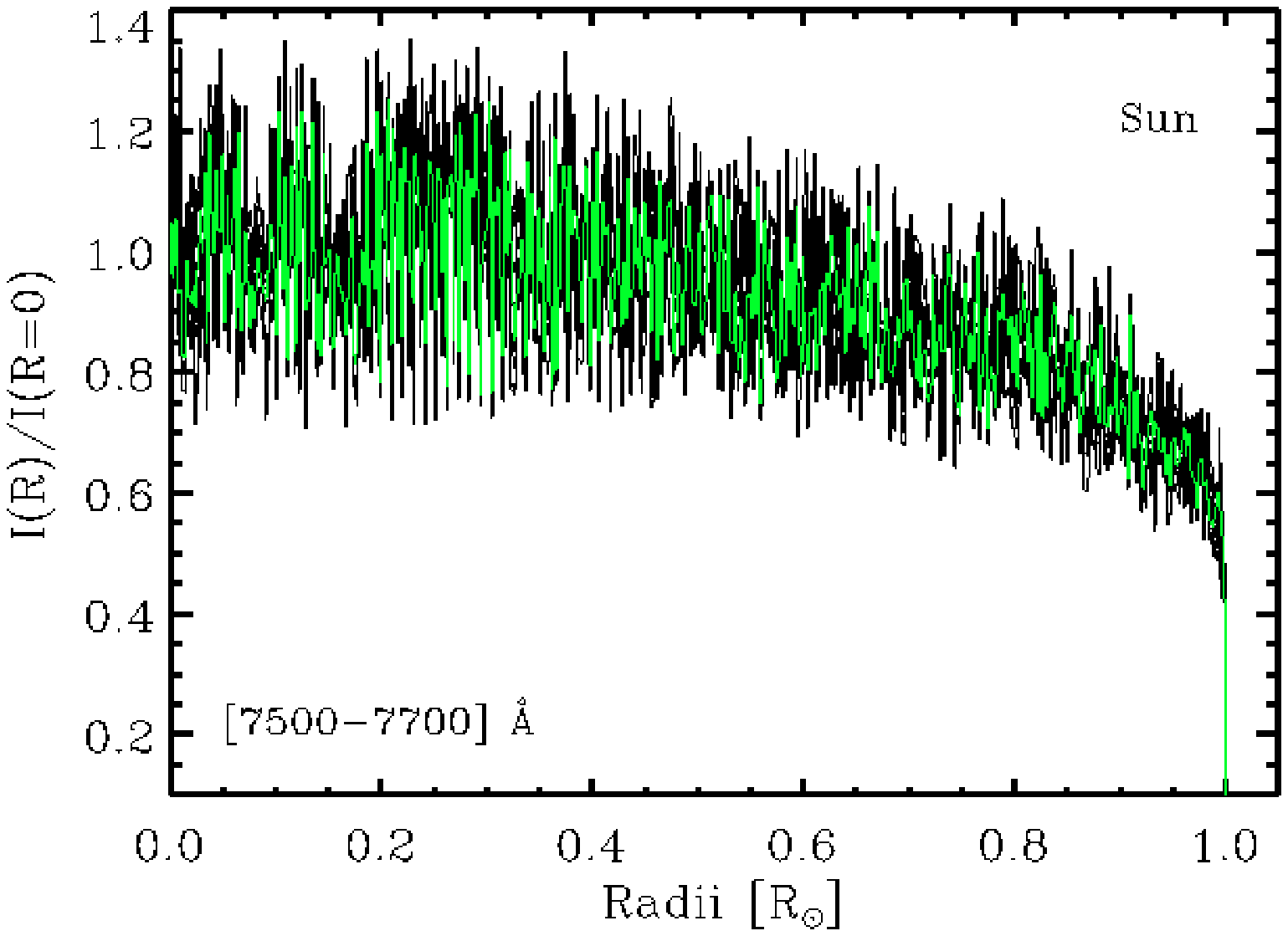} 
                        \includegraphics[width=0.5\hsize]{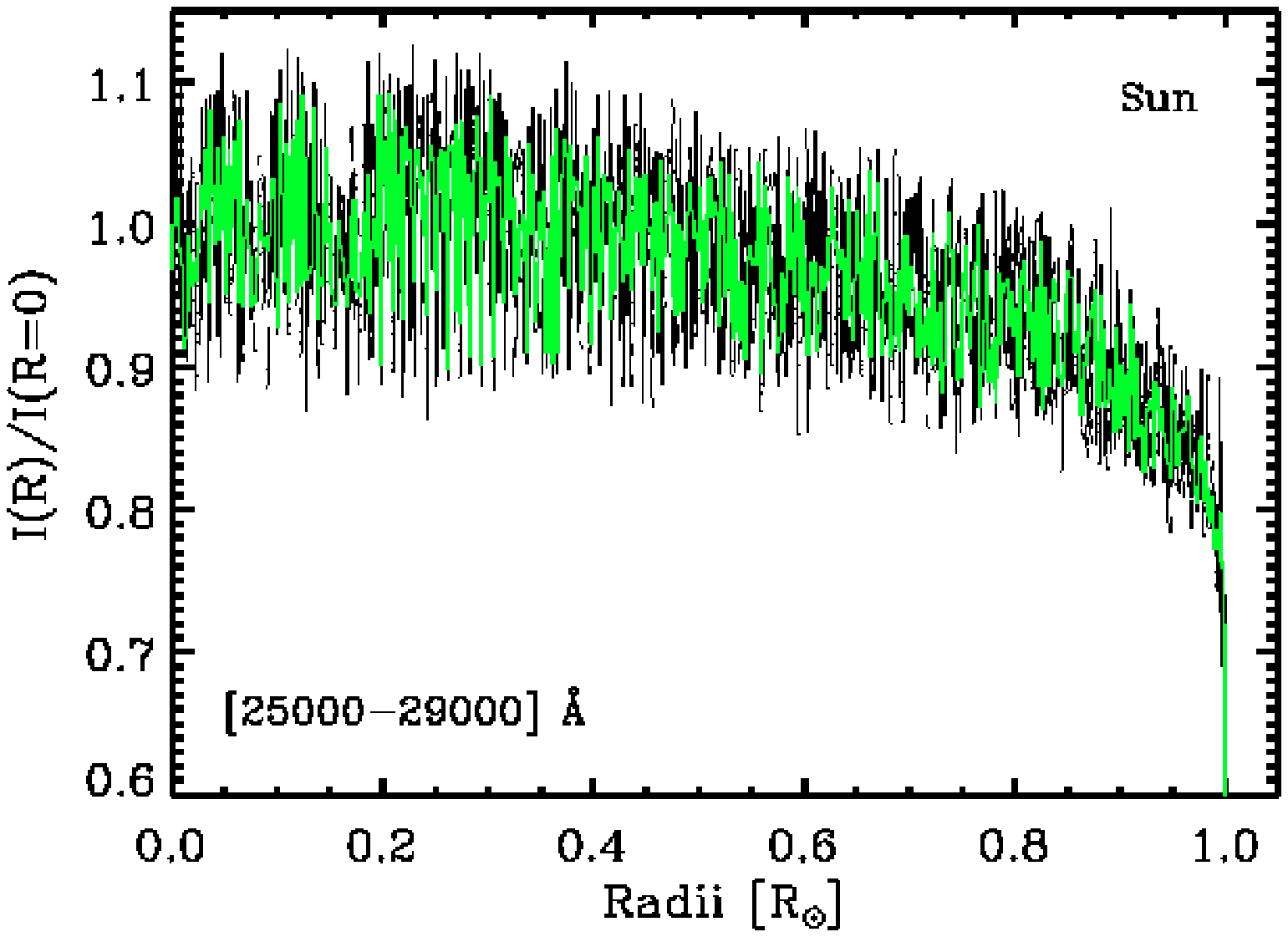}\\
                        \includegraphics[width=0.5\hsize]{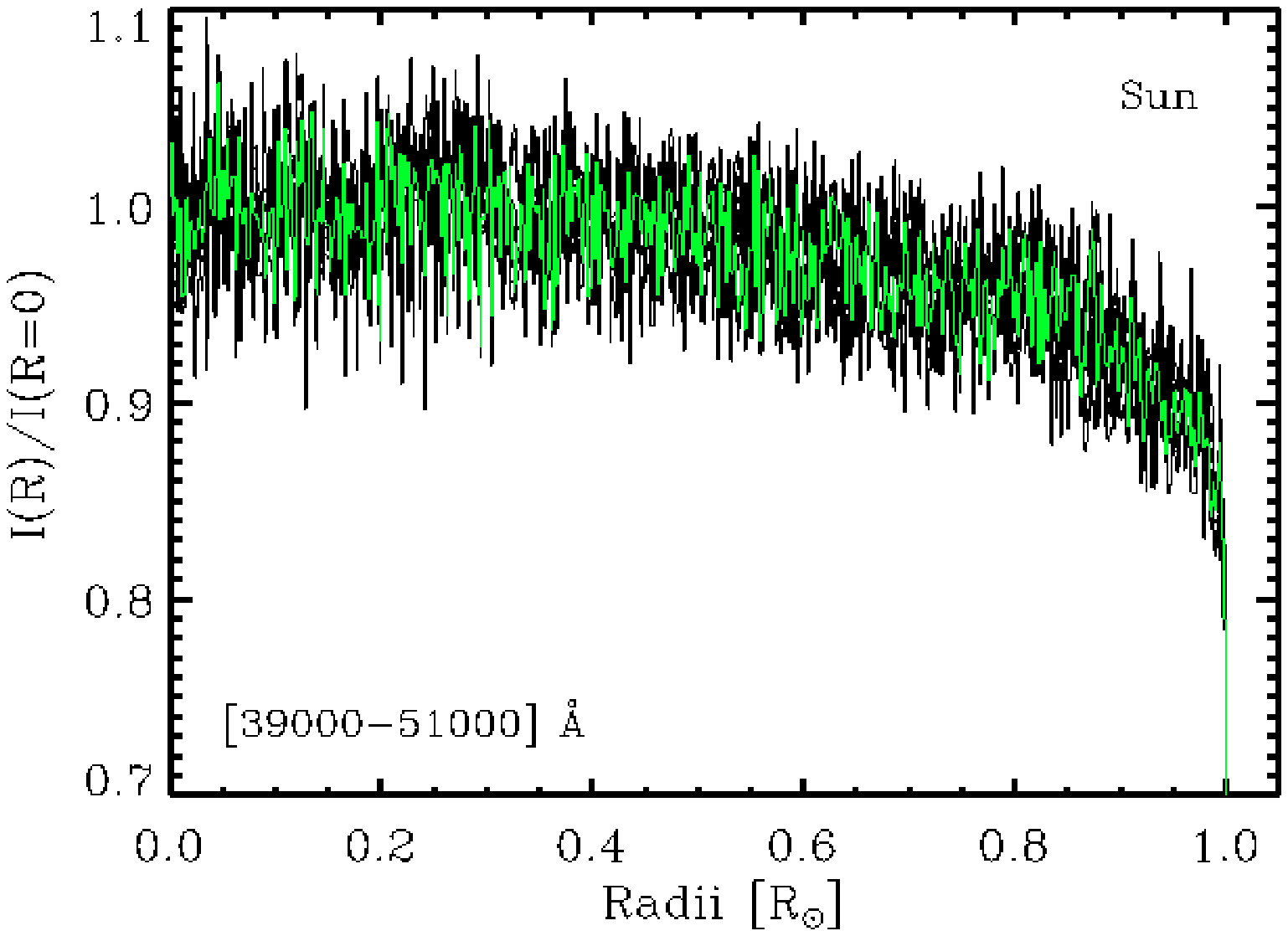}
                        \includegraphics[width=0.5\hsize]{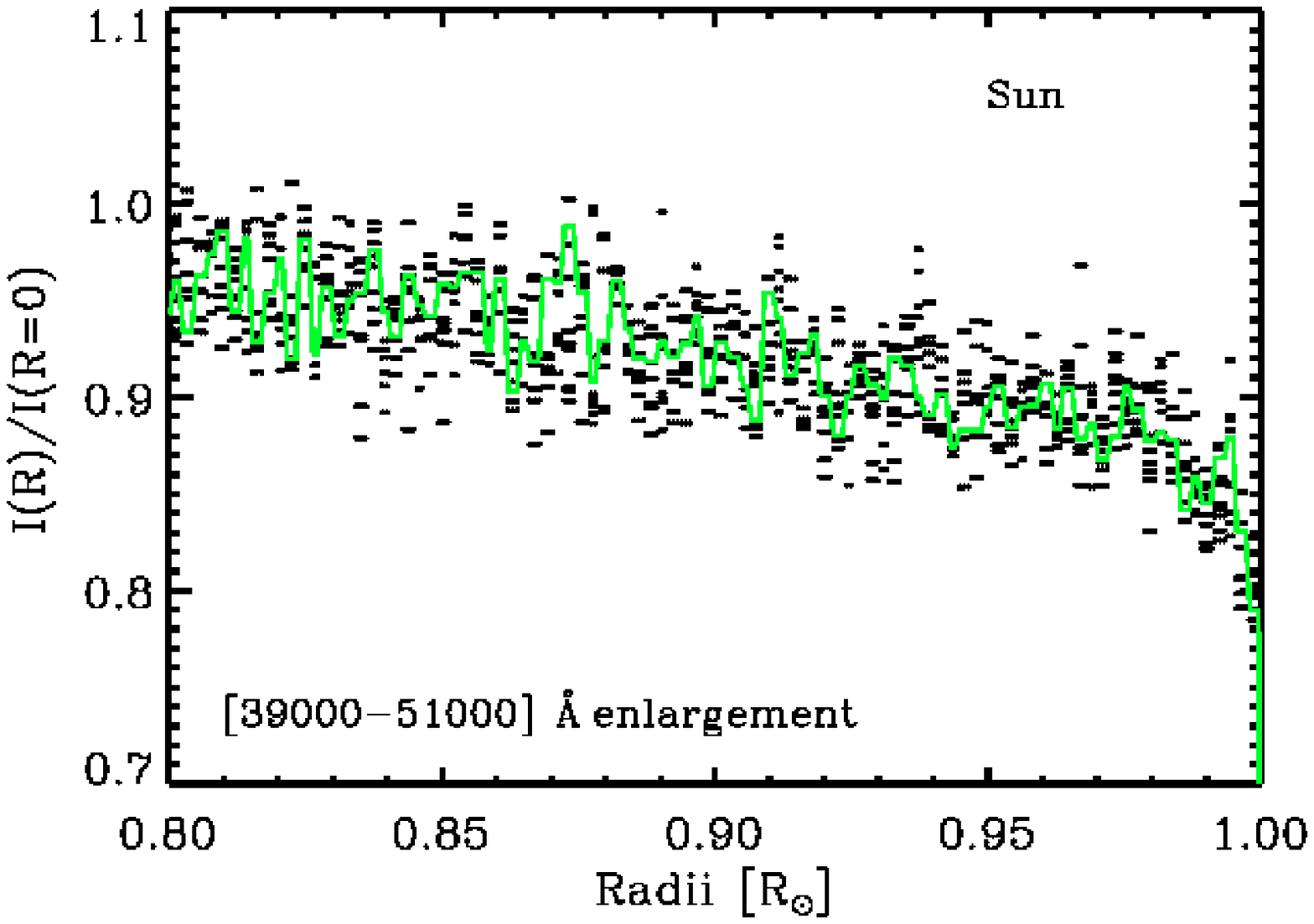}\\
                        \includegraphics[width=0.5\hsize]{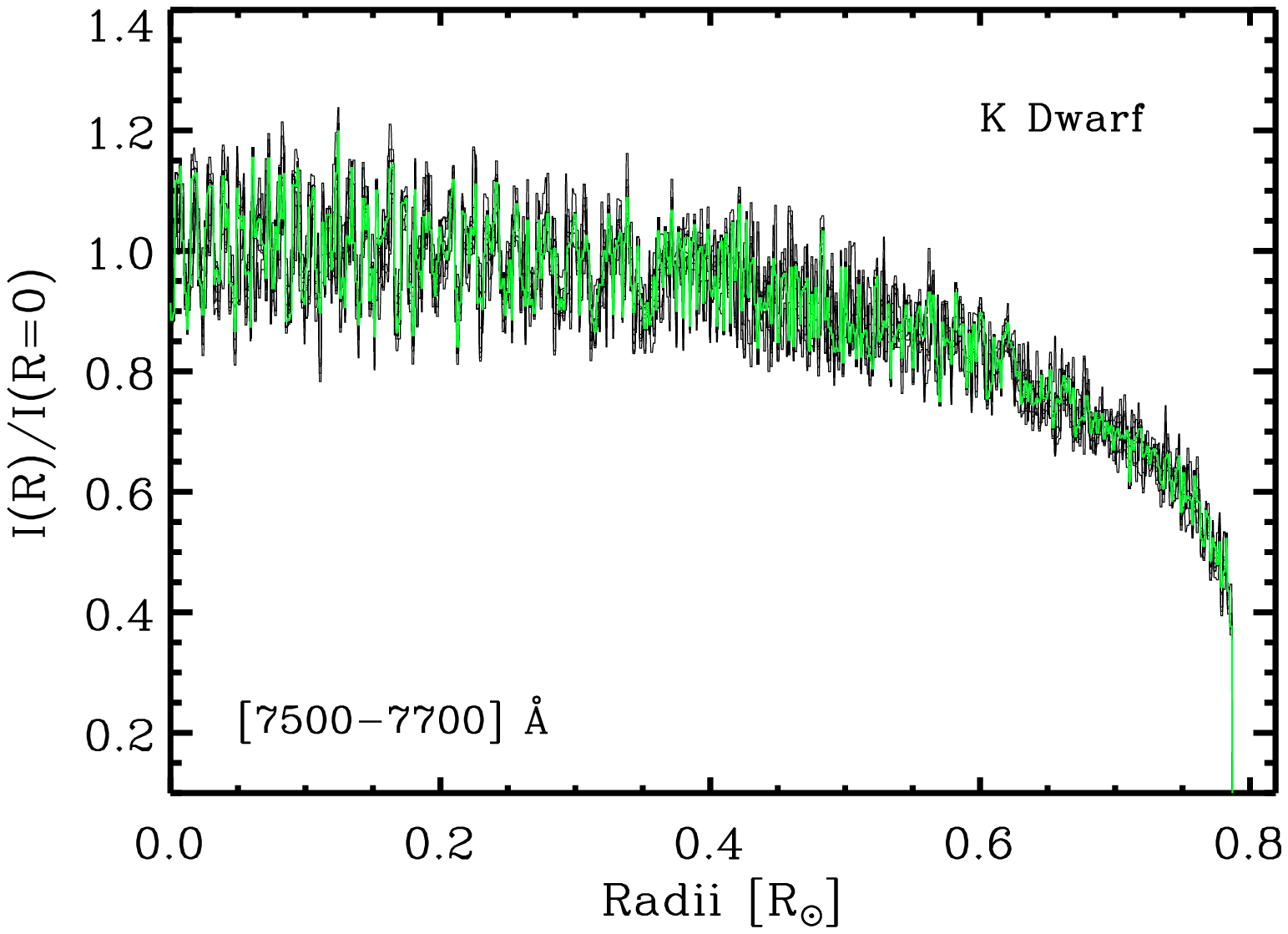} 
                        \includegraphics[width=0.5\hsize]{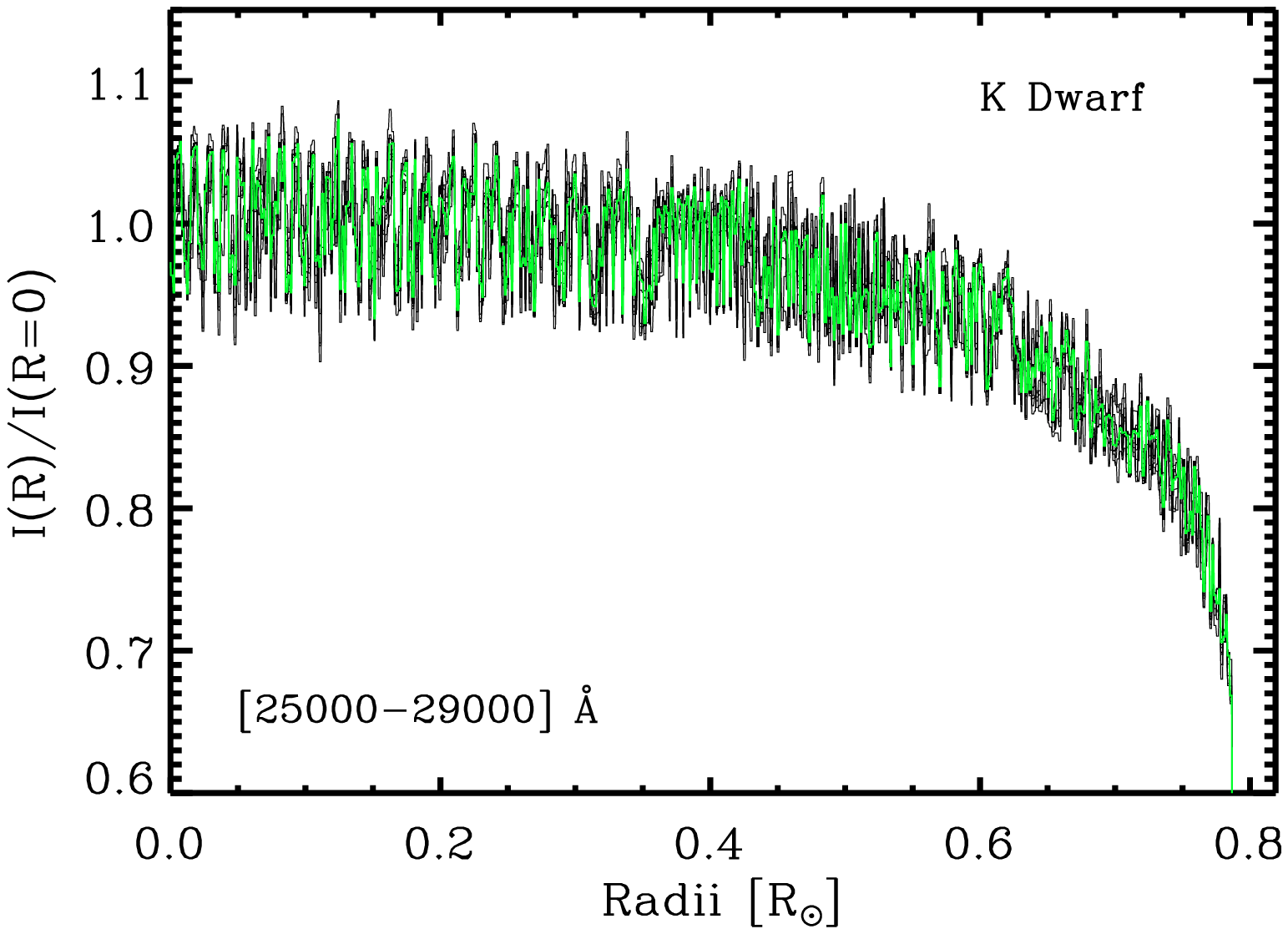}\\
                        \includegraphics[width=0.5\hsize]{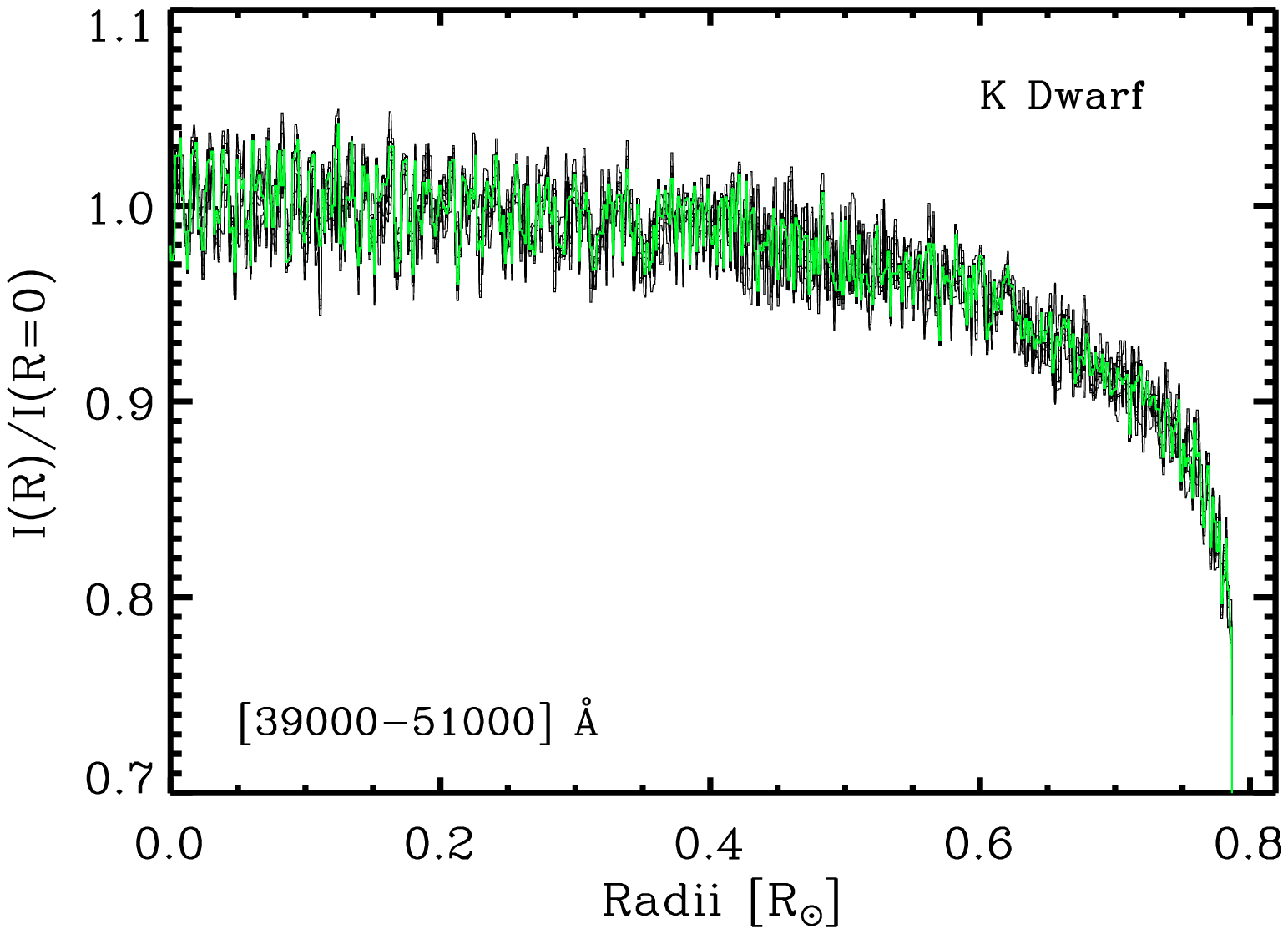}
                        \includegraphics[width=0.5\hsize]{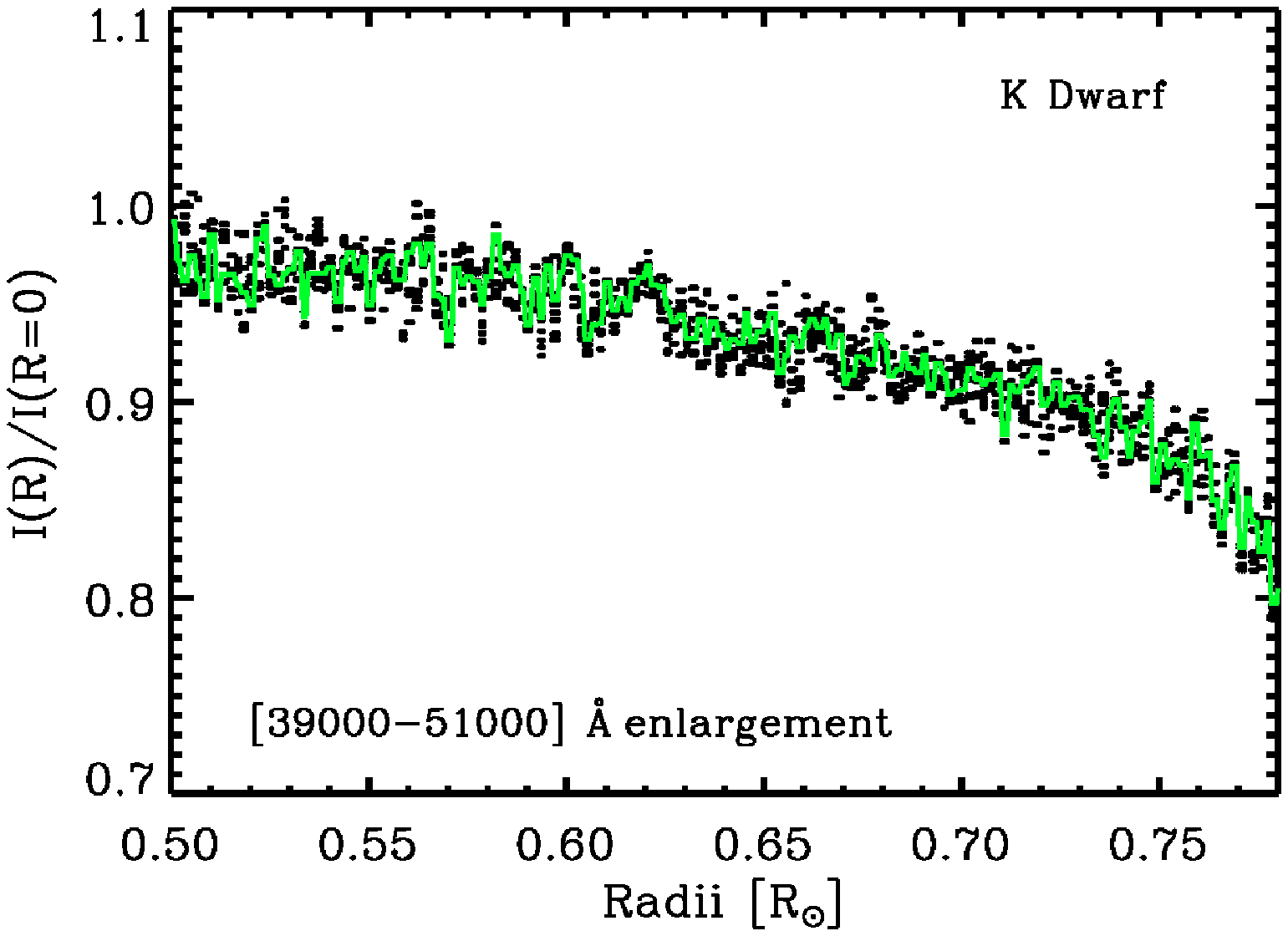}
        \end{tabular}
         \caption{Intensity profiles obtained from the cut at x = 0, y > 0 in the synthetic disk images of Fig.~\ref{starnoplanet}. The black curves are the overlap of $N=42$ different images. The number $N$ depends on the duration of the transit of Kepler 11-f (7 hours, Table~\ref{planets}) and the observed granulation timescale for the Sun is $\sim$10 minutes (see text). The intensity profiles are normalized to the intensity at the disk center (R = 0.0). The green line is the temporal average profile.} 
        \label{intensityprofiles1}
   \end{figure}  

\begin{figure}
   \centering
   \begin{tabular}{ccc}  
                         \includegraphics[width=0.9\hsize]{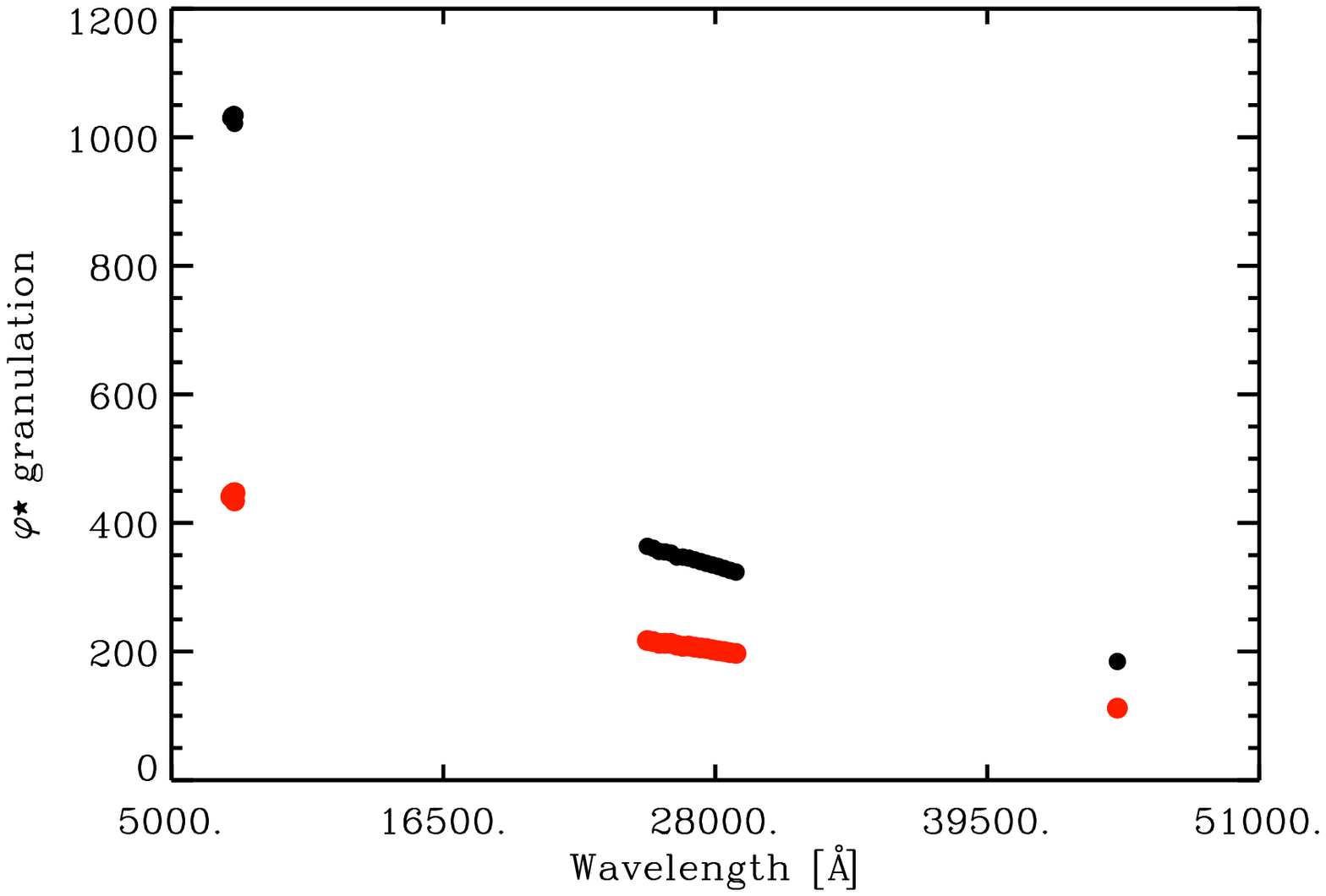} \\
                        \includegraphics[width=0.9\hsize]{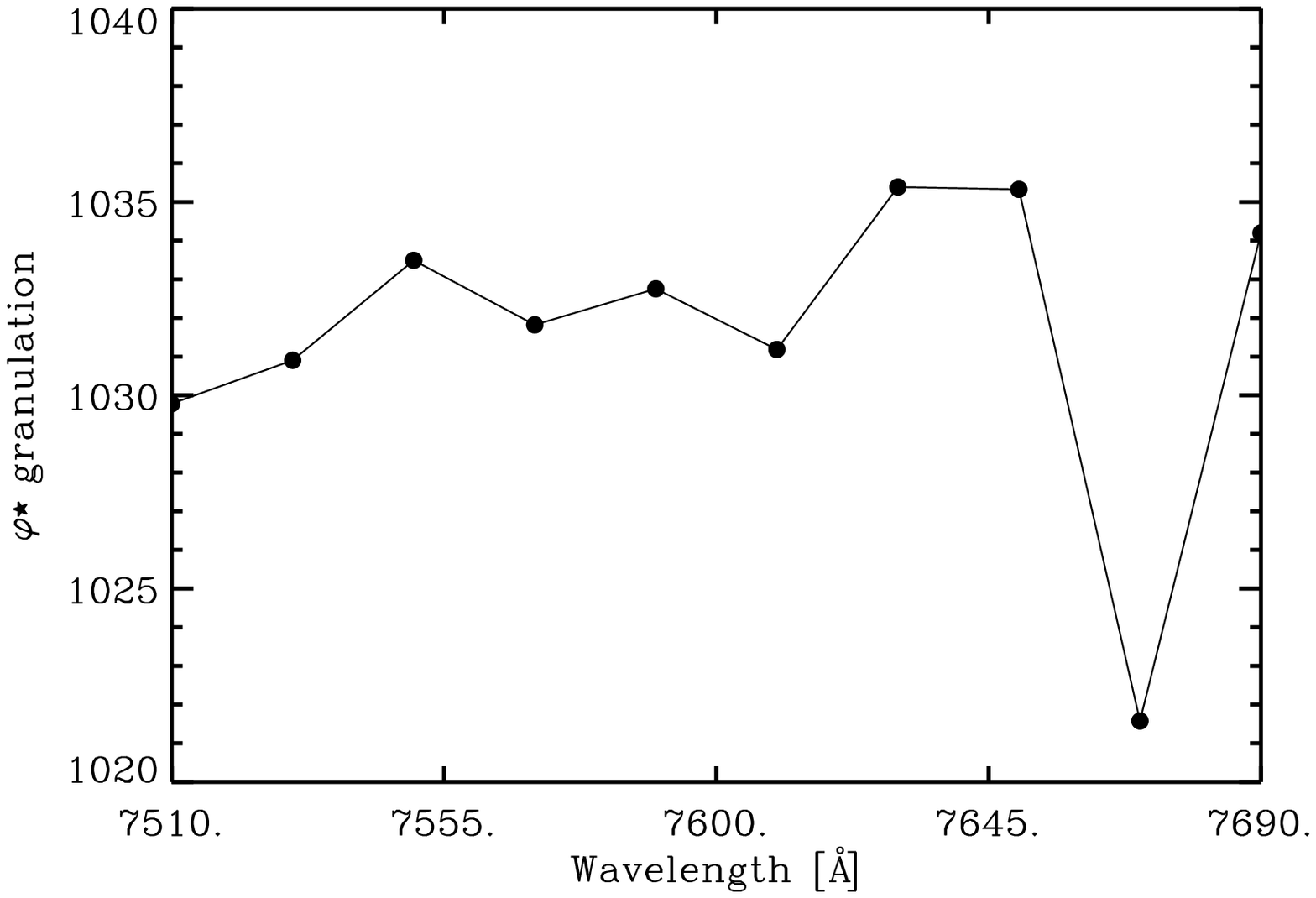} \\
                        \includegraphics[width=0.9\hsize]{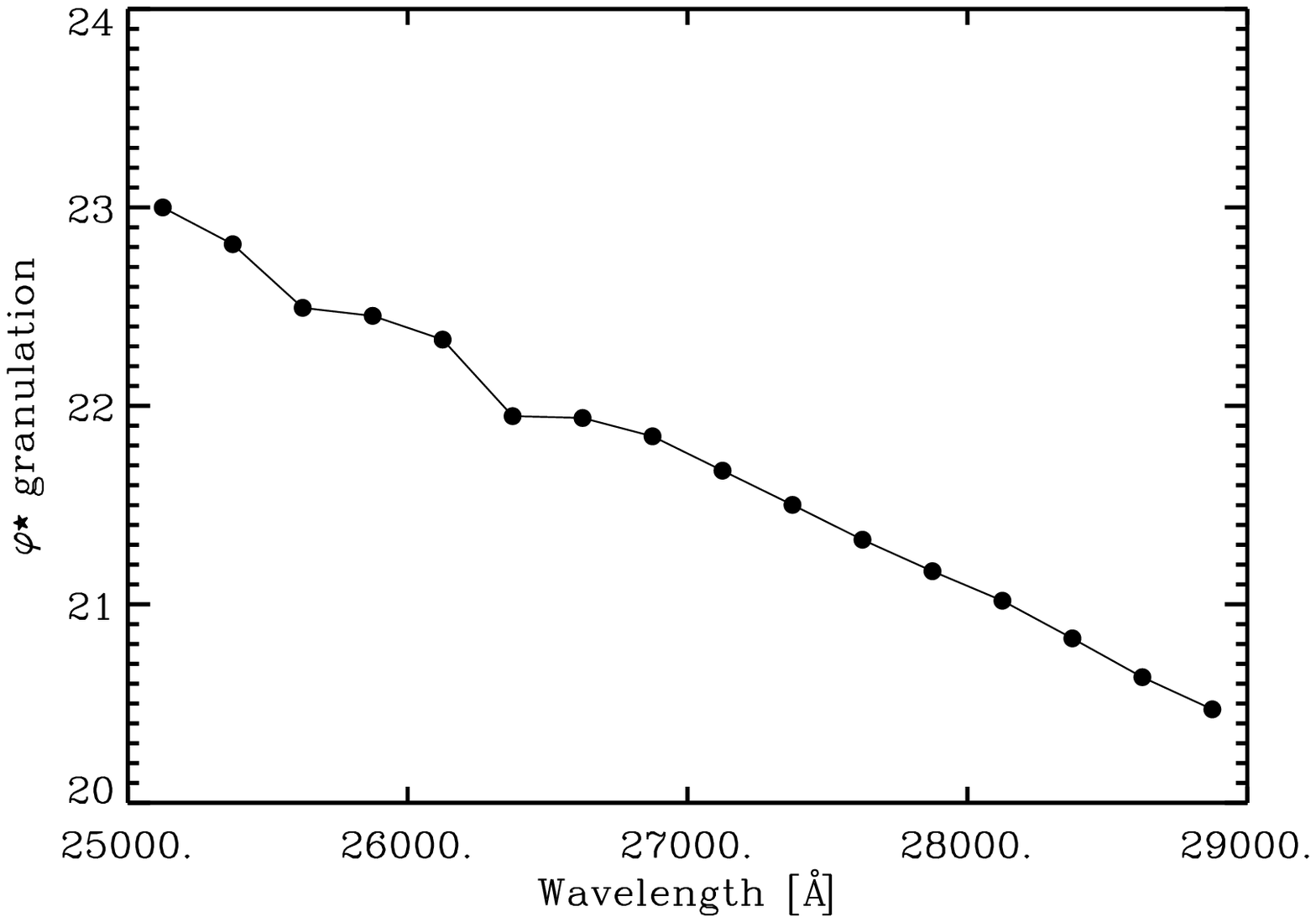}%
        \end{tabular}
         \caption{\emph{Top panel:} Number of photons ($\varphi_{\star granulation}$ for the synthetic Sun (black) and K~dwarf (red) calculated as described in the text and for the wavelengths of Table~\ref{wavelengths}. \emph{Central and bottom panels:} Enlargements for the Sun in the optical and infrared.}
       \label{photon1}
   \end{figure}  
   
We performed the calculations for 30 snapshots of the 3D RHD simulations of Table~\ref{simus}, adequately spaced apart so as to capture several convective turnovers, and for 43 different inclination angles ($\theta$) with respect to the line of sight (vertical axis): $\mu{\equiv}\cos(\theta)$ ranging from 1.000 to 0.1045 with a step of 0.0174. These synthetic images have been used to map them onto spherical surfaces to account for distortions especially at high latitudes and longitudes by cropping the square-shaped intensity maps when defining the spherical tiles. The total number of tiles ($N_{\rm{tile}}$) needed to cover half a circumference from side to side on the sphere is $N_{\rm{tile}} = \frac{\pi \cdot \rm{R}_\star}{x,y\rm{-dimension}}$, where $\rm{R}_\star$ (transformed into Mm), and the  $x,y\rm{-dimension}$ are taken from Table~\ref{simus}. The computed value of the $\theta$-angle used to generate each map depended on the position (longitude and latitude) of the tile on the sphere. \\
We aim to emulate the temporal variation of the granulation intensity, which shows a timescale on the order of $\sim$10 minutes \citep{2002A&A...396.1003N} for the Sun. We generated 80 different synthetic stellar-disk images (Fig.~\ref{starnoplanet}) using, for each tile, synthetic maps chosen randomly among the snapshots in the time-series. This resulted in a simulated granulation time-series of 800 minutes (13.3 hours). 
Since more tiles are necessary to cover the sphere than there are representative snapshots of the 3D RHD simulations, tiles randomly appear several times at different inclination angles $\theta,$ and adjacent tiles are not completely uncorrelated. However, we assumed that this statistical representation is good enough to represent the changing granulation pattern during the planet transit.

\section{Granulation noise}

The granulation pattern of the star may affect the photometric measurements during planet transit with two different types of noise: (i) the intrinsic timescale of the changes in granulation pattern (e.g., 10 minutes for solar-type stars assumed in this work) is shorter than the usual planet transit ($\sim$hours as in our prototype cases of Table~\ref{planets}), and (ii) the fact that the transiting planet occults isolated regions of the photosphere that differ in local surface brightness as a result
of convection-related surface structures. These sources of noise act simultaneously during the planet transit, and we analyze them in the next sections.

\subsection{Photon noise of the synthetic stellar disk with convection-related structures}

The granulation pattern changes with time. Figure~\ref{intensityprofiles1} displays the fluctuation of the intensity profiles for a particular cut in the synthetic disk images during a period corresponding to the  transit duration of the prototype terrestrial planet (7 hours, Table~\ref{planets}). In our approach, the solar disk intensity fluctuates during the transit by between [2.68-2.80] $\times10^6$\,erg\,cm$^{-2}$\,s$^{-1}$\,{\AA}$^{-1}$, and for the K~dwarf by between [1.63-1.66] $\times10^6$\,erg\,cm$^{-2}$\,s$^{-1}$\,{\AA}$^{-1}$.

We calculated the number of photons from the granulation synthetic images ($\varphi_{\star granulation}$) and compared it with the image produced by a black body ($\varphi_{\star BB}$) with an effective temperature of 5768 K for the Sun and 4516 K for the K~dwarf (same $T_{\rm{eff}}$ as in the RHD simulations of Table~\ref{simus}) at the different wavelength ranges of Table~\ref{wavelengths}. The granulation images were averaged over the $N_{\rm{terrestrial}}=42$ different realizations, where $N=\frac{\Delta t}{\sigma_{\rm{Sun}}}$, $\Delta t$ is the transit duration and $\sigma_{\rm{Sun}}$ is the observed granulation fluctuation timescale for the Sun, which is $\sim$10 minutes \citep{2002A&A...396.1003N}. The number of photons reaching a telescope with a collecting area $S$, a net efficiency $\epsilon$, and an integration time $\Delta t$ is equal to $\varphi_\star=I_\star\left(\lambda\right)\cdot\lambda/\left(h_{Planck}c\right)\cdot R\cdot S\cdot\epsilon\cdot\Delta t,$ where $\lambda$ is the central wavelength and $R$ the spectral resolution of the ranges in Table~\ref{wavelengths}, $I_\star\left(\lambda\right)$ the stellar intensity (either the averaged intensity of the granulation maps or the black body) at a certain wavelength and for a star at 100 pc, $h_{Planck}$ is the Planck constant, and $c$ the speed of light. In this work, we assumed a 100 cm diameter and thus a collecting area of $S=7854$ cm$^2$; a net efficiency $\epsilon= 0.15$ electron/photon; and an integration time, $\Delta t$ equal to the transit time of 7 hours from Table~\ref{planets}. Fig.~\ref{photon1} (top panel) shows the number of photons is larger for the visible region with respect to the infrared, as can be expected by the behavior of the Planck function at these wavelengths. Moreover, the central and bottom panels show a clear dependence of the intensity with respect to wavelength ranges used. 

\begin{figure}
   \centering
   \begin{tabular}{ccc}  
                        \includegraphics[width=0.95\hsize]{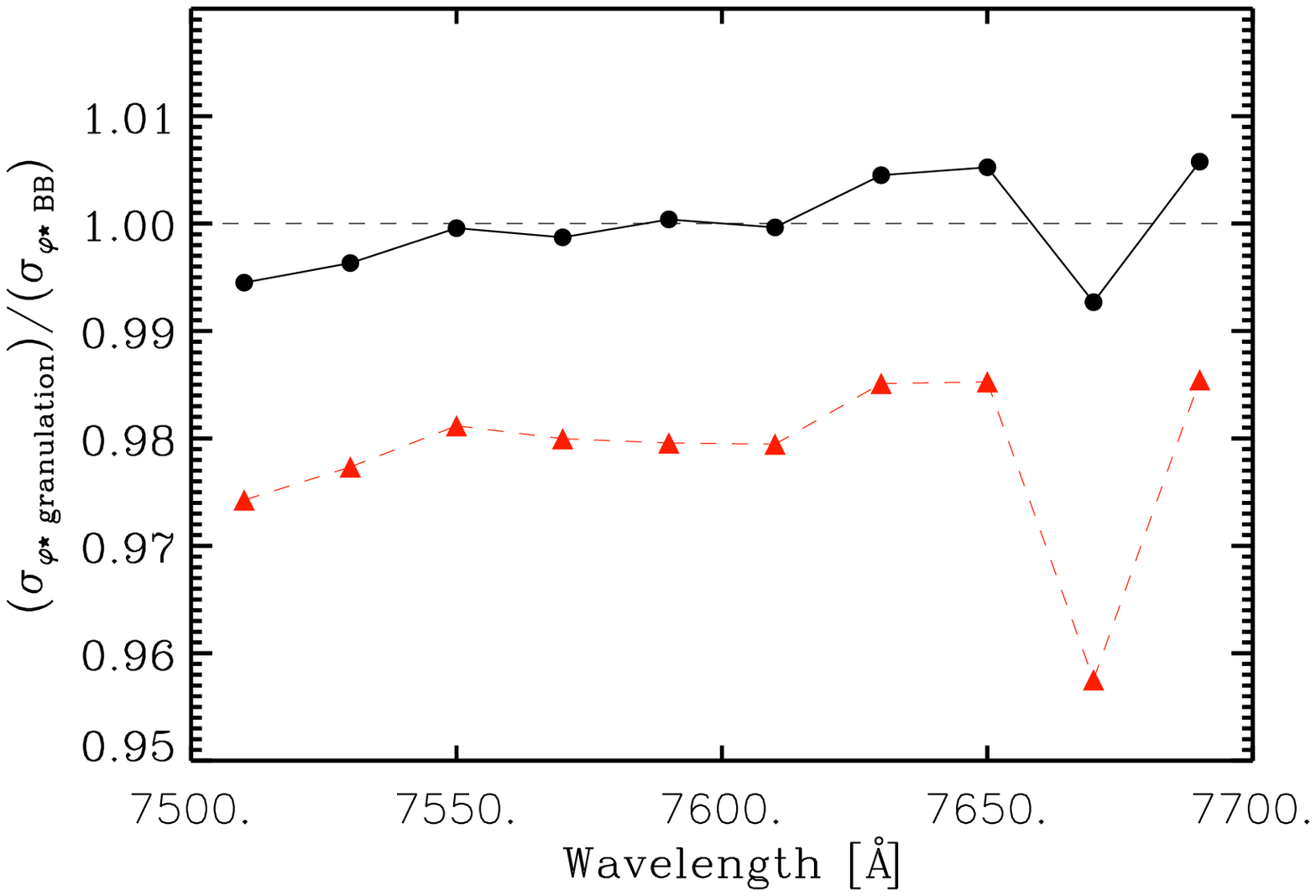} \\
                        \includegraphics[width=0.95\hsize]{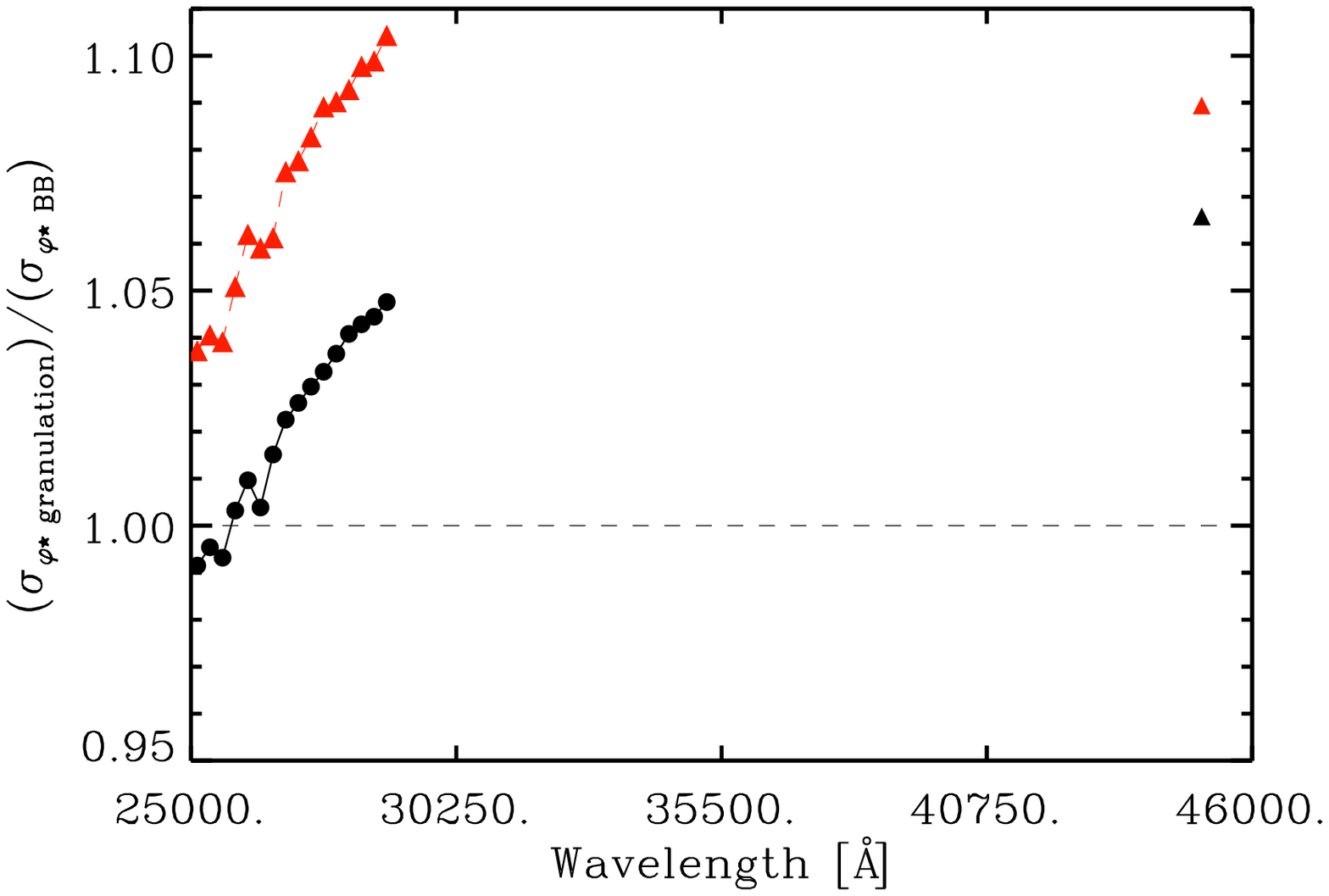}%
        \end{tabular}
         \caption{Ratio between the photon noise computed for the synthetic images with granulation ($\sigma_{\varphi_{\star granulation}}$) and for the correspondent black body ($\sigma_{\varphi_{\star BB}}$) with the same effective temperature of the RHD simulations of the Sun (black circles) and K~dwarf (red triangles). The wavelengths are taken from Table~\ref{wavelengths}.}
       \label{photon2}
   \end{figure}

The noise is the fluctuation in the total number of detected photons, and it is $\sigma_{\varphi_{\star granulation}} = \sqrt{\varphi_{\star granulation}}$ for the granulation synthetic images and $\sigma_{\varphi_{\star BB}} = \sqrt{\varphi_{\star BB}}$ for the corresponding black body. Figure~\ref{photon2} shows the ratio between the photon noise of the granulation images and the one from the black body. In the optical, the value of $\sigma_{\varphi_{\star granulation}}$ of the Sun is alternatively lower and higher than $\sigma_{\varphi_{\star BB}}$, in particular, the granulation signal becomes important at the wavelength bin of [7620-7640] \AA. On the other hand, the K-dwarf granulation photon noise is systematically lower than the corresponding black body, even if it follows the same trends as the solar one. \\
In the infrared, the situation is different: for both the Sun and the K~dwarf, the photon noise is greater than the black-body noise for all wavelength ranges; moreover, K~dwarf values are higher than the Sun owing to the lower effective temperature of the star and the consequent displacement of the radiation peak. Furthermore, increasing  the number of $N$ realizations (up to $N=80$) for the granulation average (i.e., $\Delta t = 13.3$ hours) returns values very similar to Fig.~\ref{photon2}. \\
Granulation significantly affects the photon noise in various wavelength ranges compared to the black-body approximation, so that transit uncertainties based on the black-body approximation can overestimate or underestimate the uncertainties, depending on the wavelength range considered. Furthermore, it is important to consider the change in the granulation pattern during the photometric measurements of transits like the one considered in this work, as developed in the next section.

\subsection{Flux variations caused by the transiting planet}

 \begin{figure}[!h]
   \centering
   \begin{tabular}{c}  
                        \includegraphics[width=0.99\hsize]{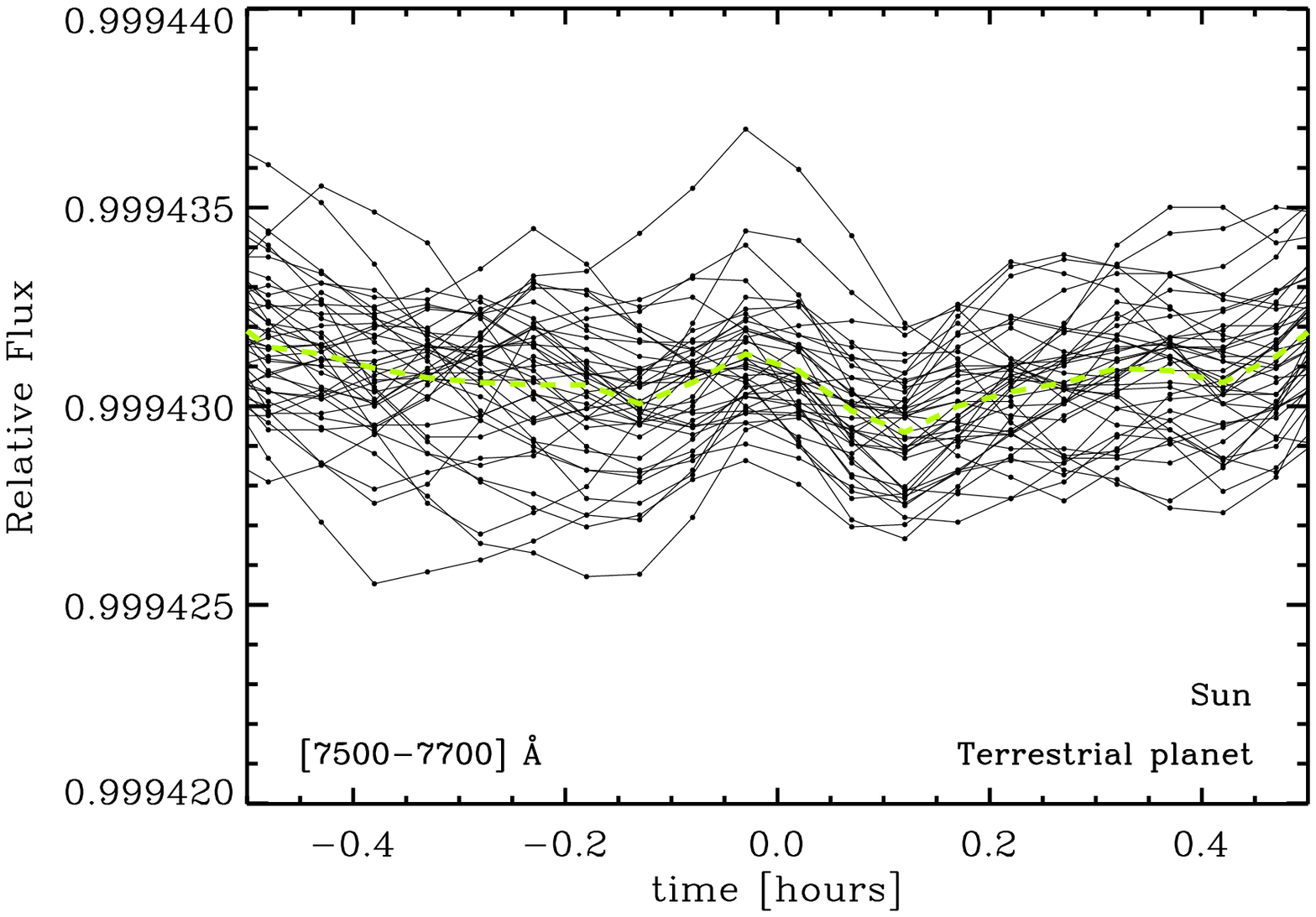} \\
                        \includegraphics[width=0.99\hsize]{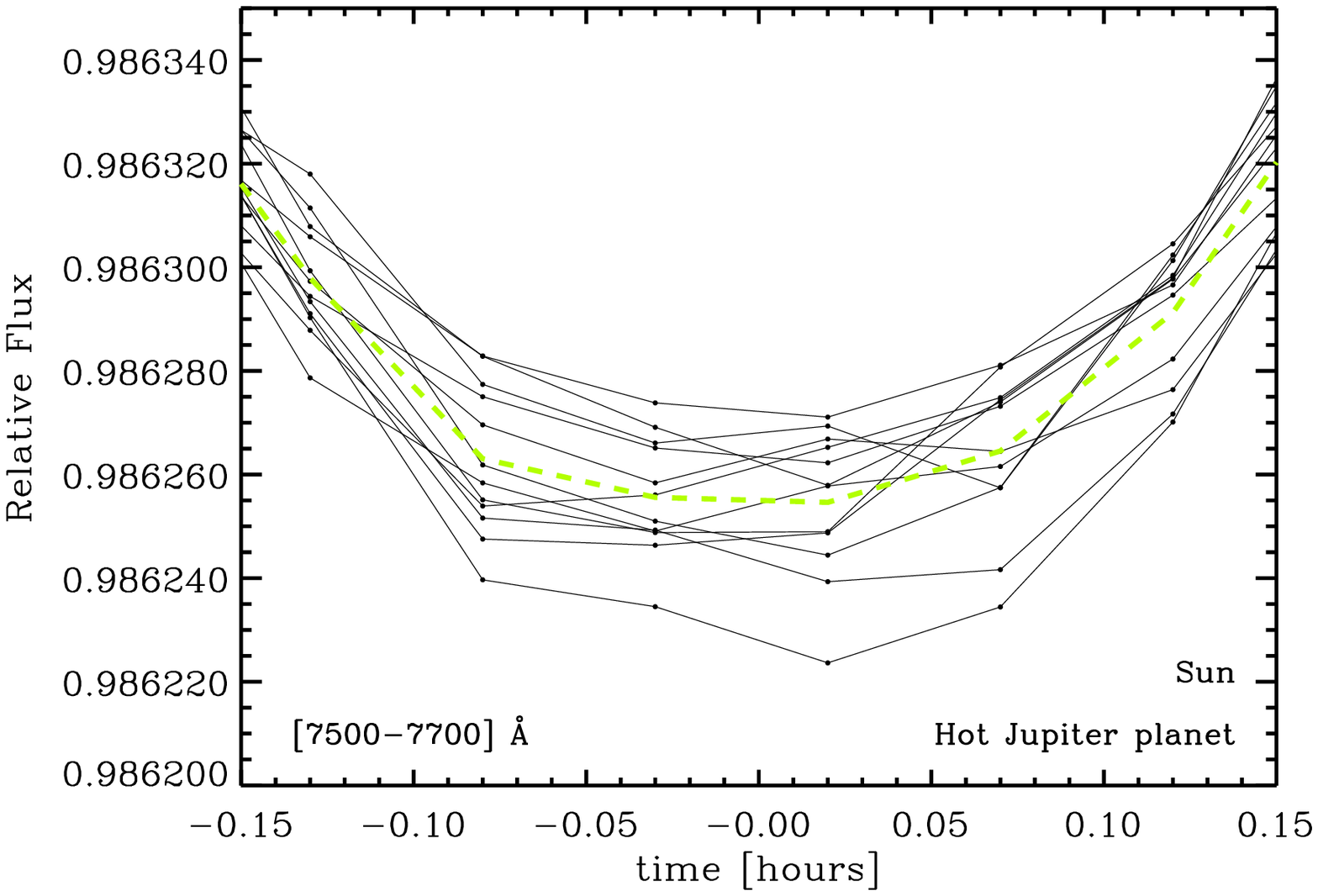} 
                \end{tabular}
         \caption{Scatter plot of 42 (top) and 12 (bottom) different transit light curves for a terrestrial and hot Jupiter planet of Table~\ref{planets}, respectively. The green dashed line is the average light curve profile.} 
        \label{stellardisk1bis}
   \end{figure}

\begin{table*} 
\begin{minipage}{\textwidth}
\caption{Prototypes of planets chosen to represent the planet transits.  }             % title of Table
\label{planets}      % is used to refer this table in the text
\centering                          % used for centreing table
\renewcommand{\footnoterule}{} 
\begin{tabular}{c c c c c c c c c }        % centreed columns (4 columns)
\hline\hline                 % inserts double horizontal lines
Planet  & Mass  &  Radius\tablefoottext{a} & Equilibrium  temperature  & semi-major  & $inc$\tablefoottext{b} & Transit  duration \\
Type   & $M_p$ [$M_{Jup}$]   & $R_p$ [$R_{Jup}$]  &  $T_p$ [K]     & axis [AU]  &  [$^{\circ}$] &  $\Delta t$ [hours]  \\
            &    &  Sun/K dwarf  & &   & & & &  \\
            \hline
Terrestrial planet&  0.006 & 0.219/0.173 &  400  & 0.2504 & 90 & $\sim$7  \\     
\hline
Hot Neptune  & 0.360 & 0.654/0.510 & 1540  &  0.0431 & 90 & $\sim$2  \\
\hline
Hot Jupiter & 7.570 & 1.090/0.850 & 1952 &  0.0269 & 90 & $\sim$3 \\ 
\hline\hline                          % inserts single horizontal line
\end{tabular}
\tablefoot{\tablefoottext{a}{$R_p$ is different for the RHD simulation of the Sun and of the K~dwarf, to keep the same ratio $R_p/R_\star$ for the transits. Moreover, the radius of the terrestrial planet is larger than Earth's radius.}\\
\tablefoottext{b}{The inclination orbit ($inc$) has been arbitrarily chosen to be equal to 90$^{\circ}$ to have a planet crossing at the center of the star.}}
\end{minipage}
\end{table*}

 \begin{table}
  \small
%\begin{minipage}[t]{\textwidth}
\caption{Typical values of the stellar (either Sun or K~dwarf) intensities, $I_{\rm{star}}$, at its center ($\mu=1$) for the synthetic images of the simulations (Fig.~\ref{starnoplanet}), and the planet integrated intensity, $I_{\rm{planet}}$, for a few representative wavelength bands (Table~\ref{wavelengths}).}             % title of Table
\label{fluxratio}      % is used to refer this table in the text
\centering                          % used for centreing table
\renewcommand{\footnoterule}{} 
\begin{tabular}{c | c c c  }        % centreed columns (4 columns)
\hline\hline                 % inserts double horizontal lines
$\lambda$  & [7620-7640] & [25000-25250]  &[39000-51000] \\
band [\AA ] & & & \\
            \hline
             $ I_{\rm{Sun}}/I_{\rm{terrestrial}}$         & $4\times10^15$ & 360000  &  6066 \\
             $ I_{\rm{Sun}}/I_{\rm{neptune}}$     & 75270 & 187 & 86  \\
            $ I_{\rm{Sun}}/I_{\rm{jupiter}}$      & 4487 & 68 & 42  \\
            \hline
            $ I_{\rm{Kdwarf}}/I_{\rm{terrestrial}}$       & $10^16$ & 262000  & 4560  \\
             $ I_{\rm{Kdwarf}}/I_{\rm{neptune}}$  & 28130 & 136  & 65 \\
            $ I_{\rm{Kdwarf}}/I_{\rm{jupiter}}$           & 7510 & 50  & 32  \\\hline\hline                          % inserts single horizontal line
\end{tabular}
%\end{minipage}
\end{table}

   \begin{table}
%\begin{minipage}[t]{\textwidth}
\caption{Transiting curve data for the different prototype planets of Table~\ref{planets} and 3D RHD simulations of Table~\ref{simus}. The values reported in Col. 4 are the maximum transit depth and
those in Col. 5 are the RMS of a set of values covering the central part of the transit periods: [-1,+1] hours for the terrestrial planet, [-0.15,+0.15] hours for hot Neptune, and [-0.1,+0.1] hours for the hot Jupiter. These values are representative for all the wavelength bands from Table~\ref{wavelengths}.}
%[-3,+3] hours for the terrestrial planet, [-1.1,+1.1] hours for hot neptune planet, and [-0.8,+0.8] hours for the hot jupiter planet. }             % title of Table
\label{transitdata}      % is used to refer this table in the text
\centering                          % used for centreing table
\renewcommand{\footnoterule}{} 
\begin{tabular}{c c c c c }        % centreed columns (4 columns)
\hline\hline                 % inserts double horizontal lines
Planet  & Star  & Wavelength  & Depth\tablefootmark{a} & RMS  \\
            &    & [$\AA$] &               &   [ppm] \\
            \hline
terrestrial   &  Sun & [7600-7700] & 0.999431 & 3.5 \\%33 \\
Neptune  &  &  & 0.995046 & 7.4 \\% 454  \\
hot Jupiter  &  &  & 0.986251  & 15.9 \\ %682  \\
\hline
terrestrial   &   & [25000-29000] & 0.999480 & 1.1 \\%13 \\
Neptune  &  &   & 0.995487 & 2.1 \\% 232  \\ 
hot Jupiter  &  &  & 0.987509 & 4.6 \\%238  \\
\hline
terrestrial   &   & [39000-51000]  & 0.999489 & 0.8 \\ %10 \\
Neptune  &  &  & 0.995557 &  1.8 \\% 66 \\
hot Jupiter  &  & & 0.987660  & 3.5 \\ %173  \\
\hline
\hline
terrestrial   &  K~dwarf & [7600-7700] & 0.999635 &  2.7 \\%31 \\
Neptune  &  & & 0.996822 & 6.3 \\ %351  \\
hot Jupiter  &  & &  0.991180 & 9.8 \\%576 \\
\hline
terrestrial   &   & [25000-29000] & 0.999679 &  0.8 \\ %11 \\
Neptune  &  &   & 0.997194 & 2.1 \\%154 \\
hot Jupiter  &  &   & 0.992244 & 2.7 \\ %196 \\
\hline
terrestrial   &   & [39000-51000]  & 0.999684 &  0.7 \\ %7 \\
Neptune  &  &  & 0.997261 & 1.6 \\%119\\
hot Jupiter  &  &   &  0.992391 & 2.1 \\ %136 \\
\hline\hline                          % inserts single horizontal line
\end{tabular}
\tablefoot{
\tablefoottext{a}{The depth is defined as the normalized flux minimum during transit.}}
%\end{minipage}
\end{table}

\begin{table}
\tiny
%\begin{minipage}[t]{\textwidth}
\caption{Photometric accuracy of different telescopes.}             % title of Table
\label{observations}      % is used to refer this table in the text
\centering                          % used for centreing table
\renewcommand{\footnoterule}{} 
\begin{tabular}{c c c c }        % centreed columns (4 columns)
\hline\hline                 % inserts double horizontal lines
Name  & ground- or  &  Accuracy & Filter     \\
            &  space-based  & [part-per-million] &  [$\AA$]               \\
            \hline
HATNet\tablefootmark{a} & ground    & $\sim$5000 & V and I bands \\
NGTS\tablefootmark{b} & ground & $<$1000 & [6000-9000]  \\
TRAPPIST\tablefootmark{c} & ground &  $\sim$300  & V band  \\
WASP\tablefootmark{d} & ground &   $\sim$4000 & [4000-7000]   \\
SPITZER\tablefootmark{e} & space & 29-143 & [36000-80000]   \\
CHEOPS\tablefootmark{f} & space & $\sim$10 & V band   \\
Kepler\tablefootmark{g} & space & 20-84 & [4230-8970]   \\
CoRoT\tablefootmark{h} & space & $\sim100$  & [5000-10000]   \\ 
TESS\tablefootmark{i}  & space &  $\sim$60  & [6000-10000]   \\
PLATO\tablefootmark{j} & space & $\sim$27 & [5000-10000]   \\
\hline\hline                          % inserts single horizontal line
\end{tabular}
\tablefoot{
\tablefoottext{a}{\cite{2004PASP..116..266B}}
\tablefoottext{b}{\cite{2013EPJWC..4713002W}}
\tablefoottext{c}{\cite{2011Msngr.145....2J}}
\tablefoottext{d}{\cite{2006PASP..118.1407P}}
\tablefoottext{e}{\cite{2010AAA...518A..25G,2010Natur.464.1161S}}
\tablefoottext{f}{\cite{2013EPJWC..4703005B}}
\tablefoottext{g}{\cite{2010ApJ...713L..79K}}
\tablefoottext{h}{\cite{2009AAA...506..411A}}
\tablefoottext{i}{\cite{2015JATIS...1a4003R}}
}
%\end{minipage}
\end{table}

\begin{figure*}
   \centering
   \begin{tabular}{ccc}  
                        \includegraphics[width=0.32\hsize]{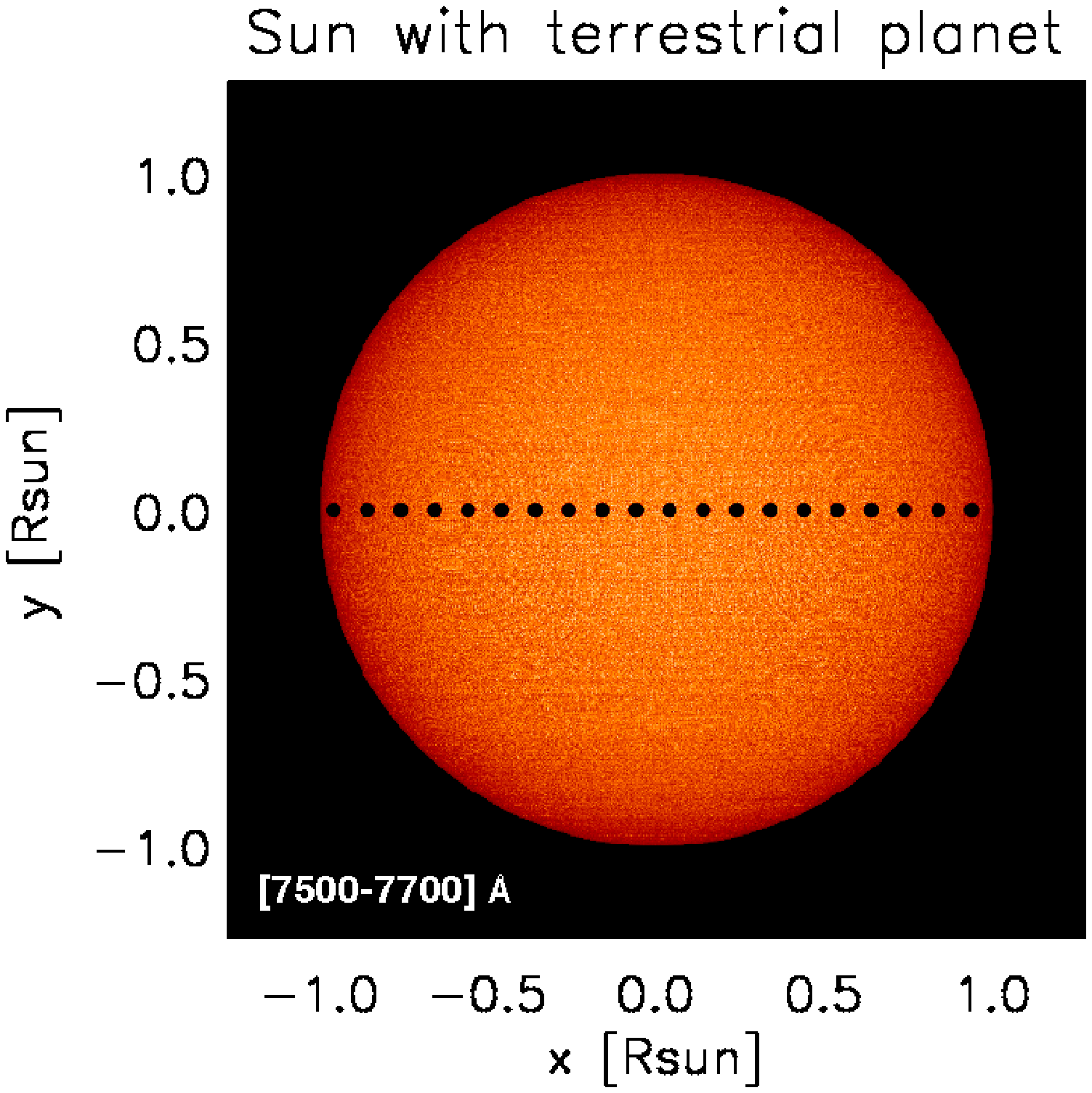} 
                        \includegraphics[width=0.32\hsize]{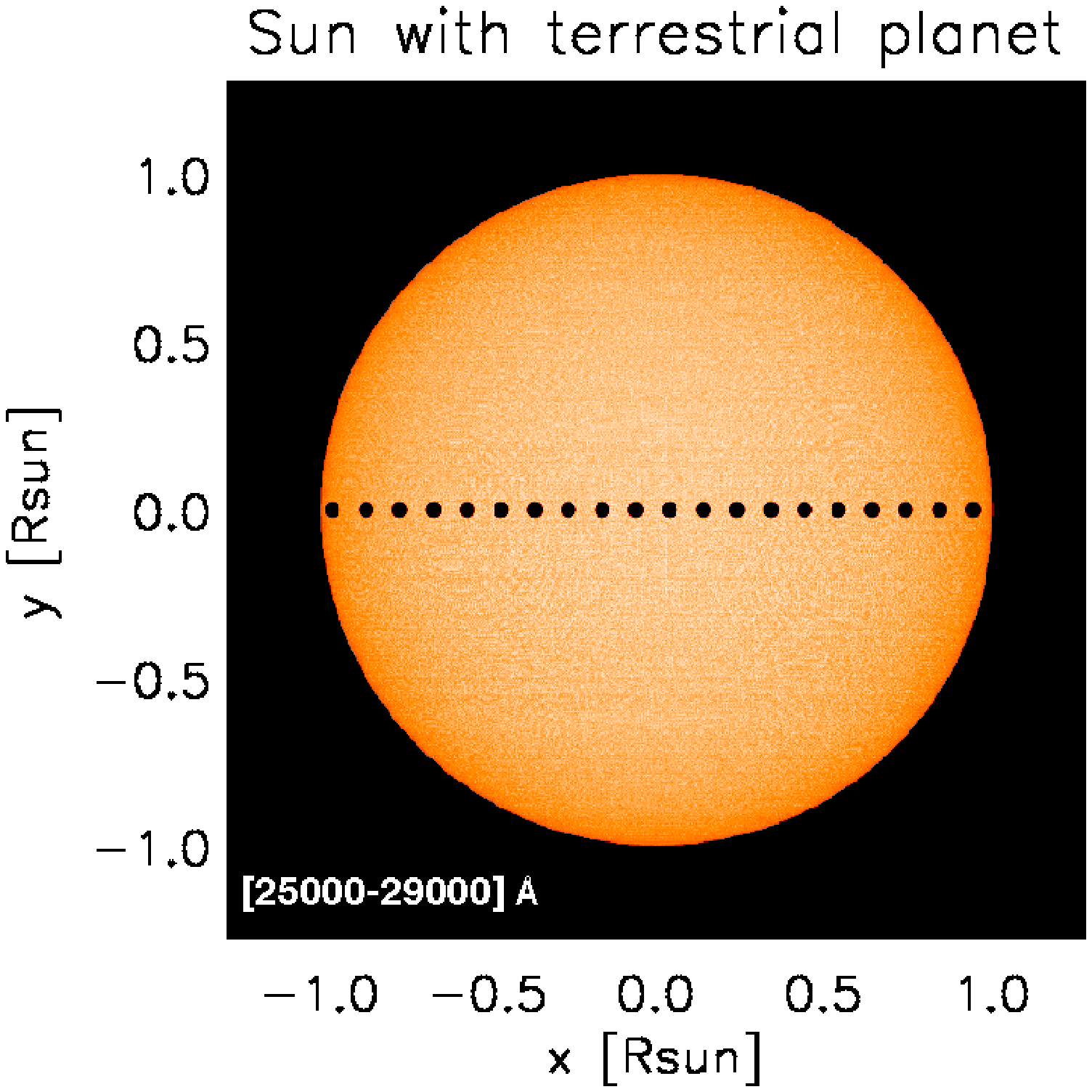} 
                        \includegraphics[width=0.32\hsize]{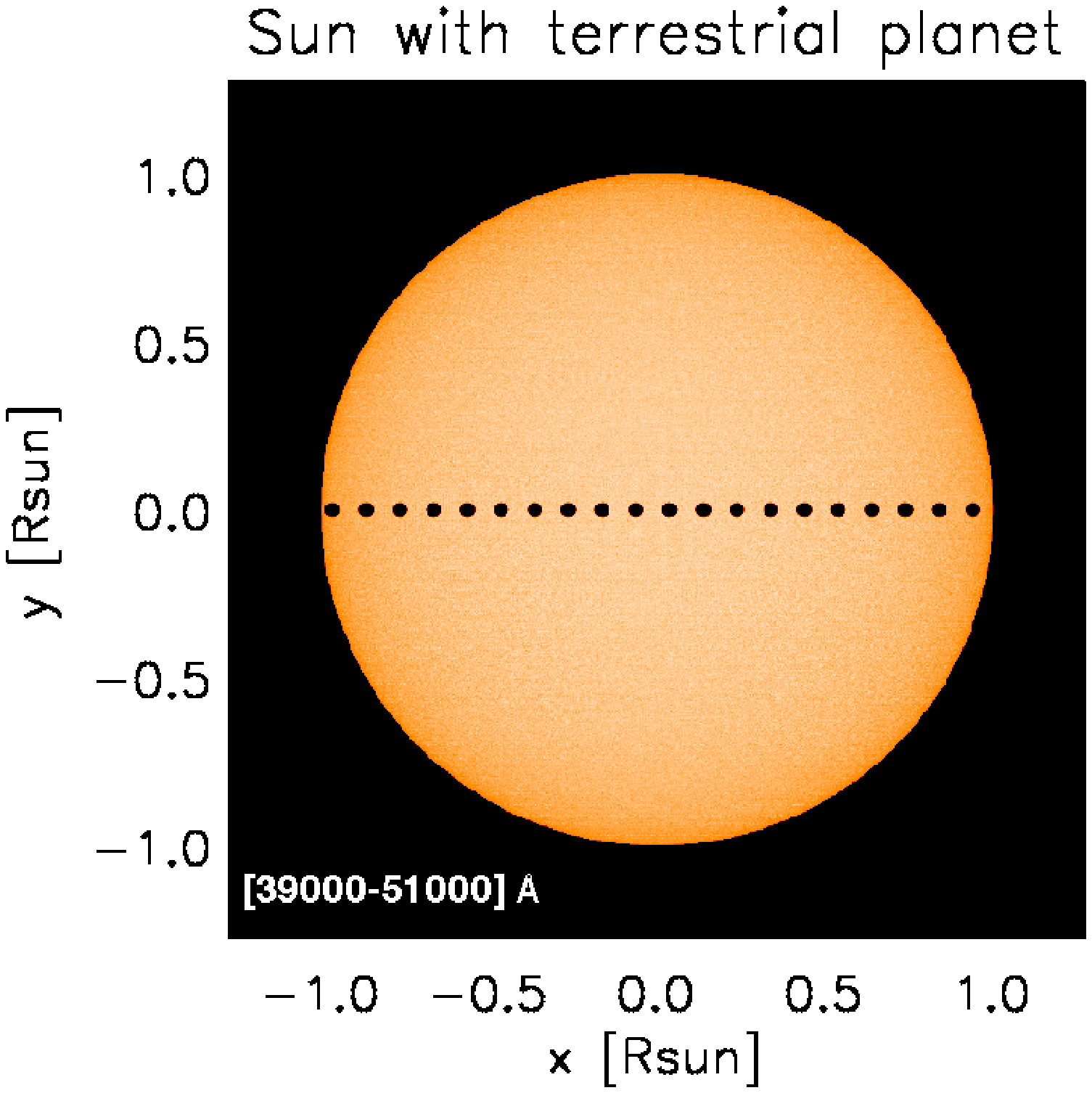}\\
                        \includegraphics[width=0.32\hsize]{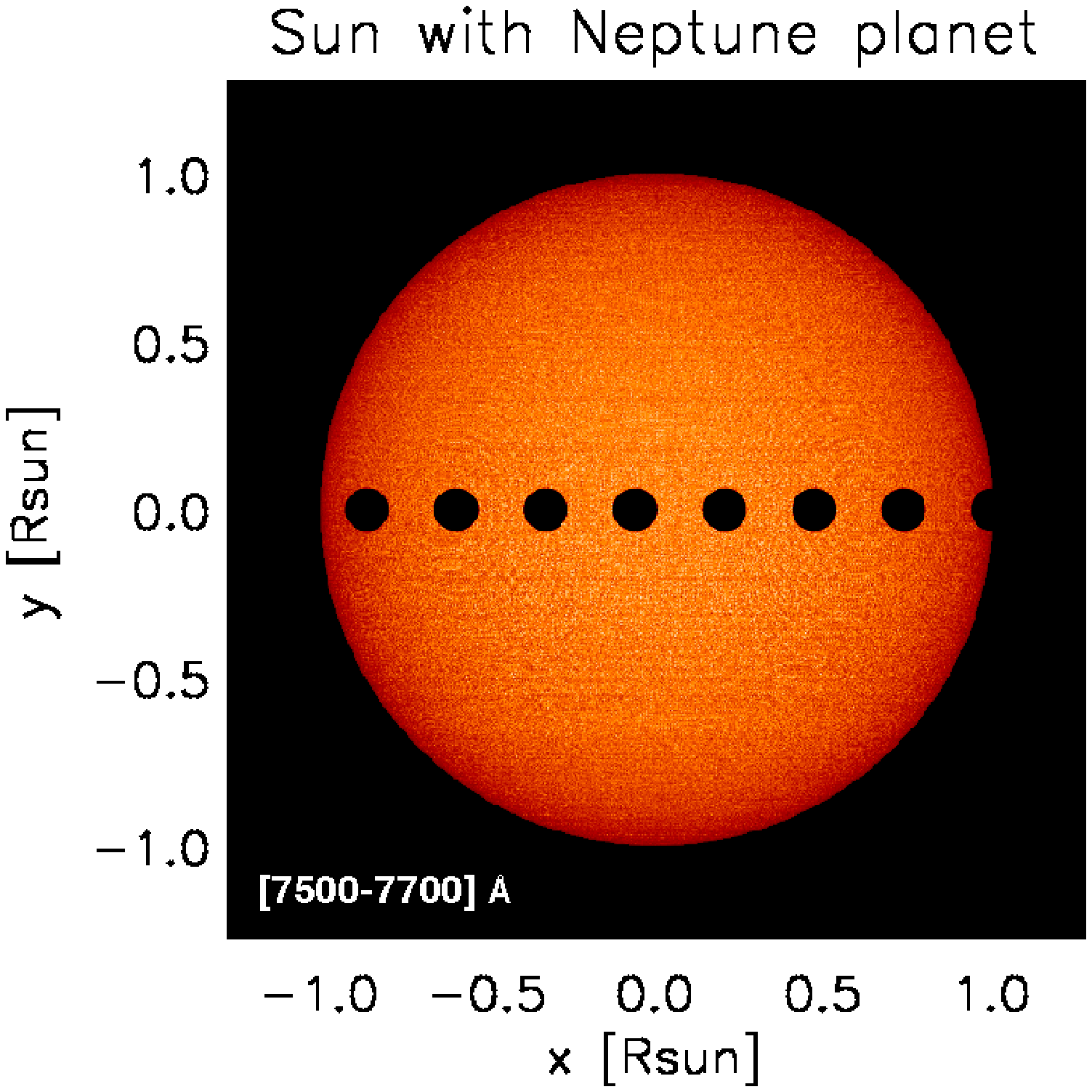} 
                        \includegraphics[width=0.32\hsize]{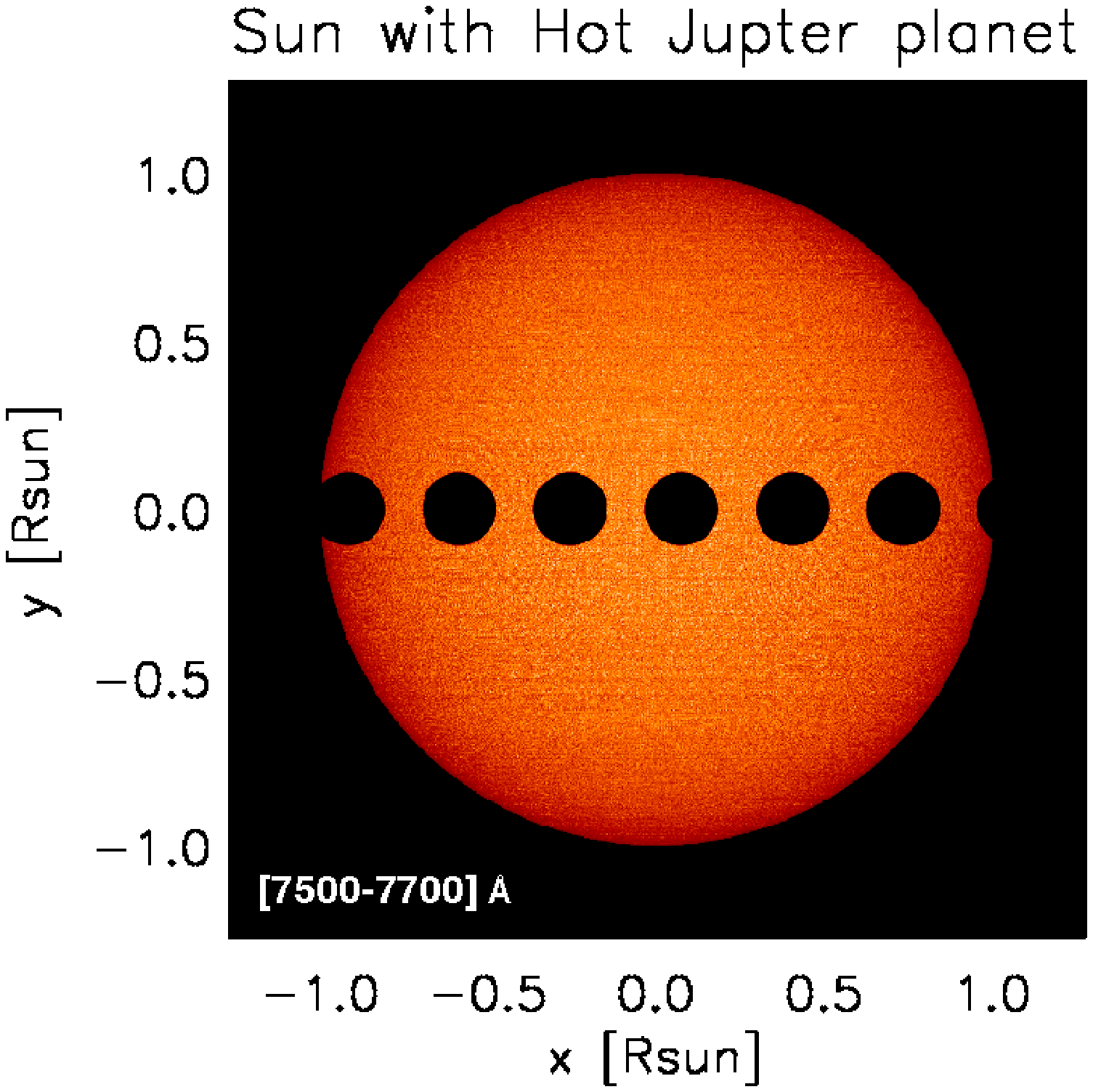}   
                \end{tabular}
         \caption{\emph{Top row:} Synthetic solar disk images with transiting planet computed at [7620-7640] $\AA$ (left, see Table~\ref{wavelengths}), [25000-25250] $\AA$ (center), [39000-51000] $\AA$ (right) for the Sun (Table~\ref{simus}). The prototype planet for the transit is the terrestrial planet (Table~\ref{planets}). For the sake of clarity, we assumed transit data measurements every 20 minutes in these plots. However, in the analysis (Fig.~\ref{transit1} and Fig.~\ref{transit2}), we simulated transits measurements every 3 minutes. \emph{Bottom row:} same as above with solar images computed at [7620-7640] $\AA$ and the prototype Neptune (left) and the hot Jupiter (center).} 
        \label{stellardisk1}
   \end{figure*}

\cite{2015A&A...576A..13C} modeled the transit light curve of Venus in 2004 assuming the 3D RHD simulation of the Sun (the same as we used here) for the background solar disk. They showed that in terms of transit depth and ingress/egress slopes as well as the emerging flux, the RHD simulation is well adapted to interpret the observed data. Furthermore, they reported that the granulation causes intrinsic changes in the total solar irradiance over the same time interval as the Venus transit, arguing that the granulation is a source of an intrinsic noise that may affect precise measurements of exoplanet transits. \\
In this work, we extended their analysis to the simulations of Table~\ref{simus} (i.e., adding more calculations for the Sun and the K~dwarf) and to the large set of wavelength bands of Table~\ref{wavelengths}. We used the following procedure:

\begin{itemize}
\item we chose three prototypes of planets with different sizes and transit time lengths corresponding to a hot Jupiter, a hot Neptune, and a terrestrial planet (Table~\ref{planets}) with the purpose of studying the resulting noise caused by the granulation on simulated transits. We did not aim to reproduce the exact conditions of the planet-star systems detected;
\item we used the synthetic disk images as the background-emitting source for all the wavelength bands of Table~\ref{wavelengths};
\item to model the flux of the planet, we used a black body with the planet equilibrium temperature reported in Table~\ref{planets}. The typical flux ratios are reported in Table~\ref{fluxratio};
\item we simulated the transits using the exoplanet data reported in Table~\ref{planets}, and collected data points every 3 minutes. Synthetic images  are reported in Fig.~\ref{stellardisk1} for the Sun;
\item the emerging intensity was collected for every transit step;
\item we accounted for the variation of the granulation intensity using a set of $N$ different synthetic stellar-disk images. We chose randomly $N_{\rm{terrestrial}}=42$, $N_{\rm{neptune}}=18$, and $N_{\rm{jupiter}}=12$ synthetic realizations from our 80 realizations. 
\end{itemize}

Following this procedure, since we collected points every 3 minutes, that is, about one-third of 10 minutes (the granulation timescale for the Sun), to build a transit there are three adjacent time
steps for which the same image was used (i.e., this corresponds to three different positions of the transiting planet). Then, a new synthetic image was chosen randomly and used for about three more times, and so on. This leads for a total of, for instance, $N_{\rm{terrestrial}}=42$ or $N_{\rm{jupiter}}=12$ synthetic realizations as in the example of different transit curves superimposed in Fig.~\ref{stellardisk1bis}. The figure shows that the central phase of the terrestrial transit  wiggles more strongly than that of the hot Jupiter, which looks smoother. Since the number of tiles needed to cover the sphere is smaller than the number of representative snapshots of the RHD simulations, tiles randomly appear several times along the transit trajectory and, consequentially, a correlated behavior between the different transit curves arises. This effect is more important for the terrestrial planet, which is more sensitive to the stellar inhomogeneities because of the convection-related surface structures because its apparent size is comparable to the RHD simulation box.\\
Figure~\ref{stellardisk1} shows the simulated transits at different wavelength bands and prototype planets. The top row panel displays pronounced center-to-limb variations in the stellar disk from the optical toward the infrared bands, which is principally caused by the Planck function behavior at different wavelengths. 

 \begin{figure*}
   \centering
   \begin{tabular}{ccc}  
                        \includegraphics[width=0.33\hsize]{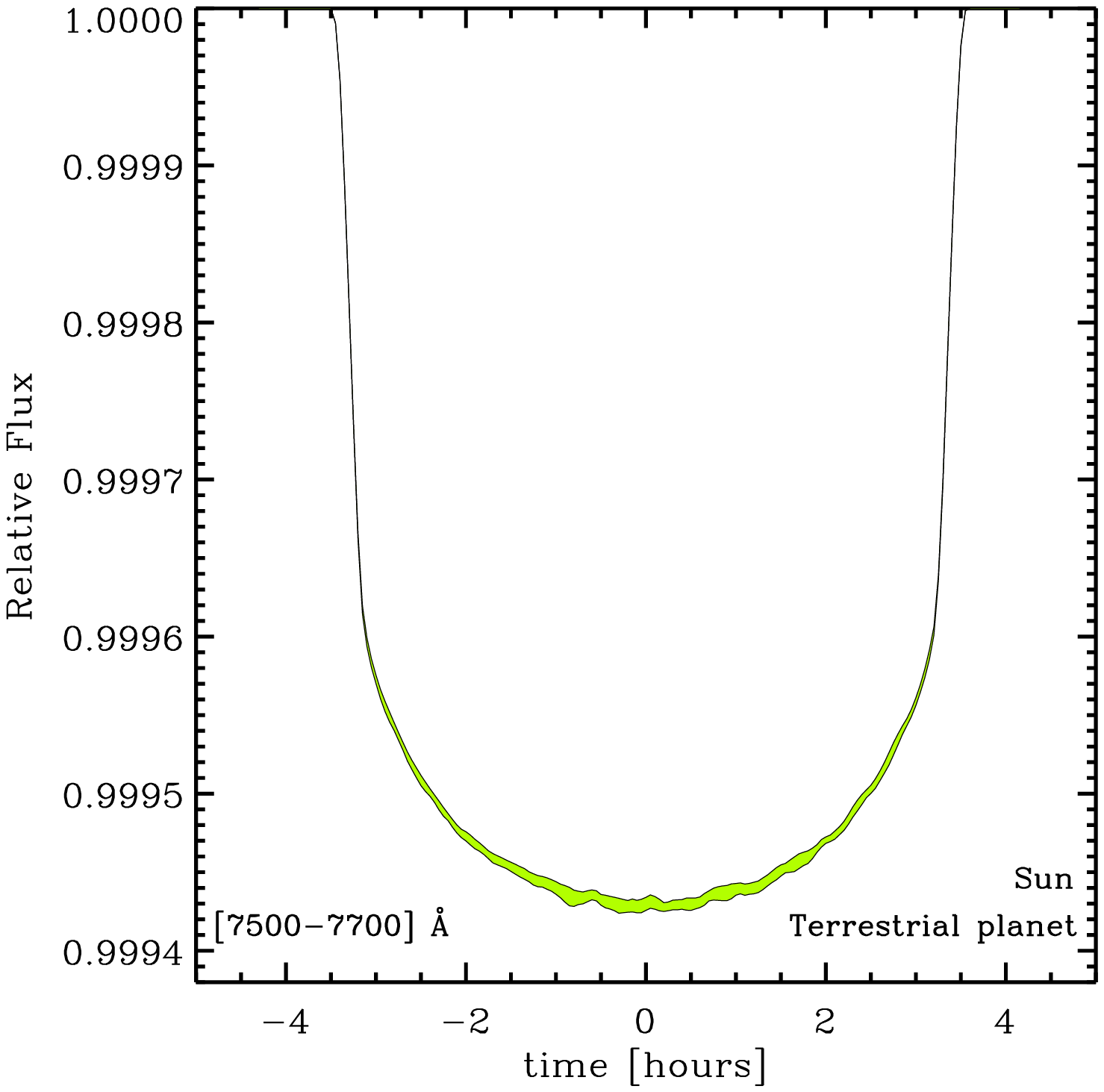} 
                        \includegraphics[width=0.33\hsize]{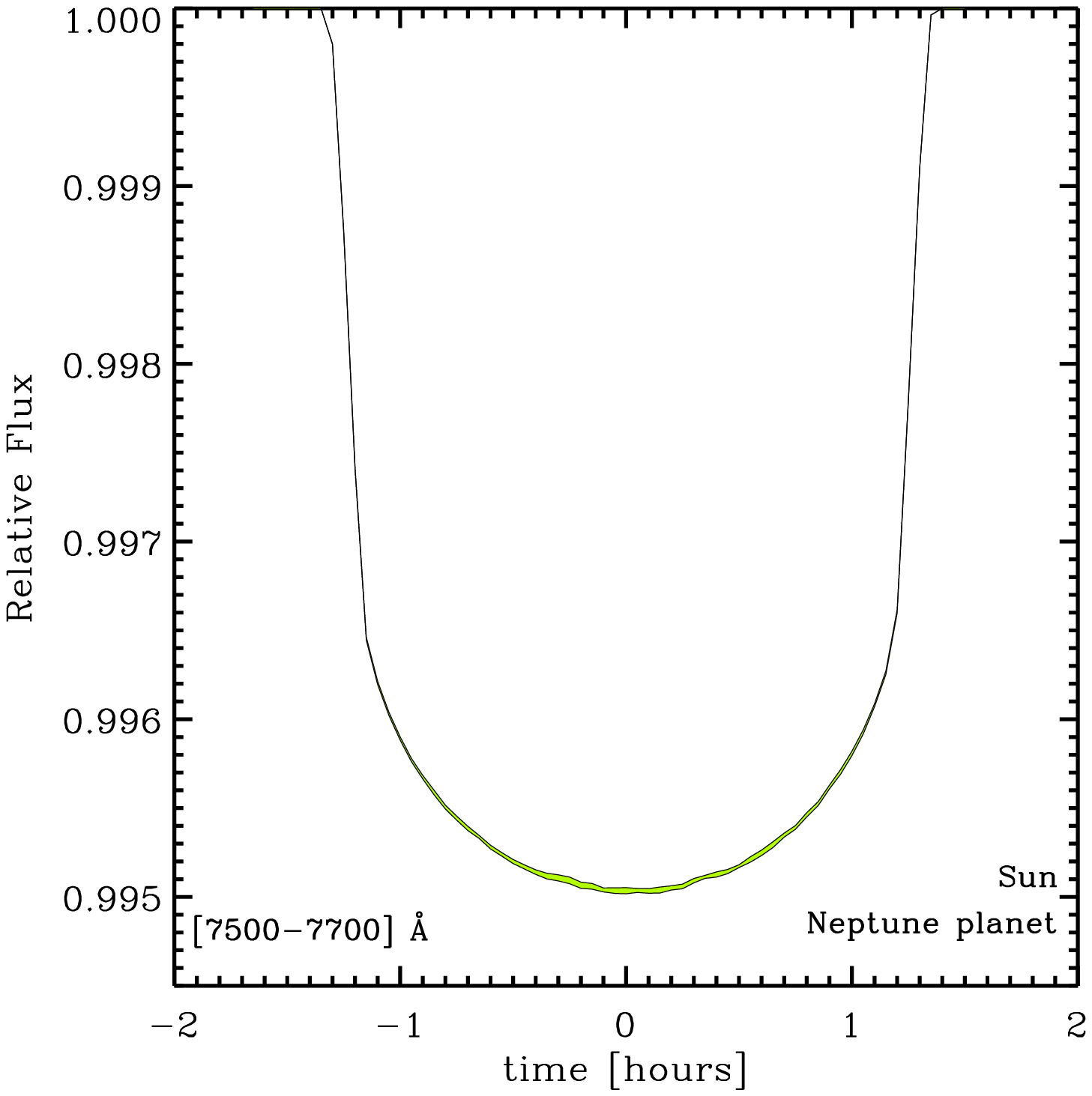}
                        \includegraphics[width=0.33\hsize]{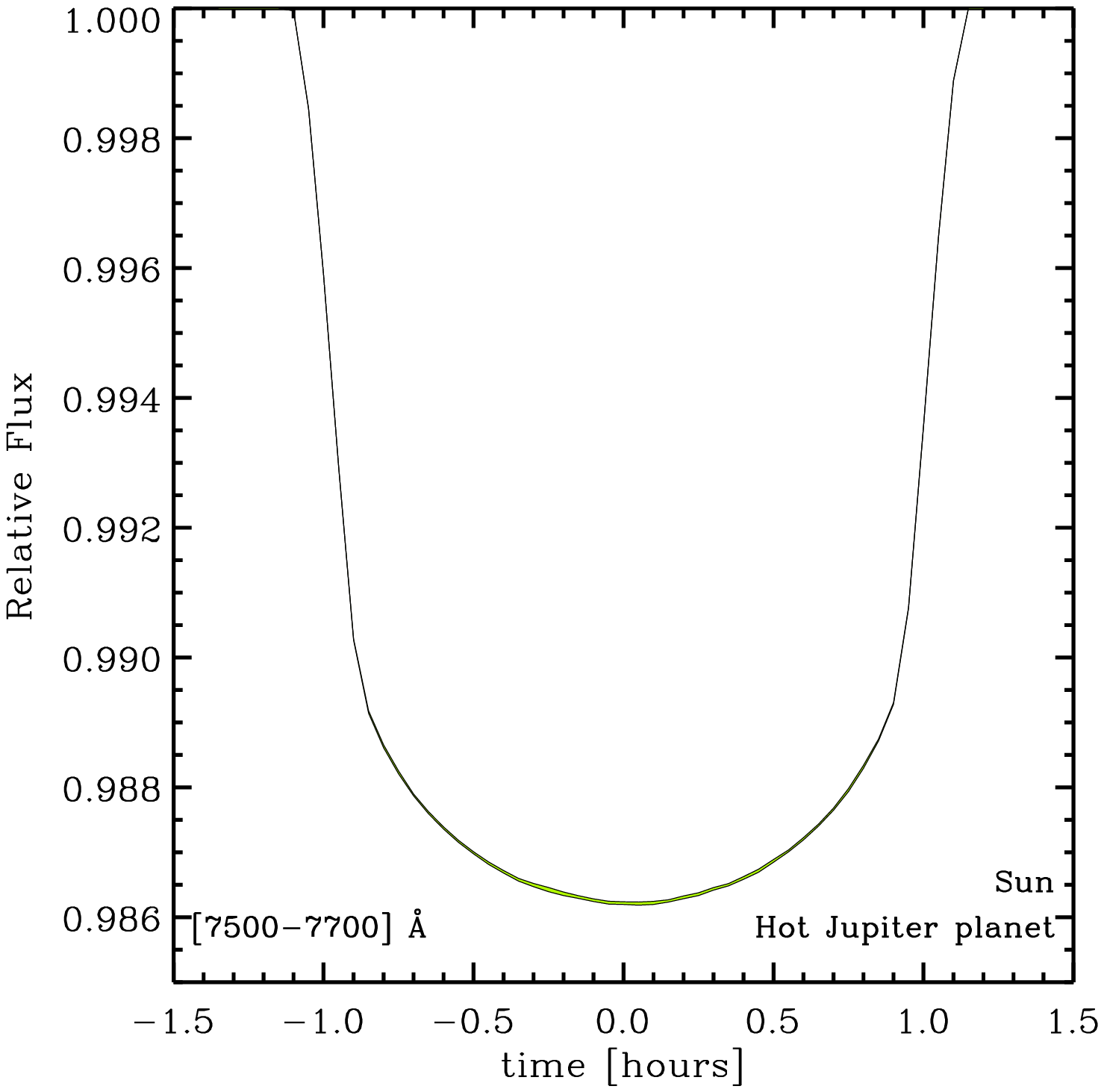}\\
                        \includegraphics[width=0.33\hsize]{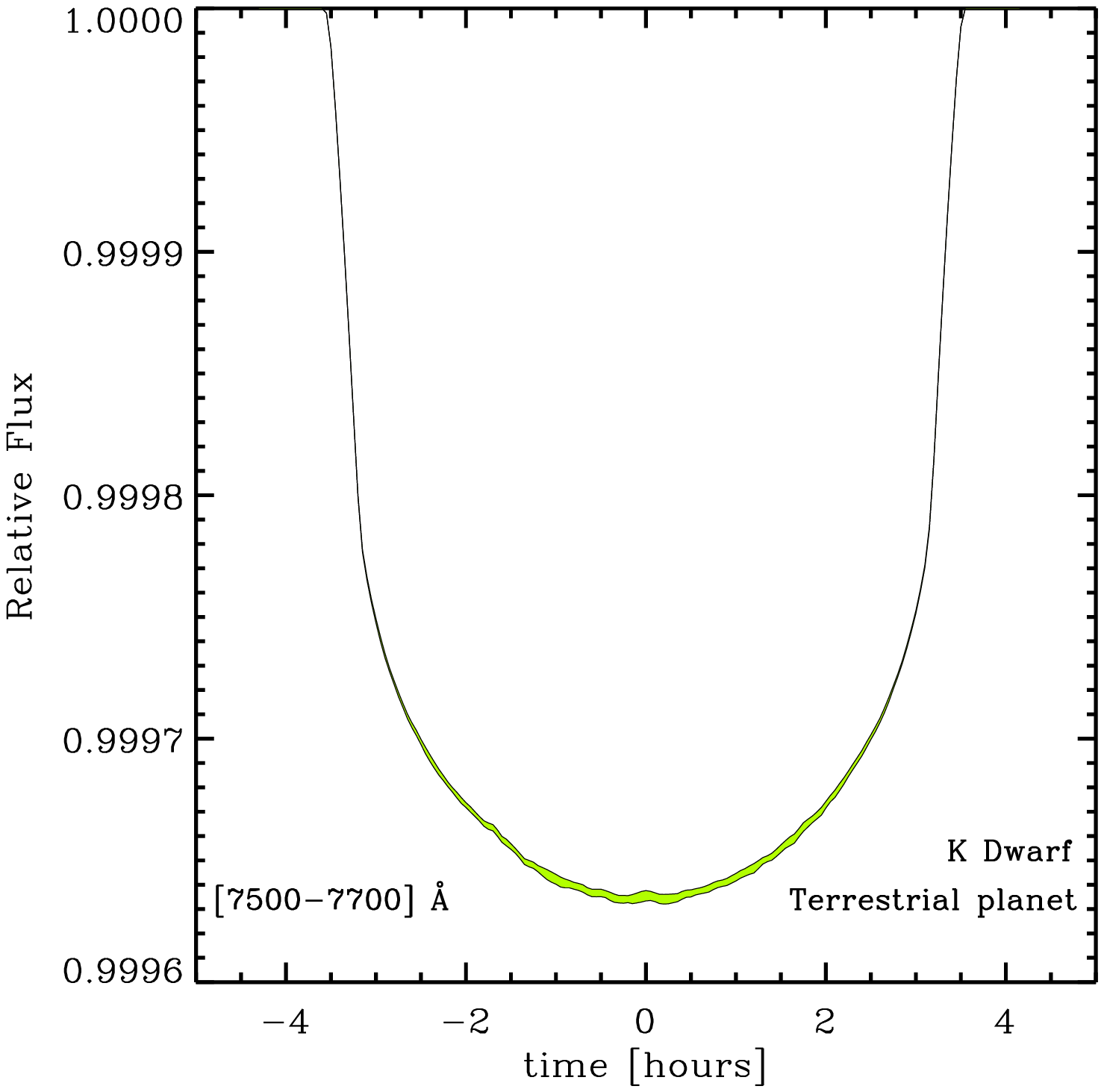} 
                        \includegraphics[width=0.33\hsize]{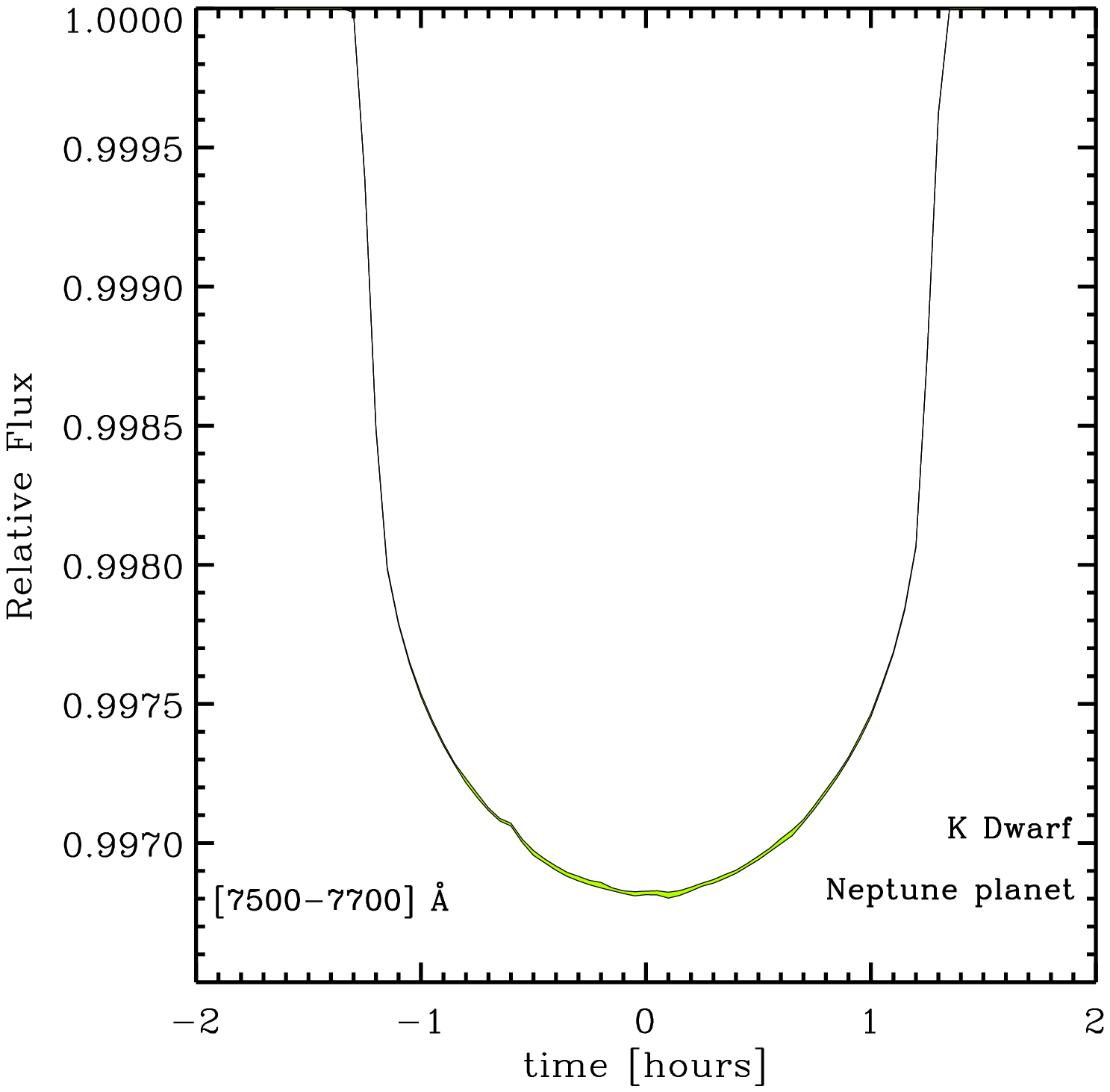}
                        \includegraphics[width=0.33\hsize]{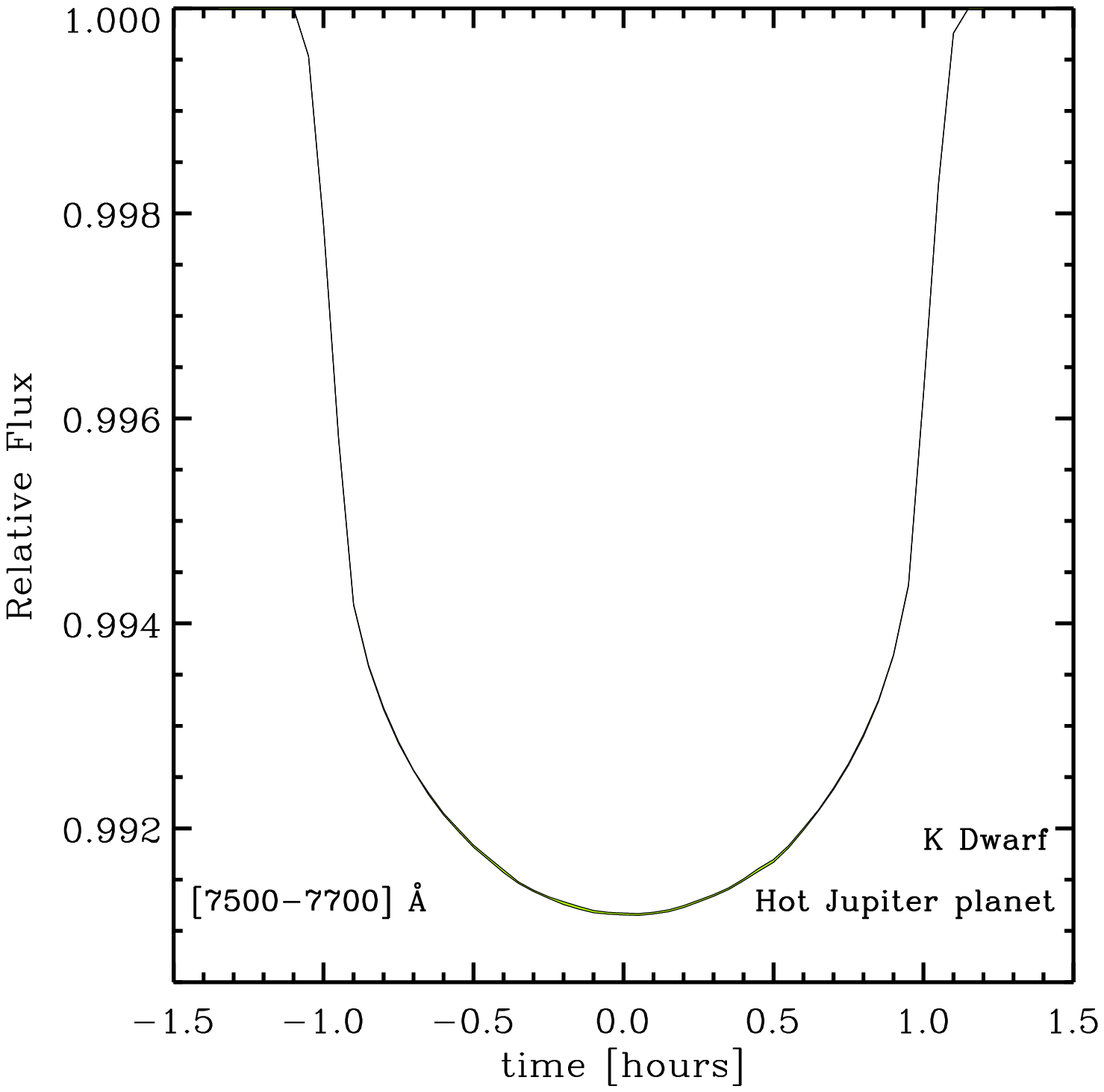}

        \end{tabular}
         \caption{Transit light curves (with the green shade denoting highest and lowest values) of 42 different synthetic Sun (top row) and K-dwarf star (bottom row) images to account for granulation changes during the transit time length for every planet (Table~\ref{planets}) and considering that the granulation timescale for the Sun is $\sim$10 minutes. The wavelength band shown is [7620-7640] $\AA$ (Table~\ref{wavelengths}).} 
        \label{transit1}
   \end{figure*}   
   
 \begin{figure*}
   \centering
   \begin{tabular}{ccc}  
                        \includegraphics[width=0.33\hsize]{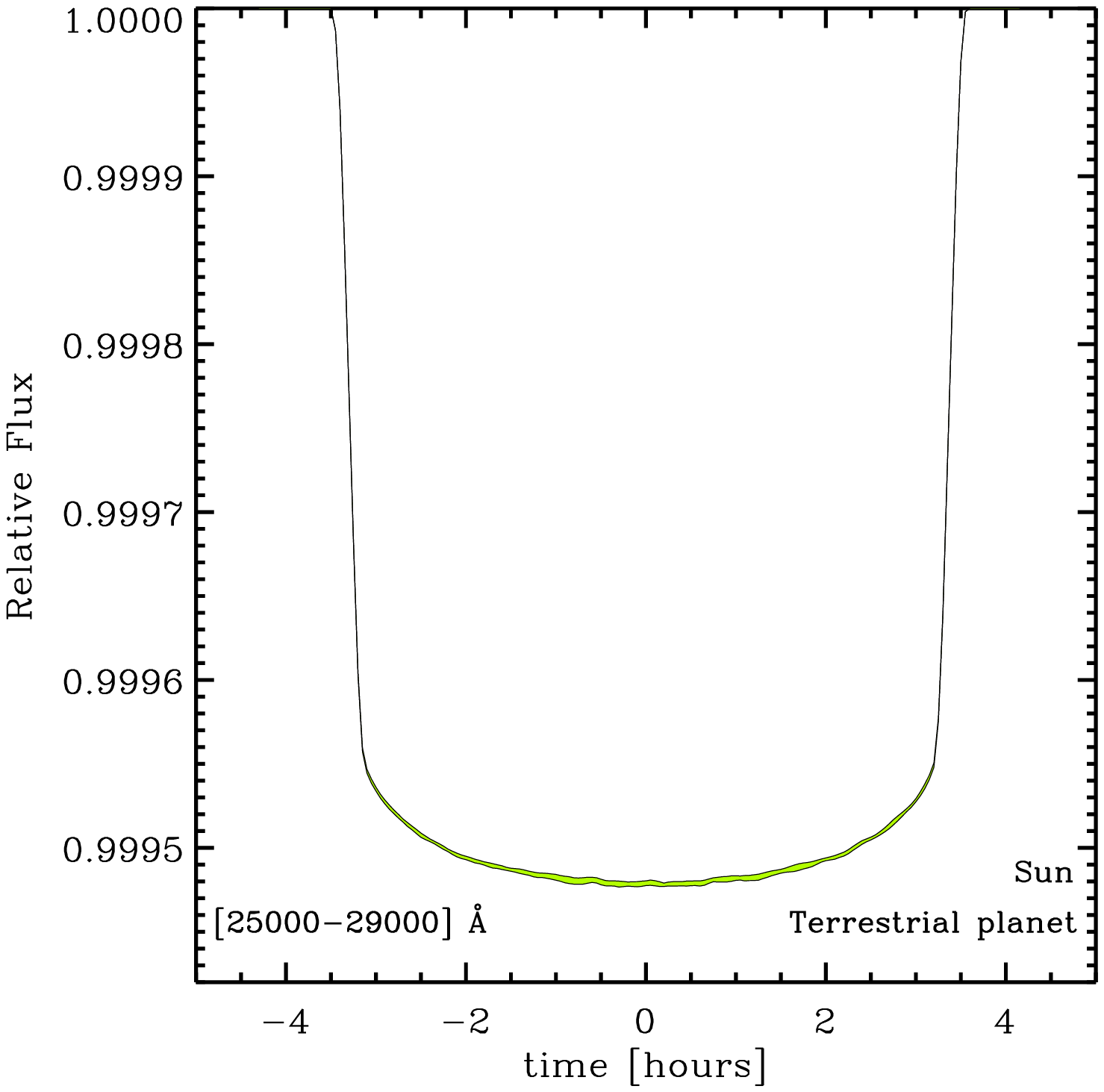} 
                        \includegraphics[width=0.33\hsize]{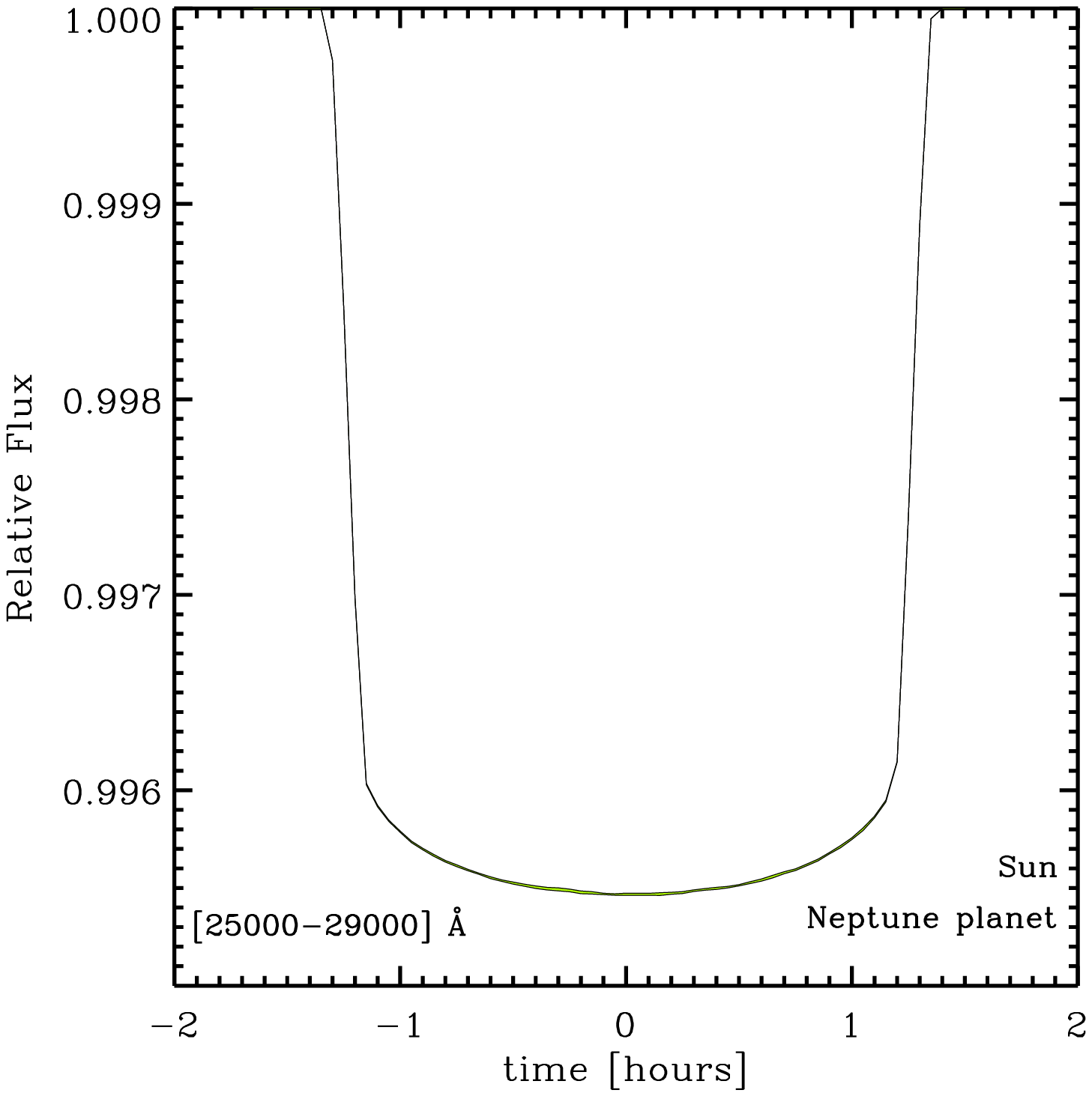}
                        \includegraphics[width=0.33\hsize]{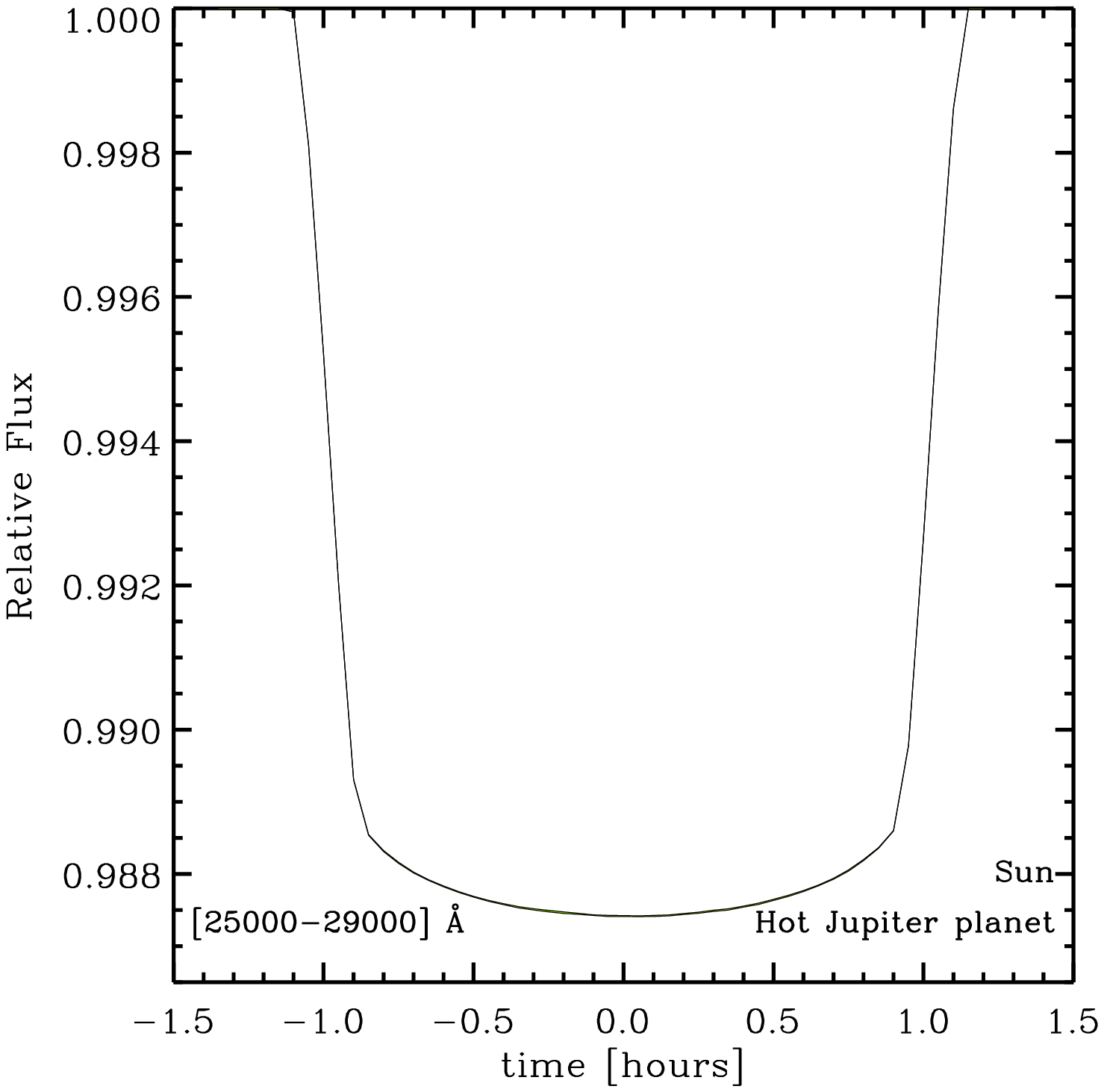}\\
                        \includegraphics[width=0.33\hsize]{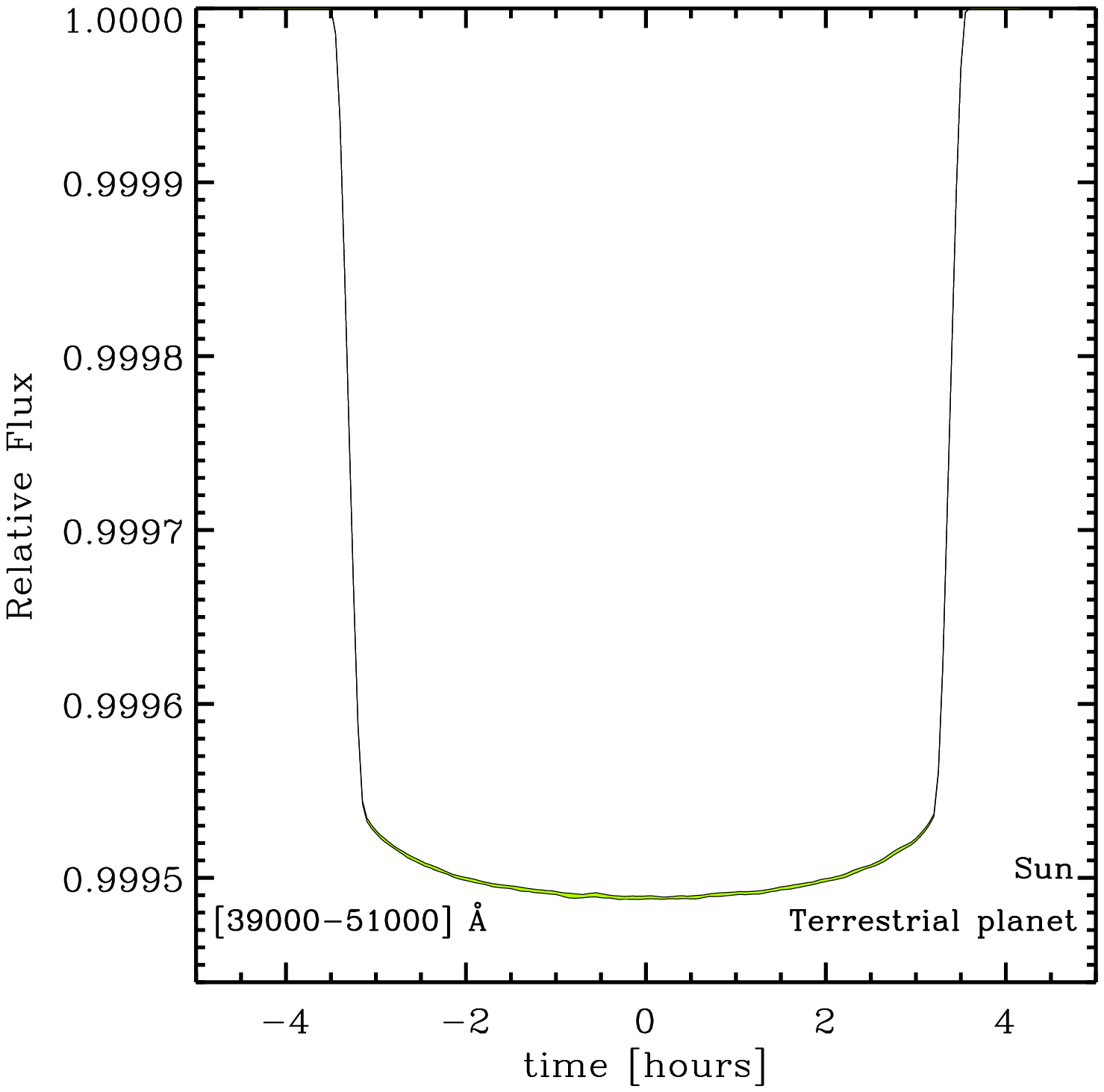} 
                        \includegraphics[width=0.33\hsize]{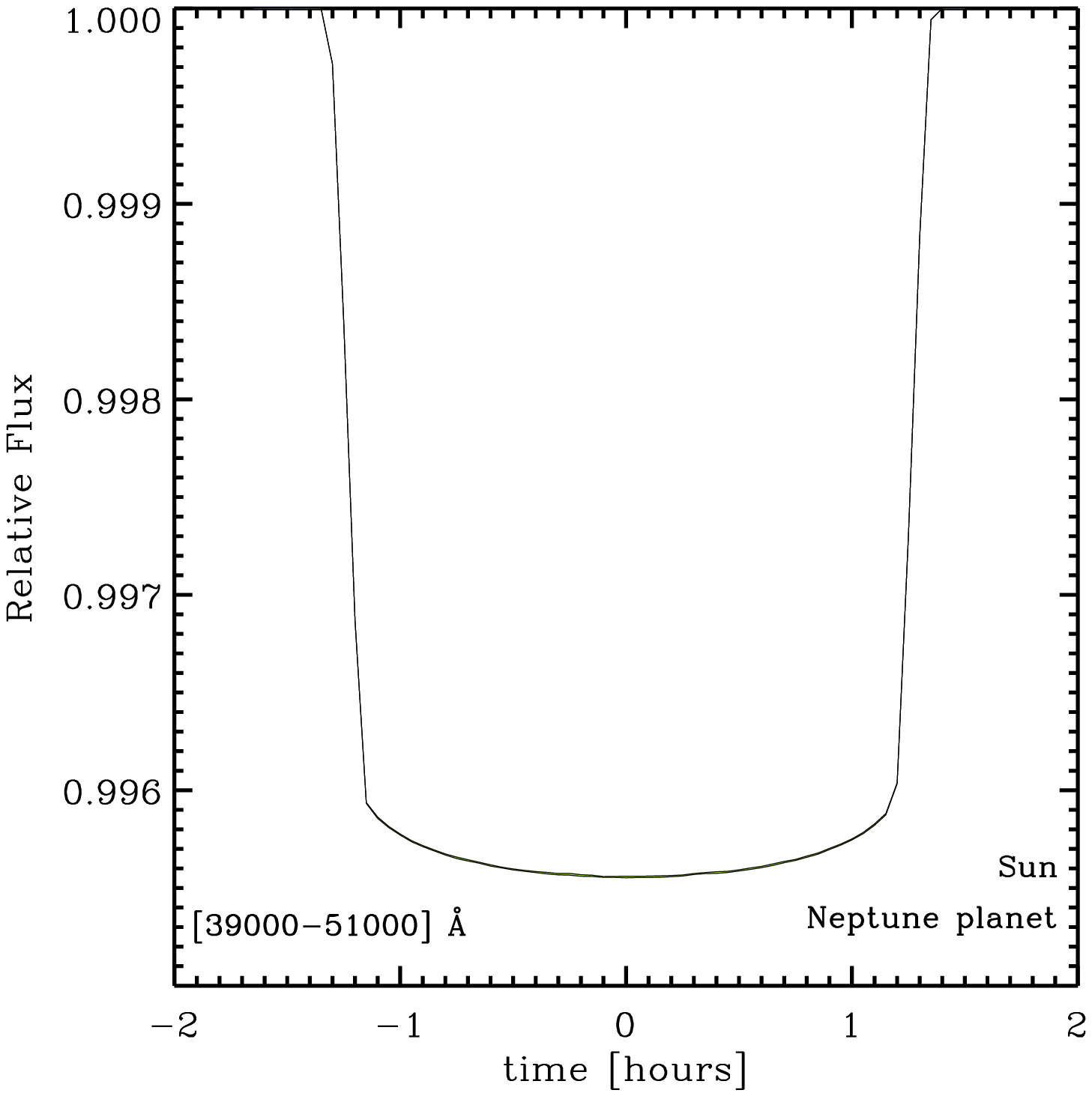}
                        \includegraphics[width=0.33\hsize]{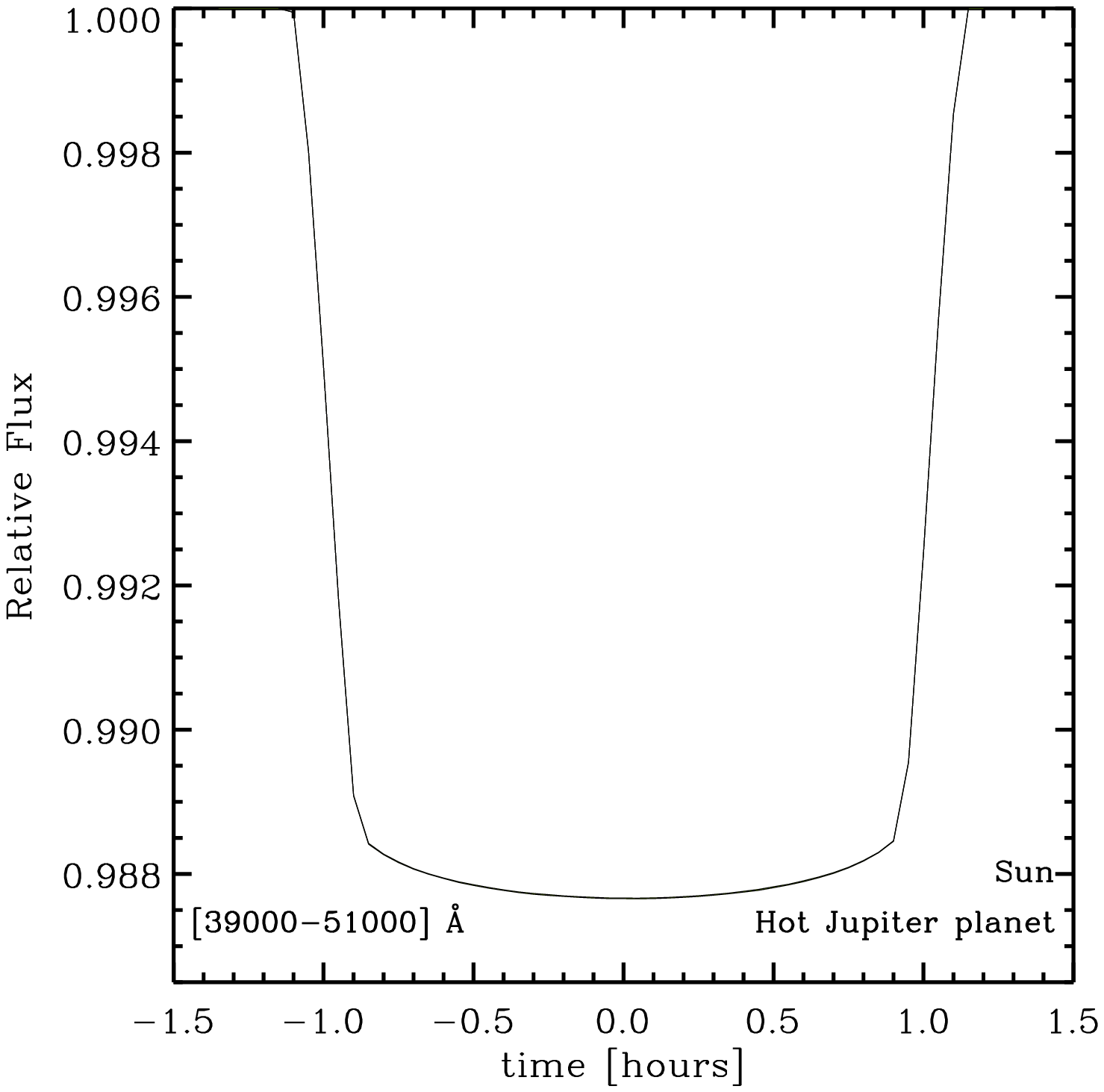}

        \end{tabular}
         \caption{Same as in Fig.~\ref{transit1} for the Sun and for the wavelength bands [25000-25250] $\AA$ (top row), and [39000-51000] $\AA$ (bottom row).} 
        \label{transit2}
   \end{figure*}

We computed light curves for the prototype planets of Table~\ref{planets} and for all the wavelength bands of Table~\ref{wavelengths}. Representative examples are shown in Figs.~\ref{transit1} and~\ref{transit2}. During a transit, the planet blocks part of the radiation of its host star. The observed dim of light is directly proportional to the ratio of the planetary and the stellar projected areas as well as to the ratio of the brightness contrast. The latter depends on the wavelength probed, owing mainly to the different Planck functions in the optical and the infrared wavelength ranges \citep[e.g., ][]{2012A&A...540A...5C}, and on the temporal variation of the granulation pattern (green shades in the transit plots and Fig.~\ref{stellardisk1bis}). 
Moreover, the depth of the curves depends of the size of the transiting planets \citep[e.g., ][]{1984Icar...58..121B} with the largest prototype planet (hot Jupiter) causing the largest transit depth (Table~\ref{transitdata}). %Also the shape of the transit at ingress/egress depends on the area hidden by the planet at the stellar limb: large planets (hot jupiter planet) show an almost symmetric "U-"shaped transit while small planets (terrestrial planet) displays more inclined ingress/egress slope with tendency to a "V-"shape. \\

The envelope of the various computed transits (green shades) in Figs.~\ref{transit1} and is affected by the granulation noise, either because during the transit the planet occults isolated regions of the photosphere that differ in local surface brightness as a result of convection-related surface structures or by the photometric fluctuations of the stellar disk (as discussed in the previous section). These two sources of noise act simultaneously during the planet transit.
In Table~\ref{transitdata} we report the maximum depth value of the different transits and the root-mean-square (RMS) of the light curves for values covering the central part of the transit periods. The RMS is the direct signature of the granulation noise. It is present for all the wavelength bands used in this work and depends on the size of the planet (larger planets return stronger fluctuations) and the wavelength probed (the optical region is characterized by stronger fluctuations with respect to the infrared wavelengths). Compared to the terrestrial planet,
hot Jupiter and Neptune planets occult larger regions of the stellar disk that differ in local surface brightness: this results in greater changes in the total stellar irradiance in the same time interval. The K-dwarf star returns weaker fluctuations than the Sun, at least at the spectral resolution considered for the calculations.\\

The granulation RMS found for the terrestrial planet transit (3.5 and 2.7 ppm for the Sun and the K-dwarf star, respectively) is close to the observed photometric variability of the SOHO quiet-Sun data, which ranges between 10 to 50 ppm \citep{2002ApJ...575..493J,1997SoPh..170....1F}. It should be noted that in our case we use narrower bands (e.g., 20 \AA\ in the optical) with respect to the very broad filter of SOHO. The accuracy of ground-based telescopes (Table~\ref{observations}) is higher than the contribution of the granulation fluctuations. On the other hand, space-based telescopes show better precision, down to $\sim$10 ppm in the optical. These values are comparable to the expected RMS of the granulation (between [3.5-15.9] ppm, with stronger fluctuations in the optical). %with the exception of the TRAPPIST telescope \citep[$\sim$300 ppm, ][]{2011Msngr.145....2J} that is close to the RMS values of large size planet, such as the hot jupiter planet and  the hot neptune planet, in the optical. 

\begin{figure}
   \centering
   \begin{tabular}{c}  
                         \includegraphics[width=0.73\hsize]{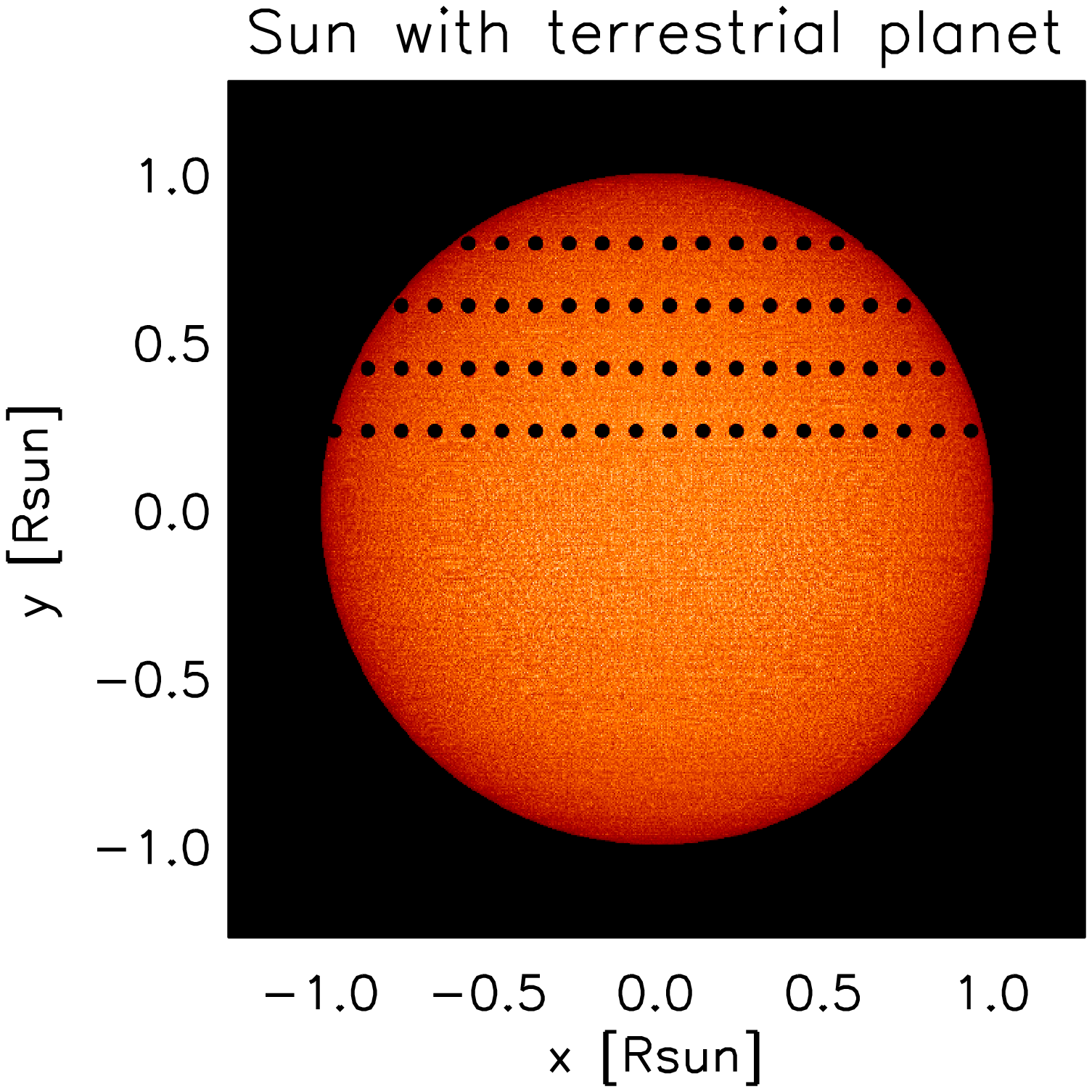} \\
                         \includegraphics[width=0.73\hsize]{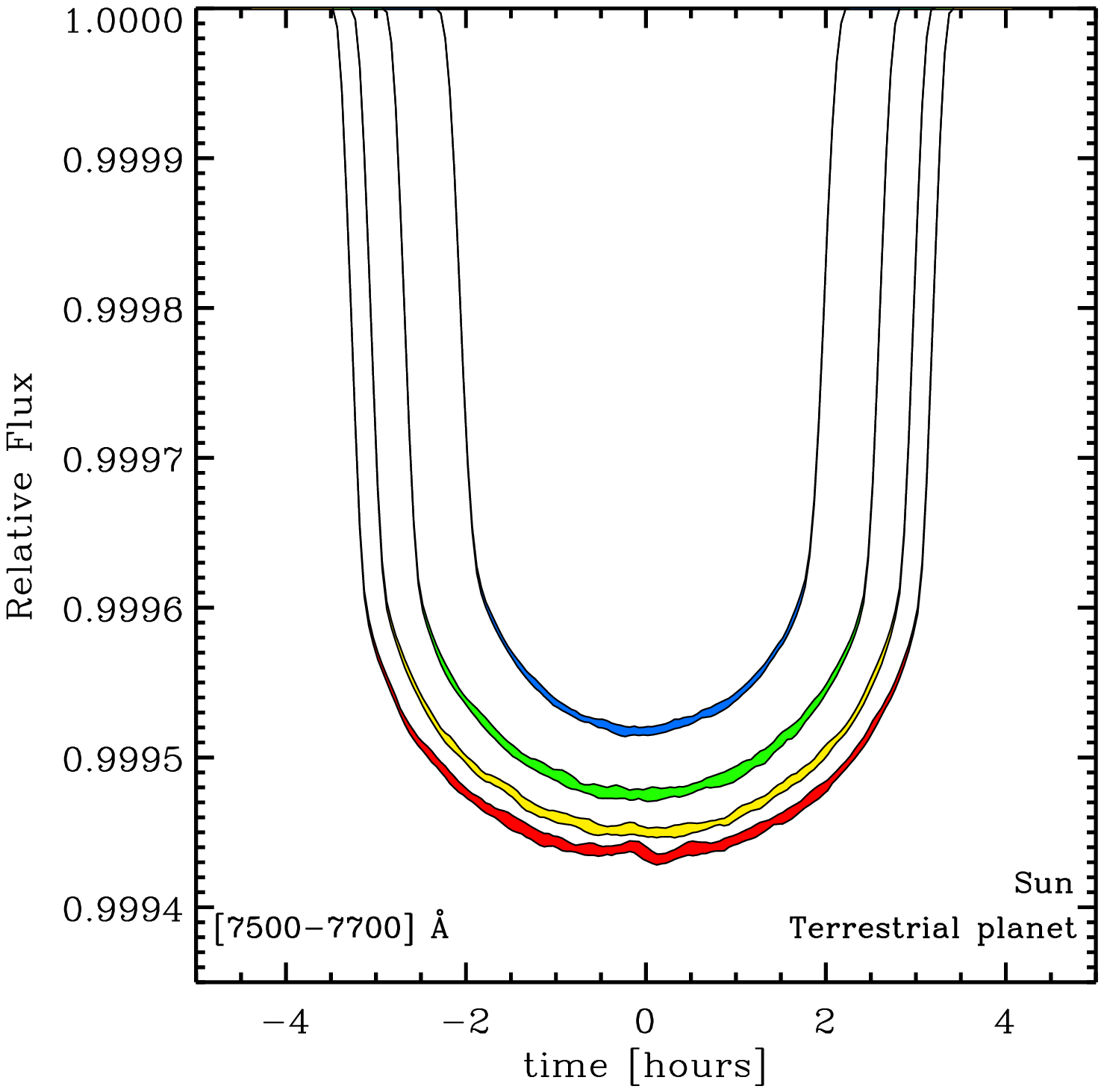}   \\
                          \end{tabular}
         \caption{\emph{Top panel:} Different transit trajectories of the prototype planet Kepler-11 on the Sun at the representative wavelength band of [7620-7640] $\AA$ and for four orbital inclination angles $inc=$[90.85, 90.65, 90.45, and 90.25]$^{\circ}$ (from top to bottom transit). An inclination angle of 90$^{\circ}$ corresponds to a planet crossing at the stellar center (Fig.~\ref{stellardisk1}). \emph{Central panel:} Transit light curves with colored shade denoting highest and lowest values of 42 different synthetic images to account for granulation changes during the transit. Blue corresponds to $inc=90.85^{\circ}$, green to $90.65^{\circ}$, yellow to $90.45^{\circ}$, and red to $90.25^{\circ}$.} %\emph{Bottom panel:} Light curves temporally averaged over the synthetic realizations from above in  the transit periods of [-3,+3] hours. The colors have the same meaning as in central panel.} 
        \label{stellardisk4}
   \end{figure}

\section{Effect of the granulation noise}                       

\subsection{Investigation on different orbital inclination angles}

In this section, we investigate the effect of the granulation pattern on different orbital inclination angles ranging from $inc$=[90.85-90.25]$^{\circ}$, with a step of 0.2$^{\circ}$. Figure~\ref{stellardisk4} (top panel) shows the orbital inclinations (decreasing $inc$ in the souther stellar hemisphere gives similar results) represented together on the same host star. Planets transiting with inclination orbits other than 90$^{\circ}$ (Fig.~\ref{stellardisk4}, bottom panel; and Table~\ref{transitdatabis}, Col. 4) have shorter transit durations, shallower transit depths, and longer ingress and egress times than the transits at 90$^{\circ}$ (i.e., transit at the stellar center). Table~\ref{transitdatabis} (Col. 4) displays the RMS of the different inclined orbits for a set of values covering the central part of the transit periods. The RMS value is correlated to the center-to-limb variation: increasing (or decreasing) the value of $inc$ amplifies the granulation fluctuations. 
%Eventually, the bottom panel of Fig.~\ref{stellardisk4} shows that there is not a clear distinction in the temporally averaged transit profiles among the diverse $inc$ chosen and for the spectral resolution used in this work.
   
   \begin{table}
%\begin{minipage}[t]{\textwidth}
\caption{Transiting curve data for the terrestrial prototype planet  of Table~\ref{planets} and the RHD simulations of Table~\ref{simus} at different inclination orbital angles ($inc$). The values reported in Col. 4 are the maximum transit depth and in Col. 5 the RMS of a set of values covering the central part of the transit period ([-1,1] hours). These values are representative for the wavelength bands in the optical (Table~\ref{wavelengths}).}             % title of Table
\label{transitdatabis}      % is used to refer this table in the text
\centering                          % used for centreing table
\renewcommand{\footnoterule}{} 
\begin{tabular}{c c c c c }        % centreed columns (4 columns)
\hline\hline                 % inserts double horizontal lines
Star  & Wavelength  & $inc$ & Depth & RMS  \\
             & [$\AA$] &    [$^{\circ}$]          & &   [ppm] \\
            \hline
Sun   & [7620-7640] & 90.85 & 0.99951 & 6.3 \\ %205 \\
                     &                      & 90.65 & 0.99947 & 4.1 \\ %157 \\
                     &                      & 90.45 & 0.99945 & 3.1 \\ %64  \\
                      &                             & 90.25 & 0.99942 & 3.0 \\ %39  \\        
\hline
K~dwarf   & [7620-7640] & 90.85 & 0.99971 & 3.7 \\ %120\NB{computation not done, these values for the k dwarf come from previous calculation of version 5. Also, the new [-1,1] RMS values are extrapoled from the above for the Sun. 6.3:205 = x : 120 ---> x = 6.3*120/205 = 6.3} \\
                     &                      & 90.65 & 0.99968 & 2.5 \\ %95 \\
                     &                      & 90.45 & 0.99966 & 2.0 \\ %41  \\
                      &                             & 90.25 & 0.99964 & 1.6  \\ % 32  \\                                 
\hline\hline                          % inserts single horizontal line
\end{tabular}
%\end{minipage}
\end{table}

\subsection{Effect on the radius of the planet}\label{limbsect}

\begin{figure}
   \centering
   \begin{tabular}{c}  
                         \includegraphics[width=1.0\hsize]{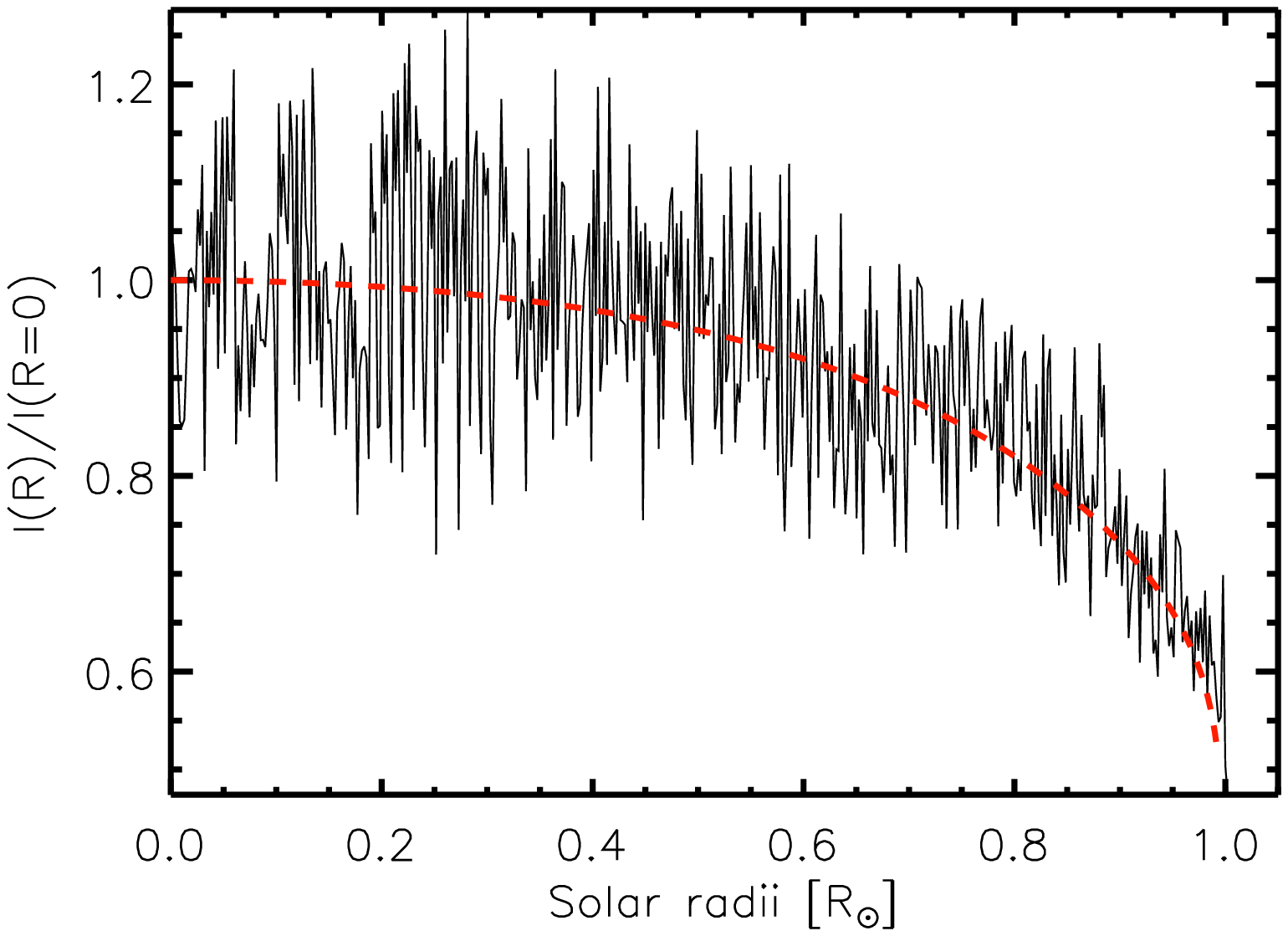}   
                          \end{tabular}
         \caption{Example of limb-darkening fit (red line) with the Claret law (see text) to the averaged intensity profiles of Fig.~\ref{intensityprofiles1}. }
        \label{limbfit}
   \end{figure}

\begin{figure}
   \centering
   \begin{tabular}{c}  
                         \includegraphics[width=1.0\hsize]{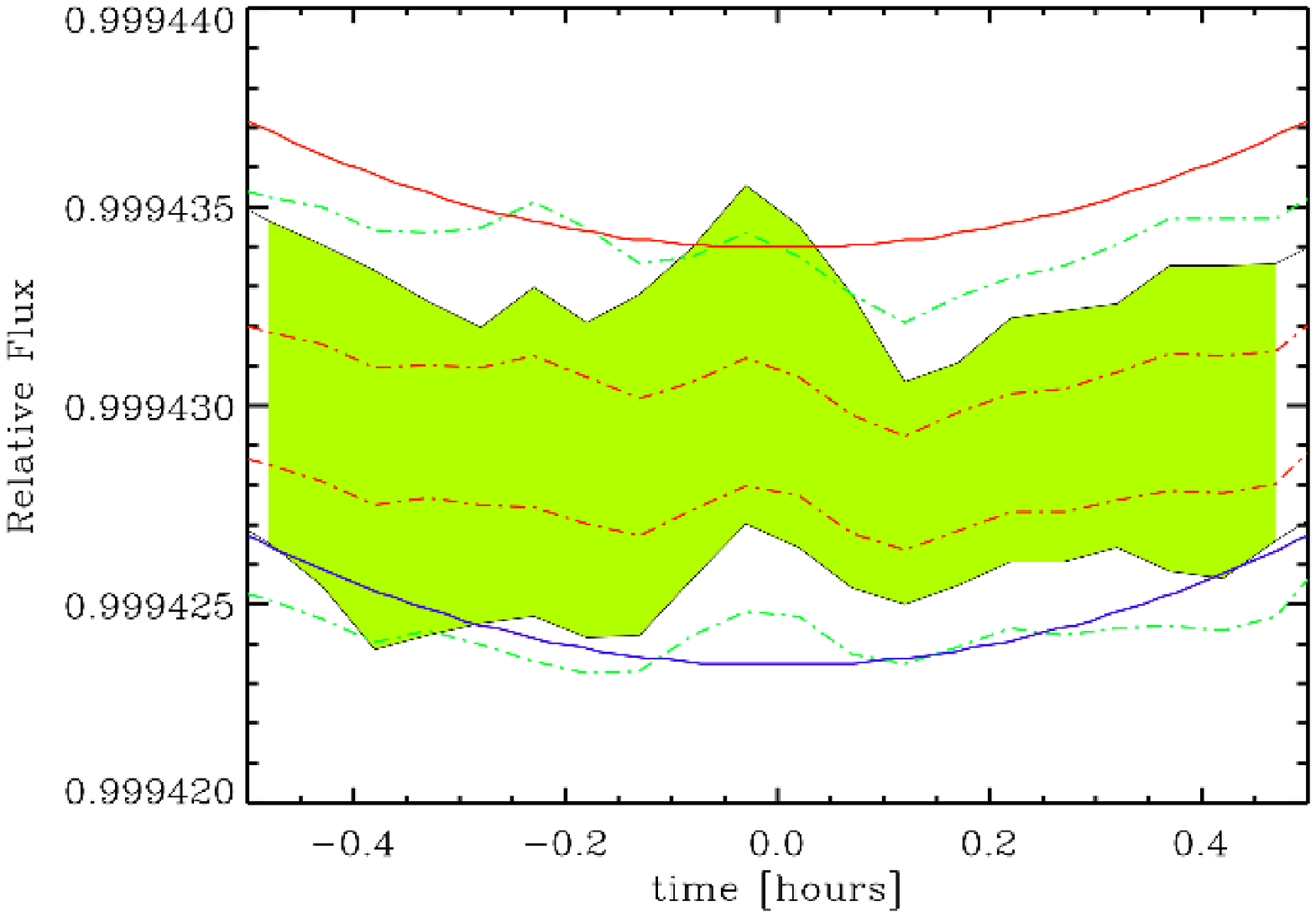}   
                          \end{tabular}
         \caption{Enlargement from Fig.~\ref{transit1} of one transit (with the green shading denoting highest and lowest values) with 42 different synthetic solar images to account for granulation changes during the transit time length for the terrestrial planet of Table~\ref{planets}. The dash-dotted red line is the 1-$\sigma$ uncertainty, while the dash-dotted light green line is the 3-$\sigma$ uncertainty on the transit distribution. The solid red and blue curves correspond to radially symmetric stellar limb-darkened disks with different planet radii to match the maximum (red)  and minimum (blue) green shading.}
        \label{radiusfit}
   \end{figure}

Our statistical approach shows that the granulation patterns of solar and K-dwarf type stars have a non-negligible effect on the light-curve depth during the transit for small and large planets. The photosphere differs in local surface brightness because of the changes in the stellar irradiance, and as a consequence, the light-curve depth varies with time. This intrinsic error affects the determination of planetary parameters such as the planet radius. To evaluate the influence of the granulation noise, we initially fitted the averaged intensity profile of Fig.~\ref{intensityprofiles1} (green line) with the limb-darkening law of  \citeauthor{2000A&A...363.1081C} \citeyear{2000A&A...363.1081C} (based on 1D model atmospheres of \citeauthor{1979ApJS...40....1K} \citeyear{1979ApJS...40....1K}): $I_\mu/I_1=1-\sum_{k=1}^4a_k\left(1-\mu^{k/2}\right)$, expressed as the variation in intensity with $\mu$-angle that is normalized to the disk-center ($I_\mu/I_1$), and Fig.~\ref{limbfit} displays an example. Then, we computed the light curves for a radially symmetric stellar limb-darkened disk by varying the planet radii of Table~\ref{planets} to match the maximum and minimum fluctuations (green shading in the transit depth of Fig.~\ref{transit1}) at the central time of the transit (Fig.~\ref{radiusfit}). In addition to this, we we also performed the same matching process to the limits set by 1-$\sigma$ uncertainty on the transit distribution. In the end, the uncertainty on the radius is calculated by dividing the maximum by the minimum matching radii. Figure~\ref{radiusfit} shows that the envelope of the various computed transits for
the terrestrial planet (green shade) falls between the 1-$\sigma$ and the 3-$\sigma$ uncertainties on the transit distribution.

\begin{table*}
%\begin{minipage}[t]{\textwidth}
\caption{Radius and uncertainty due to the granulation fluctuations for the different prototype planets of Table~\ref{planets} and 3D RHD simulations of Table~\ref{simus}. 
The values reported have been computed for transit points covering the central part of the transit periods: [-1,+1] hours for terrestrial planet, [-0.2,+0.2] hours for the hot Neptune, and [-0.15,+0.15] hours for the hot Jupiter. These values are representative for all the wavelength bands from Table~\ref{wavelengths}. Column
4 shows the radius uncertainty computed for transit depth values between the maximum and minimum transit fluctuations, while Col.
5 lists this for transit depth values between 1-$\sigma$ uncertainty  (see Fig.~\ref{radiusfit}).}             % title of Table
\label{radius_hp_corrections}      % is used to refer this table in the text
\centering                          % used for centreing table
\renewcommand{\footnoterule}{} 
\begin{tabular}{c c c c c}        % centreed columns (4 columns)1-$\sigma$
\hline\hline                 % inserts double horizontal lines
Planet  & Star  & Wavelength  & Radius Max/Min & Radius  1-$\sigma$  \\
            &    & [$\AA$] &      [$\%$]  & [$\%$] \\
            \hline
terrestrial   &  Sun & [7600-7700] & 0.90 & 0.40 \\
Neptune  &  &  & 0.43 & 0.29  \\
hot Jupiter  &  &  & 0.47 & 0.36   \\
\hline
terrestrial   &   & [25000-29000] & 0.35 & 0.13  \\
Neptune  &  &   & 0.22 & 0.11  \\ 
hot Jupiter  &  &  & 0.08 & 0.05 \\
\hline
terrestrial   &   & [39000-51000]  & 0.20 & 0.11  \\
Neptune  &  &  & 0.15  & 0.07  \\
hot Jupiter  &  & & 0.10 & 0.10  \\
\hline
\hline
terrestrial   &  K~dwarf & [7600-7700] & 0.58 & 0.22  \\
Neptune  &  & & 0.45  & 0.23 \\
hot Jupiter  &  & &  0.45 & 0.21 \\
\hline
terrestrial   &   & [25000-29000] & 0.27 & 0.18 \\
Neptune  &  &   & 0.10 & 0.07 \\
hot Jupiter  &  &   & 0.07 & 0.04 \\
\hline
terrestrial   &   & [39000-51000]  & 0.16 & 0.09\\
Neptune  &  &  & 0.10 & 0.05 \\
hot Jupiter  &  &   &  0.07 & 0.04 \\
\hline\hline                          % inserts single horizontal line
\end{tabular}
%\end{minipage}
\end{table*}

Table~\ref{radius_hp_corrections} reports the intrinsic incertitude on the planet radius  for all the prototypical planets of Table~\ref{planets} and for representative wavelengths of Table~\ref{wavelengths}. This uncertainty is given either for the envelope of the various computed transits (Col. 4) or for the 1-$\sigma$ uncertainty of the transit distribution (Col. 5). The radius uncertainty is smaller when fitting the 1-$\sigma$ uncertainty limits for all planets, but in particular for the terrestrial ones. It is strongly related to the RMS reported in Table~\ref{transitdata}: the optical region returns larger errors as well as larger RMS than those at the infrared wavelength, while the uncertainty is larger for terrestrial planets while their RMS is smaller (up to 0.90$\%$ and $\sim0.47\%$ for terrestrial and gaseous planets, respectively). The Sun returns higher values for the radius error than the K~dwarf.
%A useful quantity to describe the planet atmosphere in hydrostatic equilibrium is the scale height\footnote{the height in the planet atmosphere for which the pressure decreases by a factor $e$}, $H_p$. $H_p$ depends on the temperature as $H_p=k_b\aleph_AT_P/\left(\mu g\right)$, where $k_b$ is the Boltzmann constant, $\aleph_A$ the Avogadro number, $T_p$ the equilibrium temperature of the planet from Table~\ref{planets}, $g$ the gravity acceleration, and $\mu$ the mean molar  mass of the atmospheric gas. However, $H_p$ is also implicitly dependent on the radius of the planet ($R_p$) through the variable $g$, $H_p=3k_b\aleph_AT_p/\left(\mu G 4\pi R_p^3 \rho_p\right)$, where $G$ is the gravitational constant and $\rho_p$ the density of the planet from Table~\ref{planets}. The intrinsic incertitude on the planet radius from above, results into an incertitude on the $H_p$ reported in Table~\ref{radius_hp_corrections} (column four) that reaches up to $\sim5\%$ in the optical for the terrestrial planet and $2-2.5\%$ for the gaseous planets.

It should be noted the duration of the transits used in this work (Table~\ref{planets}) reach up to seven hours. In our analysis, longer transit durations may lead to lower but still significant
estimates for the
radius uncertainty. The effects of the granulation noise on the radius are non-negligible and should be considered for precise measurements of exoplanet transits of, in particular, planets with small diameters. The actual granulation noise is quantified in the next section. The full characterization of the granulation is essential for determining the degree of undertainty on the planet parameters. In this context, the use of 3D RHD simulations is important for estimating the amplitude of the convection-related fluctuations. This can be achieved by performing precise and continuous observations of stellar photometry and radial velocity, which are interpreted with RHD simulations, before, after, and during the transit periods.

\subsection{Light curves across wavelengths and planet sizes}

The aim of this section is to investigate how the granulation behaves across the different wavelength bands of Table~\ref{wavelengths}. For this purpose, we used the transit light curve of three representative granulation stellar disks in the optical and near-infrared region. Following the limb-darkening procedure explained in Sect.~\ref{limbsect}, we fitted the temporal averaged intensity profile (green line in Fig.~\ref{intensityprofiles1}) with the limb-darkening law of Claret 2000 and used it to generate radially symmetric stellar limb-darkened disks and, the transit light curves for a terrestrial planet (Table~\ref{planets}). For each wavelength bin, we then subtracted the light curve generated with the granulation from the smoothed limb-darkening one. Since we did not include the atmosphere in our prototype planets, the resulting signal is the noise caused by granulation.

\begin{figure}
   \centering
   \begin{tabular}{c}  
                         \includegraphics[width=1.0\hsize]{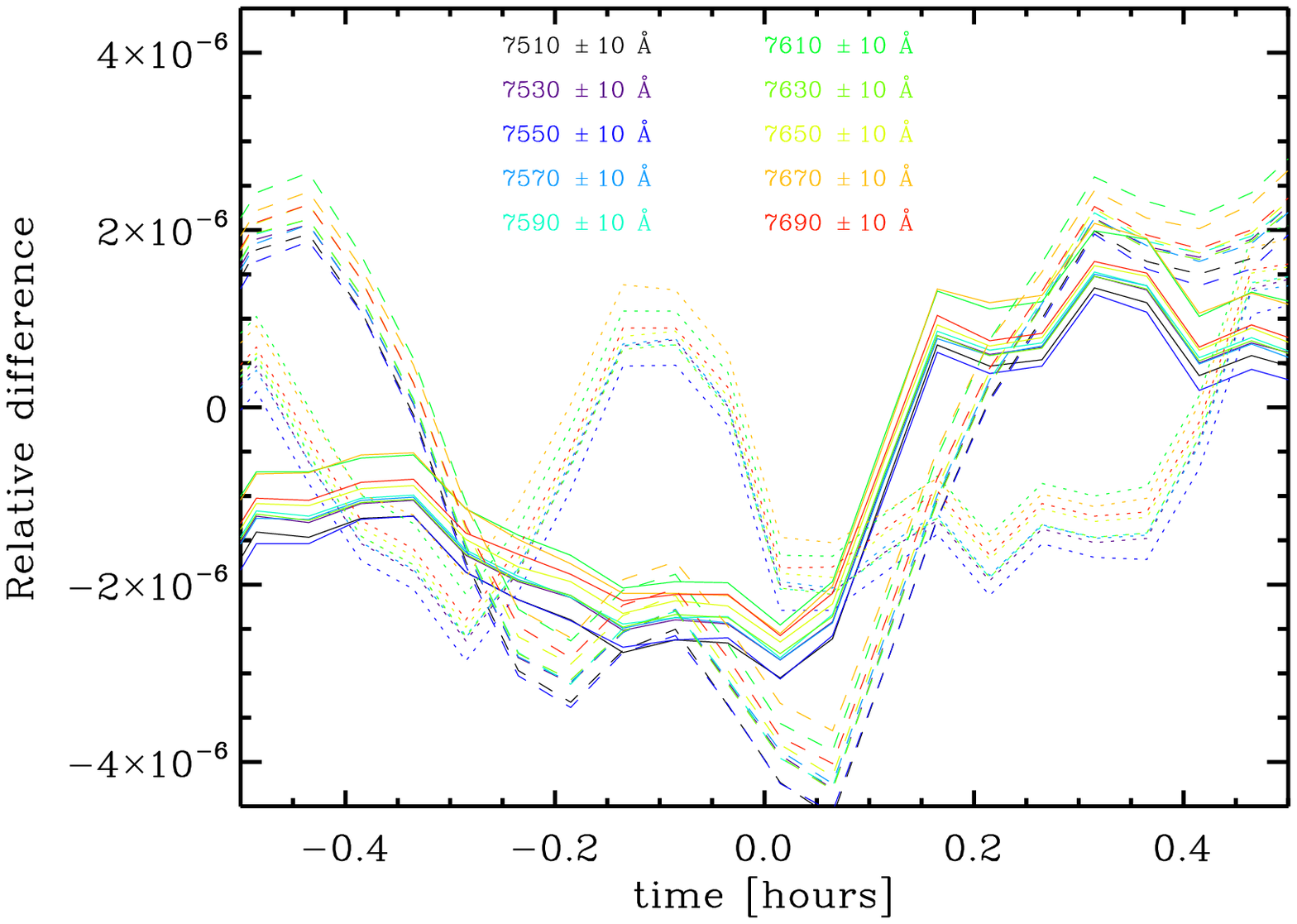}   \\
                         \includegraphics[width=1.0\hsize]{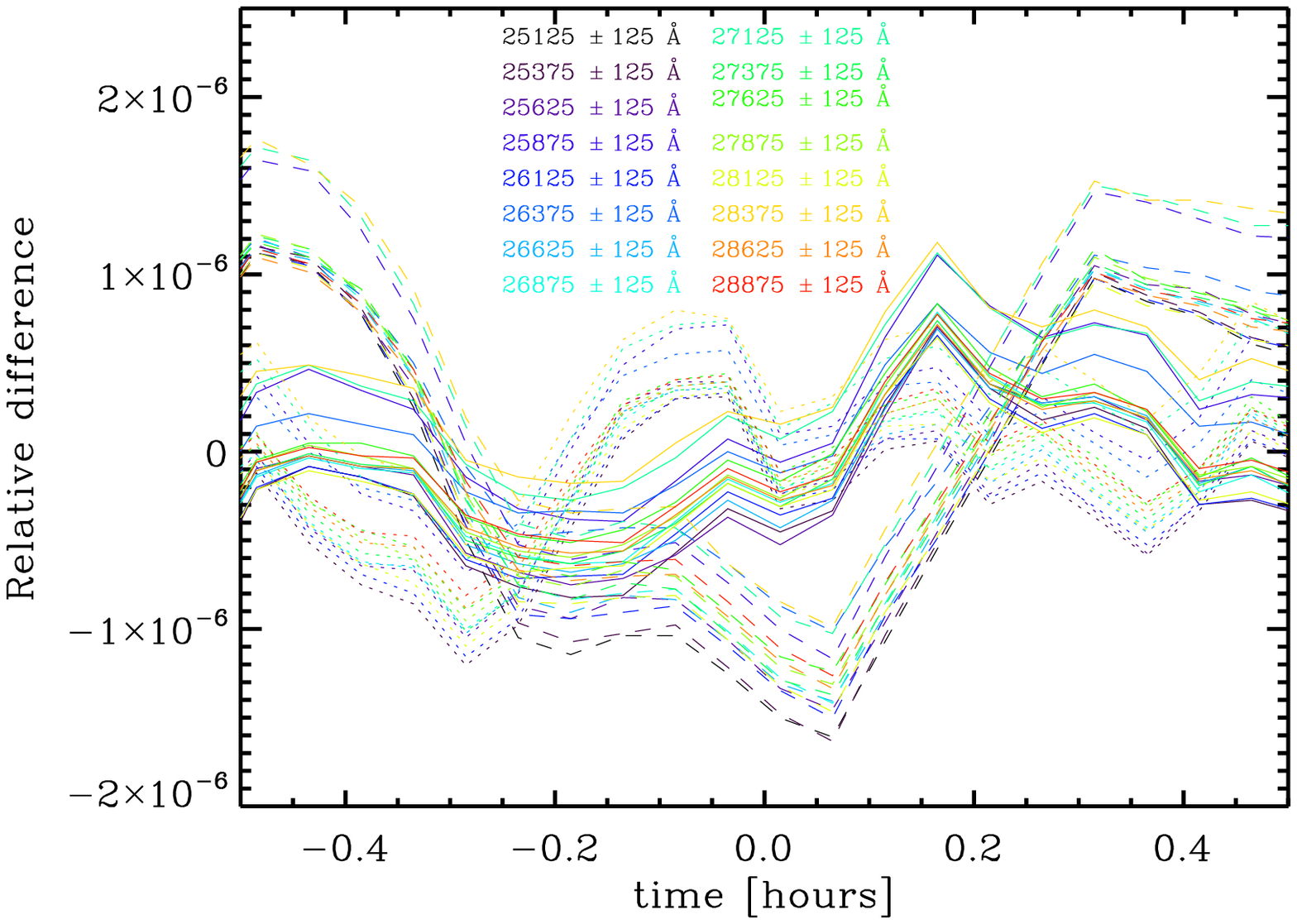} 
                          \end{tabular}
         \caption{Granulation noise for three representative granulation stellar disk realizations of the Sun affecting the central part of the light curve for a terrestrial planet (Table~\ref{planets}). The  colors denotes different wavelengths ranges (Table~\ref{wavelengths}) in the optical (top panel) and in the near infrared (bottom panel). The relative difference is obtained subtracting, for each wavelength range, the light curve generated with the granulation snapshot with the one computed with the appropriate radially symmetric stellar limb darkened disk (see text). }
        \label{wavelght_granulation}
   \end{figure}    

Figure~\ref{wavelght_granulation}  quantifies the deviations from the smoothed limb-darkening transit caused by the granulation as a function of wavelength. The amplitude of variations is $\sim 6.0 \times10^{-6}$ for the visible and larger with respect to $\sim 2.0 \times10^{-6}$ in the infrared. However, the fluctuations among the different wavelengths are stronger in the infrared (RMS $\sim 1.3 \times10^{-5}$) than in the visible (RMS $\sim 0.5 \times10^{-5}$), at least for the spectral resolution considered in this work. Figure~\ref{wavelght_granulation} also displays a  correlation among the different wavelength ranges in the visible and the infrared regions, at least for the spectral resolution used in this work. Higher spectral resolution in the wavelength bands is probably needed to isolate the contribution of the granulation effect on the stellar spectral lines. A more complete analysis on this aspect will be presented in a future work.
   
\section{Conclusion}

We used 3D RHD surface convection simulations with the \textsc{Stagger} code to provide synthetic stellar-disk images
to study the background granulation during planet transits of three prototype planets: a hot Jupiter, a hot Neptune, and a terrestrial planet.

We analyzed the effect of convection-related surface structures at different wavelengths ranging from the optical region to the far-infrared. These wavelength bands cover the range of several ground- and space-based telescopes observing planet transits and are sensitive to molecules that can give important hints on the planetary atmosphere composition. We modeled the transit light curves using the synthetic stellar-disk images obtained with the spherical-tile imaging method that was previously explained and applied in \cite{2010A&A...524A..93C,2012A&A...540A...5C,2014A&A...567A.115C,2015A&A...576A..13C}. We emulated the temporal variation of the granulation intensity, which is $\sim$10 minutes \citep{2002A&A...396.1003N} for the Sun, generating random images that cover a granulation time-series of 13.3 hours. We used the data (size, flux, and duration of the transit) of three prototype planets with the purpose of studying the resulting noise caused by the granulation on the simulated transits. From the synthetic light curves, our statistical approach shows that the granulation pattern of solar and K-dwarf-type stars have a non-negligible effect on the light-curve depth during the transit for small and large planets. This intrinsic uncertainty affects the determination of the planet transit parameters such as the planet radius (up to 0.90$\%$ and $\sim0.47\%$ for terrestrial and gaseous planets, respectively), particularly for planets with small diameters. The consequences of the granulation noise on the radius are non-negligible. The full characterization of the granulation is essential to determine the degree of uncertainty on the planet parameters. In this context, the use of 3D RHD simulations is important to estimate the amplitude of the convection-related fluctuations. This can be achieved by performing precise and continuous observations of stellar photometry and radial velocity, explained with RHD simulations, before, after, and during the transit periods.

We identified two types of noise that act simultaneously during the planet transit: the intrinsic change in the granulation pattern with timescale (e.g., 10 minutes for solar-type stars assumed in this work) is smaller than the usual planet transit ($\sim$hours as in our prototype cases), and  the noise caused by transiting planet occulting isolated regions of the photosphere that differ in local surface brightness because of convection-related surface structures. We showed that the RMS caused by the granulation pattern changes in the stellar irradiation during the transit
of the terrestrial planet (between 3.5 and 2.7 ppm for the Sun and K~dwarf, respectively) is close to what has been found by \cite{2002ApJ...575..493J} and \cite{1997SoPh..170....1F}: 10 to 50 ppm. This indicates that our modeling approach is reliable. We also showed that different orbital inclination angles with respect to transits at $inc=$90$^{\circ}$ (planet crossing at the stellar center) display a shallower transit depth, and longer ingress and egress times, as expected, but also RMS values correlated to the center-to-limb variation: granulation fluctuations increase for $inc$ different from 90$^\circ$. Finally, the granulation noise appears to be correlated among the different wavelength ranges in the visible and the infrared regions, at least for the spectral resolution used in this work. 

Three-dimensional RHD simulations are now established as realistic descriptions for the convective photospheres of various classes of stars. They have recently been employed to explain the transit of Venus in 2004 \citep{2015A&A...576A..13C}. Chiavassa and collaborators showed that in terms of transit depth and ingress/egress slopes as well as the emerging flux, a 3D RHD simulation of the Sun is well adapted to interpret the observed data. Their light-curve fit was supported by the fact that the granulation pattern changes would affect transit depth. 
Modeling the transit light curve of exoplanets is crucial for current and future observations that aim to detect planets and characterize them with this method. The good and time-dependent representation of the background stellar disk is mandatory. In this context, 3D RHD simulations are useful for a detailed quantitative analysis of the transits. 

\begin{acknowledgements}   
RC is the recipient of an Australian Research Council Discovery Early Career Researcher Award (project number DE120102940). This study has received partial financial support from the French State in the frame of the "Investments for the future" Programme IdEx Bordeaux, reference ANR-10-IDEX-03-02. The authors thank the referee for helping in finding a numerical problem during the refereeing process.
\end{acknowledgements}

%-------------------------------------------------------------------

   \bibliographystyle{aa}
\bibliography{biblio.bib}

\end{document}